\documentstyle[psfig,epsfig,graphicx,color,lscape]{mn}

\newcommand{\be}{\begin{equation}}
\newcommand{\ee}{\end{equation}}
\newcommand{\ba}{\begin{eqnarray}}
\newcommand{\ea}{\end{eqnarray}}
\newcommand{\nn}{\nonumber \\}
\newcommand{\Mpc}{\, h^{-1}\!{\rm Mpc}}

\newcommand{\thetab}{\mbox{\boldmath $\theta$}}
\newcommand{\thetabhat}{\mbox{\boldmath $\hat{\theta}$}}
\newcommand{\de}{\partial}

\newcommand{\lgl}{\langle}
\newcommand{\rgl}{\rangle}

\newcommand{\lb}{\mbox{\boldmath $\ell$}}

\newcommand{\lbh}{\mbox{\boldmath $\hat{\ell}$}}

\newcommand{\D}{\mbox{\boldmath $D$}}

\newcommand{\N}{{\cal N}}
\newcommand{\F}{\mbox{\boldmath $F$}}
\newcommand{\psib}{\mbox{\boldmath $\psi$}}

\title[The dark energy shear ratio geometric analysis]{
Probing dark energy with the shear-ratio geometric test}
\author[A.N. Taylor, T.D. Kitching, D.J. Bacon, A.F. Heavens] {
A.N. Taylor, T.D. Kitching, D.J. Bacon, A.F. Heavens
 \\
SUPA\thanks{The Scottish Universities Physics Alliance}, Institute for Astronomy, School of Physics, University of
Edinburgh, Royal Observatory, Blackford Hill, Edinburgh, EH9 3HJ,
U.K.\\
email: ant@roe.ac.uk, tdk@roe.ac.uk, djb@roe.ac.uk, afh@roe.ac.uk}
\date{}

\pagerange{\pageref{firstpage}--\pageref{lastpage}}

\pubyear{2005}

\begin{document}

\maketitle

\label{firstpage}

\begin{abstract}
We adapt the Jain--Taylor (2003) shear-ratio geometric lensing
method to measure the dark energy equation of state,
$w=p_v/\rho_v$ and its time derivative from dark matter haloes in
cosmologies with arbitrary spatial curvature. The full shear-ratio
covariance matrix is calculated for lensed sources, including the
intervening large-scale structure and photometric redshift errors
as additional sources of noise, and a maximum
likelihood method for applying the test is presented. Decomposing the
lensing
matter distribution into dark matter haloes we calculate the
parameter covariance matrix for an arbitrary experiment. Combining
with the expected results from the CMB we design an optimal survey
for probing dark energy. This shows that a targeted survey
imaging
$60$ of the largest clusters in a hemisphere with 5-band optical
photometric redshifts to a median galaxy depth of $z_m=0.9$ could
measure $w_0\equiv w(z=0)$ to a marginal $1$-$\sigma$ error of 
$\Delta w_0=0.5$. We marginalize over all other parameters including
$w_a$, where the
equation of state is parameterized in terms of scale factor $a$ as
$w(a)=w_0+w_a(1-a)$. For higher accuracy a large-scale
photometric redshift survey is required, where the largest gain in signal
arises from
the numerous $\approx 10^{14}\rm M_\odot$ haloes
corresponding to medium-sized galaxy clusters. Combined with the
expected Planck Surveyor results, such a near-future
5-band survey covering 10,000 square degrees to $z_m=0.7$ could
measure $w_0$ to $\Delta w_0=0.075$ and $\Delta w_a=0.33$. A stronger
combined constraint is put on $w$ measured at the pivot redshift
$z_p=0.27$ of $\Delta w(z_p)=0.0298$. We compare and combine the
geometric test with the cosmological and dark energy parameters
measured from planned Baryon Acoustic Oscillation (BAO) and
supernova Type Ia experiments, and find that the geometric test results
combine with a significant reduction in errors due to different
degeneracies. A combination of geometric lensing, CMB and BAO
experiments could achieve $\Delta w_0=0.047$ and $\Delta
w_a=0.111$ with a pivot redshift constraint of $\Delta w(z_p)=0.020$ at
$z_p=0.62$. Simple relations are presented that 
show how our lensing results can be scaled to other telescope
classes and survey parameters.
\end{abstract}


\section{Introduction}

Over the last decade, gravitational lensing has emerged as the
simplest and most direct way to probe the distribution of matter
in the Universe (Bartelmann \& Schneider 2001, Refregier 2003).
More recently it has become apparent that it can also be used as a
probe of the mysterious, negative-pressure ``dark energy''
component of the Universe which gives rise to the observed
acceleration of the expansion of the Universe (Hu \& Tegmark,1999;
Huterer, 2002; Jain \& Taylor, 2003; Hu, 2003; Takada \& Jain, 2003;
Song \& Knox, 2004; Ishak, 2005; Ma, Hu \& Huterer, 2006; Heavens et
al., 2006) 

The dark energy exerts its influence by its effect on the expansion
history of the Universe. If the current expansion of the Universe is
accelerating, the Universe must be older than if it was
decelerating, since the expansion was slower in the past. This
changes the distance traveled by a photon, $r(z)$, for a given
expansion factor of the Universe, $a(z)=1/(1+z)$, as photons have
had more time to travel further than in the decelerating case. The
accelerated expansion will also slow the rate of growth of matter
perturbations. The simplest phenomenological model of the dark
energy can be constructed by simply parameterizing the equation of
state of the vacuum,
 \be
        p_v= w\rho_v,
 \ee
 where $p_v$ is the dark-energy/vacuum-pressure and $\rho_v$  is its
 energy-density, and $w=w(a)$ may vary with scale factor.

Gravitational lensing depends upon both the geometry of the
Universe, via the observer-lens-source distances, and on the
growth of structure which will lens distant galaxies, and so lensing
probes both effects. Gravitational lensing is an integral effect
and so for a given line of sight these effects are degenerate with
each other and other parameters. In order to disentangle the
effects of the dark energy we require redshift information for the
source images. It has already been shown that such information can
be used to reconstruct the 3-D distribution of dark matter (Taylor,
2001; Taylor et al., 2004). For large-scale imaging surveys, the 
most practical way to get redshifts for each image is from multi-band
photometric redshift surveys. The COMBO-17 imaging and photometric
survey (Wolf et al., 2003) has already shown the power of combining
lensing with photometric redshifts (Brown et al., 2003; Taylor et
al., 2004; Gray et al., 2004; Bacon et al., 2005; Semboloni et al.,
2006). 

The parameters of the dark energy can be extracted from weak
gravitational shear measurements by taking correlations of galaxy
ellipticities at different redshifts (e.g. Bacon et al., 2005; Hu,
2003; Heavens, 2003; Heavens et al., 2006; Semboloni,
2006), where the expansion history enters both the 
lens geometry and the dark matter evolution rate. Jain \& Taylor
(2003) proposed an alternative approach, taking the ratio of the
galaxy-shear correlation functions at different redshifts. In this
case the mass of the lens dropped out leaving behind a purely
geometric quantity useful for measuring cosmological parameters.
This had the advantages of allowing the analysis to extend into the
nonlinear clustering r\'egimes where modeling the nonlinear matter
power spectrum can be inaccurate, and where the shear signal will
also be stronger, i.e. in the vicinity of galaxy clusters. In
addition, as this relies upon the correlation between galaxies and
shear many systematic effects will be averaged over, as in
galaxy-galaxy lensing. Following this a number of papers have suggested
variations on this theme (Bernstein \& Jain, 2003; Hu \& Jain, 2004;
Zhang, Hui \& Stebbins, 2005).

Geometric tests of dark energy not only complement other methods
based on the clustering of matter, but directly probe the global evolution
of the Universe via the redshift-distance relation, $r(z)$.
Other methods measure the combined effect of the growth rate of
perturbations and the global geometry. Comparison of the two can
be used to test the Einstein-Hilbert action, and extensions and
modifications of General Relativity such as extra dimensions.

While the main focus of the Jain-Taylor (2003) paper was a
statistic given by the ratio of galaxy-shear correlations (or
equivalently power spectra), they illustrated their method with
the analysis of a single cluster. In this paper we develop this
idea further and focus on applying the geometric test behind
individual galaxy clusters. The main difference between this and
the original Jain-Taylor approach is that we do not need to first
generate galaxy-shear cross-correlation functions, or cross-power
spectra, which require large data-sets. Rather the ratios used are
just of the shears behind a given cluster at fixed redshifts. This
allows the test to be applied to noisy data, since we do not need
to estimate correlation functions before applying the ratio test.
This is similar to the approach of Bernstein \& Jain (2004), who
considered a ``template matching'' approach, cross-correlating a
foreground galaxy template with the background shear pattern. Our
approach is different in that we use the galaxies to identify the
positions of lensing haloes, and then take shear ratios. In doing
so we focus on the dark matter haloes generating the signal,
allowing a halo decomposition of the matter distribution, and ask
how to maximize the signal. The price we pay for this
approach is that we become susceptible to a sampling variance due
to lensing by other large-scale structure along the line of sight,
which we can beat down using multiple lines of sight. In addition we
generalize our
methods to non-flat cosmological models.

Zhang, Hui \& Stebbins (2005) have proposed a different geometric
method, which allows them to extend the correlation/power spectrum
method to galaxy-galaxy and shear-shear correlations as well as
galaxy-shear cross-correlations. They also point out some
inaccuracies with the analysis of Jain \& Taylor (2003) and
Bernstein \& Jain (2003), which we correct here.

In the next Section we lay out the basic lensing equations we will
need. In Section \ref{ML} we derive the statistical properties of
the shear ratios, and write down a likelihood function for measuring
the dependent cosmological parameters, we then estimate the Fisher
matrix and parameter covariance matrix for the dark energy. In
Section 4 we outline the survey design formalism, using the dark
matter halo model for the distribution of galaxy clusters and group
haloes, outlining a realistic photometric redshift analysis and
discuss bias and intrinsic ellipticity issues. In Section 5 we
discuss survey strategies, considering targeted, wide-field and area
limited designs. Using the parameter covariance matrix and a model
of photometric redshifts we optimize a weak lensing photometric
redshift survey for measuring dark energy parameters from cluster
lensing in Section 6. We forecast the expected accuracy of
cosmological parameters in Section 7 and compare and combine with
other methods. In Section 8 we discuss the
required control of systematic effects, and we present our
conclusions in Section 9. We begin by introducing the necessary
cosmological and weak lensing concepts.

\section{The dark energy shear-ratio geometric method}

\subsection{Background Cosmology}

We start with the metric
\be
d \tau^2 = (1+2\Phi)d t^2 - a^2(t)(1-2\Phi)[dr^2 + S_k^2(r) d \psi^2]
\ee
where $\tau$ is the invariant proper time, $t$ is the cosmic time,
$\Phi$ is 
the Newtonian gravitational potential, $a(t) = (1+z)^{-1}$ is the
scale factor, $r(z)$ is the comoving distance given by
\be
r(z) = \int_0^z \! \frac{dz'}{H(z')},
\ee
and
\be
S_k(r) = \cases{   r_0 \, \sin (r/r_0) \quad &($k=1$)\cr
  r \quad &($k=0$)\cr
  r_0\, \sinh (r/r_0) \quad
  &($k=-1$),\cr}
\ee
is the angular distance, where $r_0 = 1 /\sqrt{|\Omega_K|}H_0$ is
the radius of curvature of the Universe and $H_0$ is the current
value of the Hubble parameter. The time-variation of the Hubble
parameter with the cosmic scale factor, $H(a)$, is given by
\be
\frac{H(a)}{H_0}= \left\{ \Omega_m a^{-3} + \Omega_K a^{-2} + \Omega_v
e^{-3 \int_1^a d \ln a' [1+w(a')]}\right\}^{1/2},
\label{hubblepara}
\ee
which is a function of the vacuum equation of state, $w(a)$, and
the present-day density parameters; the matter density,
$\Omega_m$, vacuum density, $\Omega_v$, and the energy-density
associated with the curvature, $\Omega_K=1-\Omega_m-\Omega_v$. A
useful expansion of the time-dependence of the equation of state
in terms of the expansion parameter, $a$, is (Chevallier and Polarski,
2001; Linder, 2002)
\be
w(a) = w_0 + w_a (1-a),
\ee
which evolves from $w=w_0+w_a$ at high
redshift to $w=w_0$ at low redshift, with the transition around
$z=1$. In this case the time-dependence of the Hubble parameter is
given by 
\be
\frac{H(a)}{H_0} = [ \Omega_m a^{-3} + \Omega_K a^{-2} + \Omega_v
  a^{-3(1+w_0+w_a)}e^{-3 w_a (1-a)}]^{1/2}.
\ee

\subsection{Weak Shear}
A galaxy cluster at a redshift $z_l$ will induce a shear pattern
on the background galaxies, which can be expressed in complex
notation as
 \be
    \gamma(\thetab)= \gamma_1(\thetab) + i\gamma_2(\thetab),
 \ee
where $\gamma_1$ and $\gamma_2$ are orthogonal components of the
shear field at an angle of $\thetab$. Around lensing clusters it
is convenient to use the shear tangential around the cluster
centre. This can be projected out from the total shear by
 \be
    \gamma_t =-[\gamma_1 \cos(2\varphi)+\gamma_2 \sin (2\varphi)],
 \ee
where $\varphi$ is an azimuthal angle around the centre of the
cluster.

The amplitude of the induced tangential shear distortion behind a
cluster at redshift $z_l$ will grow with redshift as
\be
\label{gammanorm}
\gamma_t(z) = \gamma_{t,\infty} \frac{S_k[r(z)-r(z_l)]}{S_k[r(z)]},
\hspace{1.4cm} z_l<z
\ee
where $\gamma_{t,\infty}$ is the tangential shear induced on a galaxy at
infinite redshift. If we take the ratio of the shear values at
two different background redshifts, $z_i$ and $z_j$, (Jain
\& Taylor, 2003) then
\be
R_{ij}=\frac{\gamma_{t,i}}{\gamma_{t,j}} = \frac{S_k[r(z_j)] S_k[r(z_i)-r(z_l)] }{
  S_k[r(z_i)] S_k[r(z_j)-r(z_l)]},
\hspace{0.4cm}
z_l<z_i<z_j.
\ee
This is the key equation describing the geometric method.
In the last term the mass and structure of the cluster has
dropped out.

In the real Universe galaxy clusters are not isolated, and
additional large-scale structure along the line of sight between
the lens and the background source galaxies will contribute to the
observed shear in both backgrounds. If we assume that the
large-scale structure is uncorrelated with the cluster then this
effect will average out over independent clusters. Defining
\be
D_{ij} = \frac{\gamma_{t,i}}{\gamma_{t,j}}
\ee
as the observed ratio of the tangential shear between two
redshifts for a given cluster we find on average that
\be
\lgl D_{ij} \rgl = R_{ij} \equiv   \frac{S_k[r(z_j)] S_k[r(z_i)-r(z_l)] }{
  S_k[r(z_i)]
  S_k[r(z_j)-r(z_l)]},
\hspace{0.4cm}
z_l<z_i<z_j.
\ee
\begin{figure}
 \psfig{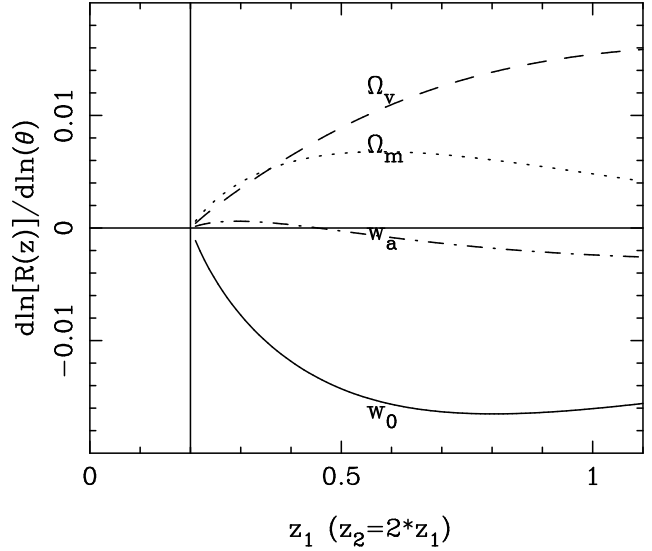}
 \caption{The response of the shear ratio, $R$, to each of the four cosmological
 parameters, $\Omega_m$, $\Omega_v$ (where $\Omega_K=1-\Omega_m-\Omega_v$), $w_0$
 and $w_a$, as a function of source redshift. We have rescaled the
 line for $w_a$ by a factor $1/w_a$ to make it finite,
 so that $\de \ln R/\de\ w_a$ is plotted.
 Here we have set $z_2=2 z_1$. The assumed fiducial model is
 $\Omega_m=0.27$, $\Omega=0.73$, $w_0=-1.0$ and $w_a=0.0$.}
 \label{R-response}
\end{figure}

\subsection{Response of Shear Ratios to Cosmological Parameters}
The intrinsic sensitivity of the shear ratio, $R_{ij}$, to a
cosmological parameter, $\theta$, can be estimated from its
logarithmic response,
 \be \frac{\Delta R_{ij}}{R_{ij}} =
        \left(\frac{\de \ln R_{ij}}{\de
        \ln \theta_i}\right) \frac{\Delta \theta}{\theta}.
 \ee
Figure \ref{R-response} shows the response of the shear ratio,
$R_{ij}$, to each of the cosmological parameters that fix the
geometry of the Universe for a lens at $z_l=0.2$, and with
backgrounds at $z_1=z$ and $z_2=2z$. From this we can see that the
response of the shear ratio to cosmological parameters is weak,
scaling roughly as
 \be
        R \propto |w_0|^{-0.02} \Omega_v^{0.01}
        \Omega_m^{0.002} e^{-0.001 w_a},
 \ee
for sources at $z_1=1$ and $z_2=2$. This weak dependence calls for
high accuracy in the shear measurements. We discuss the control of
systematics in Section 7.

The similarity of the responses of the shear ratio to different
cosmological parameters in Figure 1 also indicates their strong
degeneracies. We can expect that $w_0$ will be correlated with
$\Omega_v$ and $\Omega_m$, whilst the $\Omega_v$--$\Omega_m$ and
$w_0$--$w_a$ combinations will be anti-correlated with each other. The
similarity of the
responses of $w_0$ and $\Omega_m$ suggest these parameters will be
highly degenerate, while the differences between $w_0$ and
$\Omega_v$ at low redshift suggest these should be less
correlated. The response of $R$ to $w_0$ peaks at around $z=0.8$,
when the dark energy begins to dominate the energy-density of the
Universe.
\begin{figure}
 \psfig{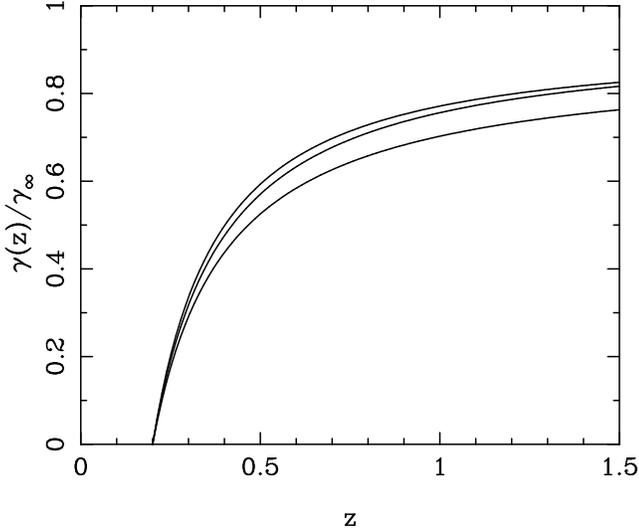}
 \caption{The tangential shear profile as a function of $w_0$ for a
 lens at $z=0.2$
 normalized relative to $\gamma(z=\infty)$, showing the effect of any shape
 changes. The lines are, from lowest to
 highest are for $w_0=-1.5,-1.0,-0.5$.}
 \label{normw0}
\end{figure}
\begin{figure}
 \psfig{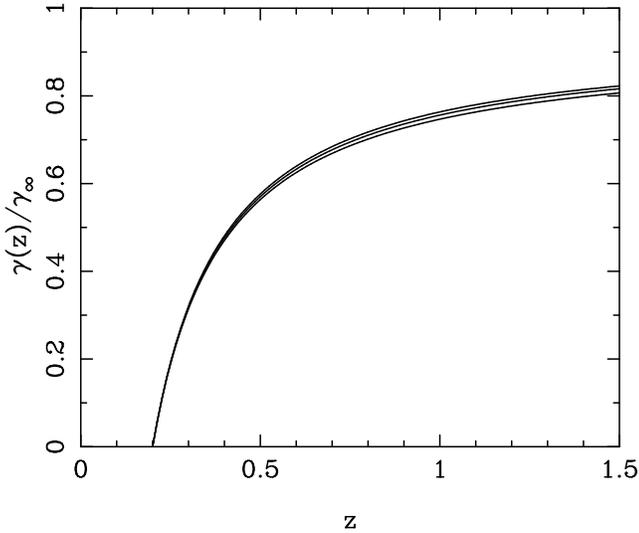}
 \caption{The tangential shear profile as a function of $w_a$ for a
 lens at $z=0.2$
 normalized relative to $\gamma(z=\infty)$, showing the effect of any shape
 changes. The lines are from lowest to
 highest are for $w_a=-0.5,0.0,0.5$.}
 \label{normwa}
\end{figure}
Interestingly, although weak, the geometric shear ratio method is
most sensitive to $w_0$. As we parameterize the dark energy equation
of state as $w(z)=w_0+w_a[z/(1+z)]$ a change in $w_0$ affects the
amplitude of $w(z)$ at all redshifts and hence affects the shape of
the tangential shear as a function of redshift at all redshifts. At
low redshift $w_a$ only changes the slope of $w(z)$ and its
amplitude at higher redshift where the effect of dark energy is less
significant.

Since we take shear ratios, we are only sensitive to changes in
the shear-redshift relation.  Figure \ref{normw0} and \ref{normwa}
show the shear as a function of $z$ normalized to unity at $z=\infty$
using 
equation (\ref{gammanorm}) for different $w_0$ and $w_a$. This
shows the effect that changes in $w_0$ and $w_a$ have on the shape
of the tangential shear. Varying both $w_0$ and $w_a$ by $0.5$, we
see that $w_0$ has a much larger effect than $w_a$ on the shape of
the shear as a function of redshift.

\section{Statistical Properties}
\label{ML}
In this Section we present a maximum likelihood approach to
measuring the geometry of the Universe from shear ratios around
individual clusters and galaxy groups. In this analysis we shall
consider shot-noise, from galaxy discreteness and intrinsic galaxy
ellipticities, the effect of lensing by large-scale structure
between the lens and the background source galaxies which will act
as an addition source of correlated and uncorrelated noise, and
photometric redshift errors.

\subsection{Likelihood Analysis}
Compressing the notation for a pair of background galaxies as
$\mu=(i,j)$ and $\nu=(m,n)$ we can write the covariance matrix for
shear ratios as
\be
C_{\nu \mu}^{RR} \equiv \lgl \Delta R_\nu\Delta R_\mu \rgl.
\ee
The log-likelihood function for the four cosmological parameters
estimated from a single cluster is then
\ba
\lefteqn{ -2 \ln L_c(\Omega_v,\Omega_m,w_0,w_a|\D) =} \nn
& & \hspace{2.cm}  \sum_{\mu,\nu} (R_\mu - D_\mu ) [C^{RR}_{\mu
    \nu}]^{-1}
(R_\nu - D_\nu).
\label{like}
\ea
Here we have further assumed that the scatter between $R_\nu$ and
$D_\nu$, due to shot-noise, photometric redshift errors and cosmic
shear from large-scale structure, is Gaussian distributed. We do
not need to assume the lensing signal from the clusters itself is
Gaussian.

If our survey contains multiple independent clusters, the total
log-likelihood is just the sum of the log-likelihoods for the individual
clusters;
\be
\ln L_{\rm TOT} (\Omega_v,\Omega_m,w_0,w_a)=\sum_{c=1}^{N_{\rm cl}}
\ln L_{c} (\Omega_v,\Omega_m,w_0,w_a|\D),
\ee
where $N_{\rm cl}$ is the number of independent clusters in the
survey.

\subsection{The Covariance of R}
The covariance matrix of shear ratios, $C^{RR}_{\nu \mu}$, is
given by
 \be
 \lgl \Delta R_{ij}\Delta R_{mn} \rgl = \lgl \Delta
    R_{ij} \Delta R_{mn} \rgl_{\rm sn} + \lgl \Delta R_{ij}\Delta
    R_{mn} \rgl_{\rm lss}, \label{cov}
 \ee
which can be decomposed into a shot-noise term due to the
intrinsic dispersion in galaxy ellipticities, and a term due to
cosmic shear induced by the intervening large-scale structure
between the lens and the two background sources.

To avoid double counting and taking ratios of the same redshift
bins, the indices in equation (\ref{cov}) are restricted to $i<j$
and $m<n$. Finally, there is a remaining degeneracy between the
shear ratios, since the ratios between any three galaxies at
redshift $z_i$, $z_j$ and $z_k$ obey the relation
 \be
        R_{ij} = R_{ik} R_{kj}.
 \ee
This reduces the total number of permutations of usable pairs of
galaxies to $(N_g-1)$, where $N_g$ is the total number of
galaxies. In practice we will bin data, in which case this also
applies to bins.

\subsubsection{Shot-Noise Covariance}
The first term in equation (\ref{cov}) is due to shot-noise,
arising from the discrete nature of galaxies and the intrinsic
dispersion in galaxy ellipticities;
\ba
\lefteqn{\frac{\lgl \Delta R_{ij}\Delta R_{mn}\rgl_{\rm sn}}{
    R_{ij}R_{mn}}=}\nn
& &  \left( \frac{\Delta \gamma_i}{\gamma_i}\right)^2
 \left(\delta^K_{im}-\delta^K_{in}\right)
+\left(\frac{\Delta \gamma_j}{\gamma_j}\right)^2
 \left(\delta^K_{jn}-\delta^K_{jm}\right),
\label{sn_cov}
\ea
where $\delta^K_{ij}$ is the Kronecker delta-function,
and
\be
\left( \frac{\Delta \gamma_i}{\gamma_i}\right)^2 =
\frac{\sigma_e^2}{2 \gamma^2_i}
\label{snvar_gal}
\ee
is the fractional variance in the tangential shear due to the
intrinsic dispersion in background galaxy ellipticity per mode,
$\sigma_e$, and $\gamma_i=\gamma(z_i)$ is the expected tangential
shear signal from the cluster for a background galaxy at redshift
$z_i$. Here we shall use $\sigma_e =0.3$ per mode.

There is a subtlety in determining the distribution of the
fractional variance for the ratio of two ellipticity measurements.
If we assume that the observed ellipticities have zero mean, and
that the distribution of intrinsic galaxy ellipticities is
Gaussian, the resulting distribution of the ratio of ellipticities
has a Cauchy/Lorentzian distribution, and so an infinite variance.
Around a lensing cluster, the mean ellipticity of the background
galaxies is non-zero, and if we assume that the mean signal is
always greater than the dispersion in the mean due to intrinsic galaxy
ellipticities the variance is finite and we can assume Gaussian
errors. This is certainly the case when we average the shear both
tangentially around a cluster and in redshift bins. Hence, instead
of working with individual galaxies we shall consider binned
galaxies, where the fractional variance in the shear is
\be
\left( \frac{\Delta \gamma_i}{\gamma_i}\right)^2 =
\frac{\sigma_e^2}{2 N_i\gamma^2_i}
\label{snvar}
\ee
per redshift bin, where $N_i$ is the number of galaxies in the
$i^{th}$ redshift bin. From hereon, the indices $i$, $j$ will
refer to bin number, rather than individual galaxies.

\subsubsection{Photometric Redshift Errors}
\label{Photometric Redshift Errors}
In current and future weak lensing surveys photometric redshifts
will also be available as an estimate of galaxy distances (see,
e.g., the COMBO-17 photometric redshift surveys, Wolf et al,
2001; CFHTLS, Semboloni et al., 2006). Here we characterize the effect of photometric redshift
uncertainty on shear ratios.

The effect of errors on the photometric estimates of galaxy
redshifts is to dilute the shear signal in each redshift bin by
randomly moving galaxies in and out of any particular bin. If we
assume that the distribution of redshift errors is a Gaussian with
width $\sigma_z(z_g)$ which depends on the true redshift of the
galaxy, $z_g$, and has a bias in the mean of the distribution
$z_{{\rm bias}}$, then
 \be
        p(z|z_g,\sigma_z) = \frac{1}{\sqrt{2
        \pi} \sigma_z(z_g)} e^{-(z-z_g+z_{{\rm
        bias}})^2/2\sigma_z^2(z_g)}.
 \ee
We shall take $z_{{\rm bias}}=0$ for all experiments, but its
effect on the marginal error of $w(z)$ will be discussed in Section
\ref{Bias in the Photometric Redshifts}, where we also discuss the
effect of a change in the variance
$\sigma_z(z)\rightarrow\sqrt{\sigma_z^2(z)+\Delta\sigma_z^2(z)}$.
We discuss the specific form for $\sigma_z(z)$ for photometric
redshift surveys in Section \ref{photozs}.

The expected shear in a redshift bin is given by
\be
\lgl \gamma_{t,i} \rgl = \gamma_{t,\infty}
\int_{z_l}^\infty \!\! dz \, n(z)
\frac{S_k[r(z)-r(z_l)]}{S_k[r(z)]}
P_{\Delta z}[z_i-z |\sigma_z(z_i)],
\label{mean_gamma}
\ee
where (e.g., Ma et al., 2005)
\ba
P_{\Delta z}[z |\sigma_z]=\frac{1}{2}\left[{\rm erf}\left(\frac{z+z_{{\rm bias}}+\Delta z/2}{\sqrt{2}\sigma_z}\right)\right]\nonumber\\
                          -\frac{1}{2}\left[{\rm erf}\left(\frac{z+z_{{\rm bias}}-\Delta z/2}{\sqrt{2}\sigma_z}\right)\right]
\ea
is the part of the redshift error distribution which lies in a
redshift bin of width $\Delta z$ centred on $z$, and ${\rm
  erf}(x)$ is the error function. The estimated shear is weighted by
the number of galaxies scattered from one redshift to another,
given by the galaxy redshift distribution, $n(z)$. Equation
(\ref{mean_gamma}) is normalized so that
\be
\int_0^\infty \! dz \, n(z) P_{\Delta z}[z_i - z |\sigma_z(z)]
=1
\label{norm},
\ee
for each redshift slice at $z_i$.

\subsubsection{Cosmic Shear Covariance}
The second term in equation (\ref{cov}), due to the cosmic
tangential shear induced by large-scale structure between the lens
and the source planes, is given by
\ba
\lefteqn{\frac{\lgl \Delta R_{ij}\Delta R_{mn}
    \rgl_{\rm lss}}{R_{ij} R_{mn}}=
  \frac{C^{\gamma_t \gamma_t}_{1,im}}{\gamma_{t,i} \gamma_{t,m}}+
  \frac{C^{\gamma_t \gamma_t}_{1,jn}}{\gamma_{t,j} \gamma_{t,n}}
  -
  \frac{C^{\gamma_t \gamma_t}_{1,in}}{\gamma_{t,i} \gamma_{t,n}} -
  \frac{C^{\gamma_t \gamma_t}_{1,jm}}{\gamma_{t,j}
    \gamma_{t,m}}}
\nn
& & \hspace{2.cm} +
\frac{C^{\gamma_t \gamma_t}_{2,i,{\rm min}(j,n)}}{\gamma^2_{t,j}}
\delta^K_{im} +
\frac{C^{\gamma_t \gamma_t}_{2,{\rm max}(i,m),j}}{\gamma^2_{t,j}}
\delta^K_{jn},
\label{lsseq}
\ea
with the same restriction on indices as for the shot-noise term. $C_1$
and $C_2$ are defined in equations (\ref{CC}), (\ref{C1}) and (\ref{C2}).
The first four terms in equation (\ref{lsseq}) are due to the
correlated distortions induced on both background galaxy images by
matter lying in front of the nearest source plane. The last two
terms arise from matter lying between the background source
planes and can be regarded as an extra noise term on the
ellipticities of the furthest background source galaxies. The
covariance of the induced tangential shear for background galaxies
at redshifts $z_i$ and $z_j$ due to large-scale structure between
the observer and the background source galaxies and averaged over
an aperture of radius $\theta$ is
\be
\label{CC}
C^{\gamma_t \gamma_t}_{\alpha,ij}(\theta)= \int_0^{\infty} \!
\frac{\ell d \ell}{ \pi} \, C^{\gamma \gamma,\alpha}_{\ell,ij}
\left\{ \frac{2[1-J_0(\ell \theta)]}{\ell^2 \theta^2}-
\frac{J_1(\ell\theta)}{\ell \theta}\right\}^2,
\ee
where $\alpha=(1,2)$. This is derived in Appendix A. Here $J_n$ is
the $n^{th}$ order Bessel function, and the angular shear-shear
power spectrum for the two source galaxies is
\be
\label{C1}
C^{\gamma \gamma,1}_{\ell,ij} =  \frac{9}{4} \Omega_m^2
H_0^4 \int_0^{r_i<r_j} \!\! dr \, P_\delta[\ell/S_k(r),r]
{\cal W}[r,r_i] {\cal W}[r,r_j],
\label{shear_power1}
\ee
and
\be
\label{C2}
C^{\gamma \gamma,2}_{\ell,ij} =  \frac{9}{4} \Omega_m^2
H_0^4 \int_{r_i}^{r_j} \!\! dr \, P_\delta[\ell/S_k(r),r]
{\cal W}^2[r,r_j]
\label{shear_power2}
\ee
for sources at redshifts $z_i$ and $z_j$, and $r_i=r(z_i)$. We
have used a nonlinear matter power spectrum, $P_\delta(k,r)$, with
a $\Lambda$CDM model with concordance parameter values,
$\Omega_m=0.27$, $\Omega_v=0.73$, $h=0.71$, using the functional
form of Eisenstein \& Hu (1999) for the linear power spectrum. The
linear power spectrum is mapped to the nonlinear r\'egime using the
fitting functions of Smith et al. (2003). The lensing weighting
function in equations (\ref{shear_power1}) and
(\ref{shear_power2}) is given by
\be
   {\cal W}[r,r_i]=  \frac{S_k(r_i-r)}{S_k(r_i) a(r)}.
\ee
In the case of binned data with photometric redshift errors,
this becomes
\be
\overline{{\cal W}}(r,r_i) =
\int_{z_l}^\infty \!\! dz \, n(z)
P_{\Delta z}[z_i - z |\sigma_z(z_i)]{\cal
  W}(r,r(z)).
\label{mean_W}
\ee
These integrals are normalized as in equation (\ref{norm}).

\subsection{Parameter Covariances}
\label{sectionpara}
This likelihood analysis can be used to extract cosmological parameter
error estimations.
The parameter covariance matrix can be calculated from the inverse
of the Fisher matrix,
\be
\lgl \Delta \theta_i \Delta \theta_j \rgl = F_{ij}^{-1},
\ee
where
\be
F_{ij}  = - \left\lgl \frac{\de^2 \ln L}{\de \theta_i \de \theta_j}
\right\rgl
\ee
is the Fisher matrix and $\thetab=(\Omega_v,\Omega_m,w_0,w_a)$ is a
vector containing our cosmological parameters (see Tegmark, Taylor
\& Heavens, 1997, for an introductory review).

For a Gaussian Likelihood function with parameters in the mean,
such as equation (\ref{like}) the Fisher matrix reduces to (e.g
Tegmark, Taylor \& Heavens, 1997) 
 \be
        F_{ij} = \frac{1}{2}\sum_{\mu,\nu} \big(\de_i R_\mu
            [C^{RR}_{\mu \nu}]^{-1} \de_j R_\nu +
                \de_j R_\mu [C^{RR}_{\mu \nu}]^{-1} \de_i R_\nu\big),\nn
 \ee
where $\de_i$ denotes differentiation in parameter space, and the
summation in $\mu$ and $\nu$ denotes summing over all
non-degenerate source configurations (see Section 3.2). The
marginalized error on the parameters is given by
 \be
        \lgl \Delta \theta_i^2 \rgl_{\rm marg} = [F^{-1}]_{ii},
 \ee
 while the conditional error is
 \be
        \lgl \Delta \theta_i^2\rgl_{\rm cond} = 1/F_{ii} \le [F^{-1}]_{ii}.
 \ee
Throughout we shall quote marginalized errors. 
Results on parameter accuracies are presented in Section \ref{paraforecast}.

\section{Survey Design Formalism}
To understand the contribution to the geometric test
signal, we use a halo decomposition of the matter density
distribution (Peacock \& Smith, 2000; Smith et al., 2003; Seljak,
2000). The full
signal then comes from integrating over all halo masses, lens
redshifts and background sources, but with the halo decomposition
we can extract information about which halo mass range contributes
most to the signal. This will help to determine optimal survey
strategies.

In this Section we shall also discuss a more detailed model for
photometric redshift errors, based on studies of photometric
redshift accuracies from the COMBO-17 survey (Wolf et al., 2003),
and the limits of ground-based measurements of galaxy
ellipticities. These elements are then factored into the
optimization of a weak lensing survey in Sections 5 and 6.

\subsection{Halo Decomposition of the Matter Density Field}
\label{Averaging over many clusters}
So far we have only considered the shear signal from a single
cluster. In practice we would want to sum over many galaxy
clusters in a weak lensing survey. In this case we need a model
for the abundance of clusters as a function of mass and redshift,
$\N(M,z)$. To apply this we must first find the relation between
mass and shear. For simplicity we shall use the singular
isothermal sphere model.

The mean shear signal inside a circular aperture of angular radius
$\theta$ for a singular isothermal sphere, and a source with
virial mass $M$ at infinity is
 \be
  \label{gammatheta}
        \overline{\gamma}_{t,\infty}(<\theta,M) =
        \frac{\theta_{\infty}(M)}{\theta},
  \ee
where
  \be
        \theta_\infty(M) = \frac{4 \pi \sigma_v^2}{c^2} = \left(
        \frac{M}{M_0}\right)^{2/3}
 \label{thetainfinity}
 \ee
is the Einstein radius for a source at infinity. In the last
expression we have made use of the constant virial velocity of the
singular isothermal sphere,
 \be
 \sigma_v^2 =\frac{3GM}{2r_v},
 \label{sigmav}
 \ee
where the virial mass, $M$, is the mass enclosed by the virial
radius, $r_v$;
 \be
 M = \frac{4 \pi}{3} r_v^3 \overline{\rho}_m \delta_v.
 \label{rv}
\ee
Here $\overline{\rho}_m$ is the mean mass-density of the Universe,
$\delta_v=340$ (Eke et al. 1996) is the virial overdensity for a
$\Lambda$CDM Universe, and
 \be
 M_0  =\frac{c^3}{ \pi^2 \sqrt{288  G^3 \overline{\rho}_m
 \delta_v}}
 \ee is a characteristic mass. Equation (\ref{rv}) defines the
virial radius, $r_v$, in terms of the virial mass, $M$, and is
given by
 \be
        r_v = 0.293 \left(
        \frac{M}{10^{13}M_\odot}\right)^{1/3} (\Omega_m h^2)^{1/3} {\rm
        Mpc}.
 \ee
Substituting this into the expression for the velocity dispersion,
$\sigma_v$, in equation (\ref{sigmav}) and then into the
expression for $\theta_\infty$ (equation \ref{thetainfinity}) we
find the shear signal scales as $\gamma_t \propto M^{2/3}$. As
more massive clusters are larger their surface mass-density, and
hence mean shear, scales more slowly than in proportion to the mass, as
would be expected for fixed sized haloes.

The shot-noise term in the shear covariance matrix for sources at
infinity is
 \be
    \frac{\Delta \gamma^2}{\overline{\gamma}_{t,
 \infty}^2} = \frac{\sigma^2_e}{n_z(>z_l) \pi \theta_\infty^2},
 \label{shear_lim}
 \ee
where $n_z(>z_l)$ is the surface density of galaxies that lie at a
redshift greater than the lens at $z_l$. The angular radius,
$\theta$, drops out of this expression since the signal-to-noise
ratio for the mean shear of an isothermal sphere is a constant for
a uniformly distribution of background sources. Hence the Fisher
matrix will scale as $F_{ij} \propto \theta_\infty^2 \propto
M^{4/3}$.

The Fisher matrix for the halo geometric test, integrating over
lensing cluster mass, $M$, and lens redshift, $z_l$, is
 \be
 F_{ij}
 = \int_0^\infty \! \!\frac{d z_l}{H(z_l)} \,
 \left[\int_{M_{\!-}(z_l)}^\infty \! dM
  \,{\cal N}(M,z) \left( \frac{M}{M_0} \right)^{4/3}\right]
 F_{ij}(M_0,z_l)\\
 \ee
where have factored out the mass-dependency of the Fisher
matrix and set $\theta_\infty=1$ in $F_{ij}(M_0,z_l)$ for a single
halo. In this expression ${\cal N}(M,z_l)$ is the number density
of clusters per $[\Mpc]^{3}$ with mass $M$ at redshift $z_l$. The
lower mass cut-off in the integral over mass, $M_-(z_l)$, is set
by the condition that a cluster shear must be measurable with a
signal-to-noise of
\be
\frac{\gamma}{\Delta \gamma}>\mu,
\ee
which sets
\be
\label{thresholdeq}
M_-(z_l) = M_0 \left[\frac{\sigma_e \mu }{ \sqrt{n_z(>z_l)
    \pi}}\right]^{3/2}.
\ee
We shall assume $\mu=1$ from hereon. Note that although we have a
low signal-to-noise threshold for measuring the shear signal from
a given halo, we assume that the detection of a halo is based on
the detection of galaxies in the halo, and therefore has a high
signal-to-noise.

The halo number density is a function of mass, $M$, and redshift,
$z$, given by
\be
\label{NMZ}
\N(M,z) = \frac{\overline{\rho}_m}{M} f(M,z),
\ee
where fraction of matter, $f(M,z)$, in haloes of mass $M$ at
redshift $z$, can be written in the universal Sheth-Tormen form
(Sheth \& Tormen, 1999),
\be
\nu f(\nu) = B \left(1+\nu/\sqrt{2}\right)^{-0.3}
(\nu/\sqrt{2})^{1/2} e^{-\nu/2\sqrt{2}},
\label{sheth-tolman}
\ee
where $B$ is a constant of normalization so that
\be
\int_0^\infty \! d \nu \,  f(\nu) = 1,
\ee
and
 \be
    \nu = \frac{\delta_c^2}{\sigma^2(M,z)}.
 \ee
The form of equation (\ref{sheth-tolman}) finds justification from
the ellipsoidal collapse model of haloes (Sheth et al, 2001). The
collapse threshold for linear matter overdensities, $\delta =
\delta \rho/\overline{\rho}$,  is $\delta_c=1.686$. The variance
of overdensities in spheres of radius $R(M)$ is
 \be
 \sigma^2(M,z) = \int_0^\infty \! \frac{d^3k}{(2
   \pi)^3}\,
 P_{\rm lin}(k,z) j_0^2[k R(M)],
 \ee
where
 \be
        P_{\rm lin}(k,z) = D^2(z) P_{\rm lin}(k,z=0)
 \ee
is the linear matter power spectrum.  This can be split into a
linear growth factor,
 \be
 \label{GROWTH}
        D(z) =H(z) \int_z^{\infty} \frac{(1+z') dz'}{H(z')^3},
 \ee
(see Heath, 1977; Carroll, Press \& Turner, 1992; Linder, 2003)
where $H(z)$ is given by equation (\ref{hubblepara}), and the
present-day linear matter power spectrum, $P_{\rm lin}(k,z=0)$. This
equation is only valid for $w_0=\{-1$, $-1/3$, $0\}$. We use
it here as $\Lambda$CDM is our fiducial cosmology.
The spherical Bessel function, $j_0(x)=\sin (x)/x$ is the
transform of a sphere. The radius of the sphere, $R(M)$, is the
linearized radius of a cluster halo of mass $M$, which will
collapse down to a nonlinear virial radius, $r_v$, and is given by
 \be
        R(M)=(1+z) \delta_{\rm v}^{1/3} r_v(M).
 \ee

\begin{figure}
 \psfig{figure=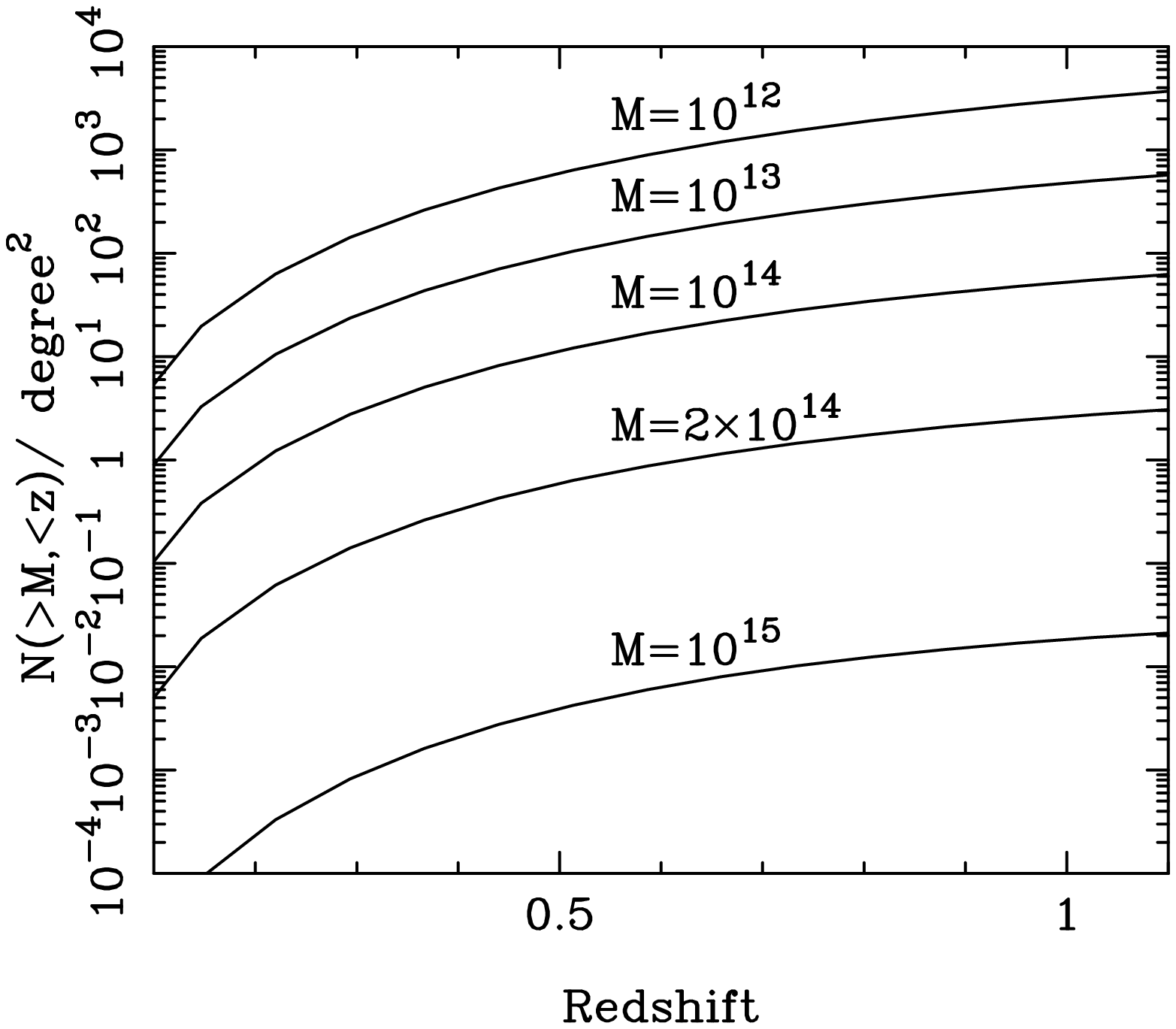,width=\columnwidth,angle=0}
 \caption{The cumulative number count of dark matter haloes
 per square degree, $\N(>M,<z)$, as a function of redshift,
 for the mass range $M=10^{12}M_\odot$ to $M=10^{15}M_\odot$.}
 \label{numberdensity_z}
\end{figure}

We plot the cumulative number count of dark matter haloes per
square degree as a function of redshift, for a range of halo
masses in Figure \ref{numberdensity_z}. Typically, for
$10^{12}M_\odot$ haloes we expect $10^3$ haloes per square degree,
while for $10^{15}M_\odot$ haloes we expect $10^{-2}$ haloes per
square degree.

\begin{figure}
 \psfig{figure=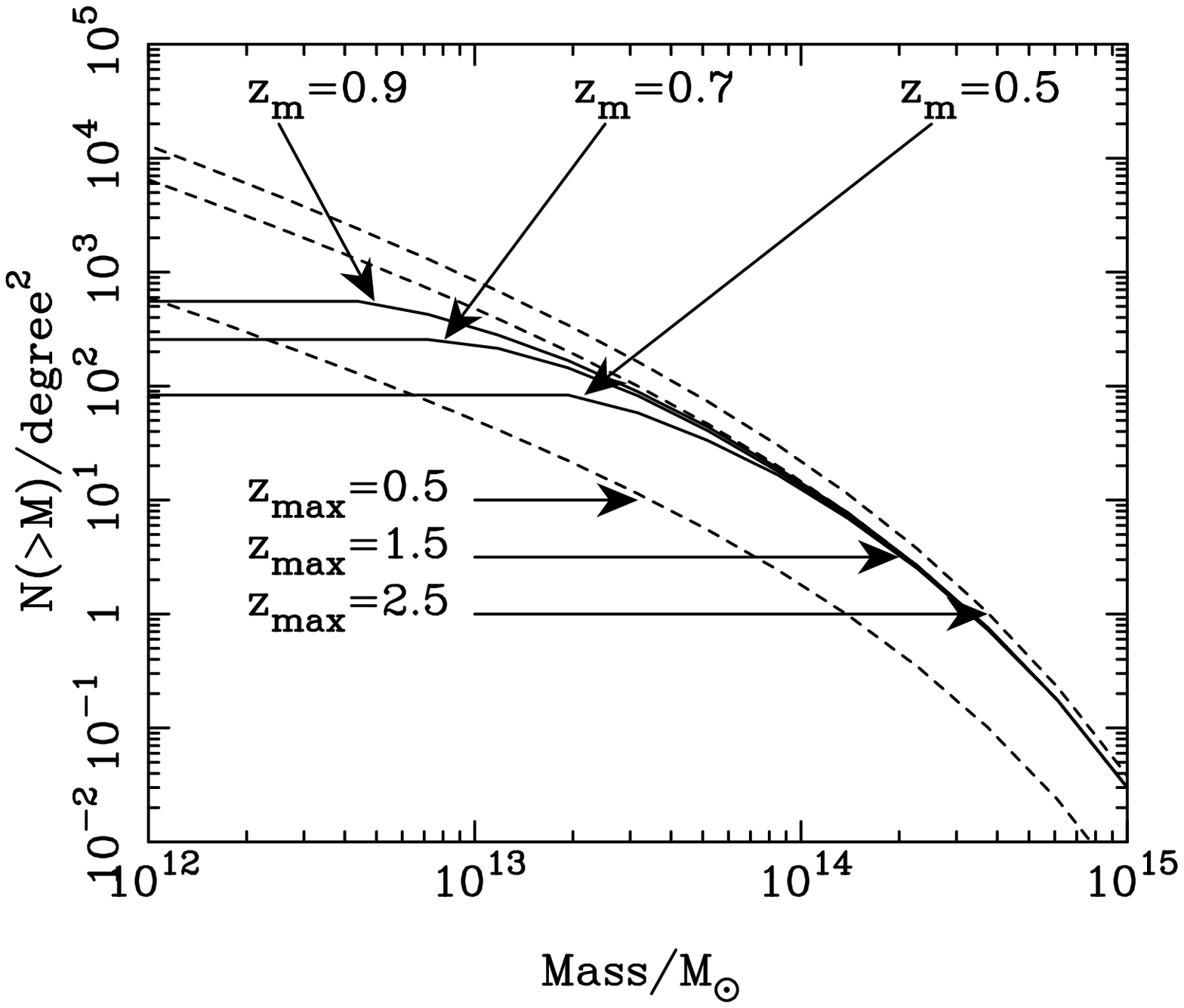,width=\columnwidth,angle=0}
 \caption{The cumulative number count of dark matter
haloes for a range of median redshift distributions, $\N(>M,z_m)$,
per square degree. The solid lines assume a maximum redshift in the
halo population of $z_{\rm max}=1.5$, while the upper dotted line
assumes $z_{\rm max}=2.5$, and lower dotted line $z_{\rm
  max}=0.5$. The cut-offs 
in halo numbers for the $z_{\rm max}=1.5$ (solid) lines are for different
median redshifts with a shear signal-to-noise limit $\mu>1$.  }
 \label{numberdensity_threshold}
\end{figure}

\begin{figure}
\centering
 \psfig{figure=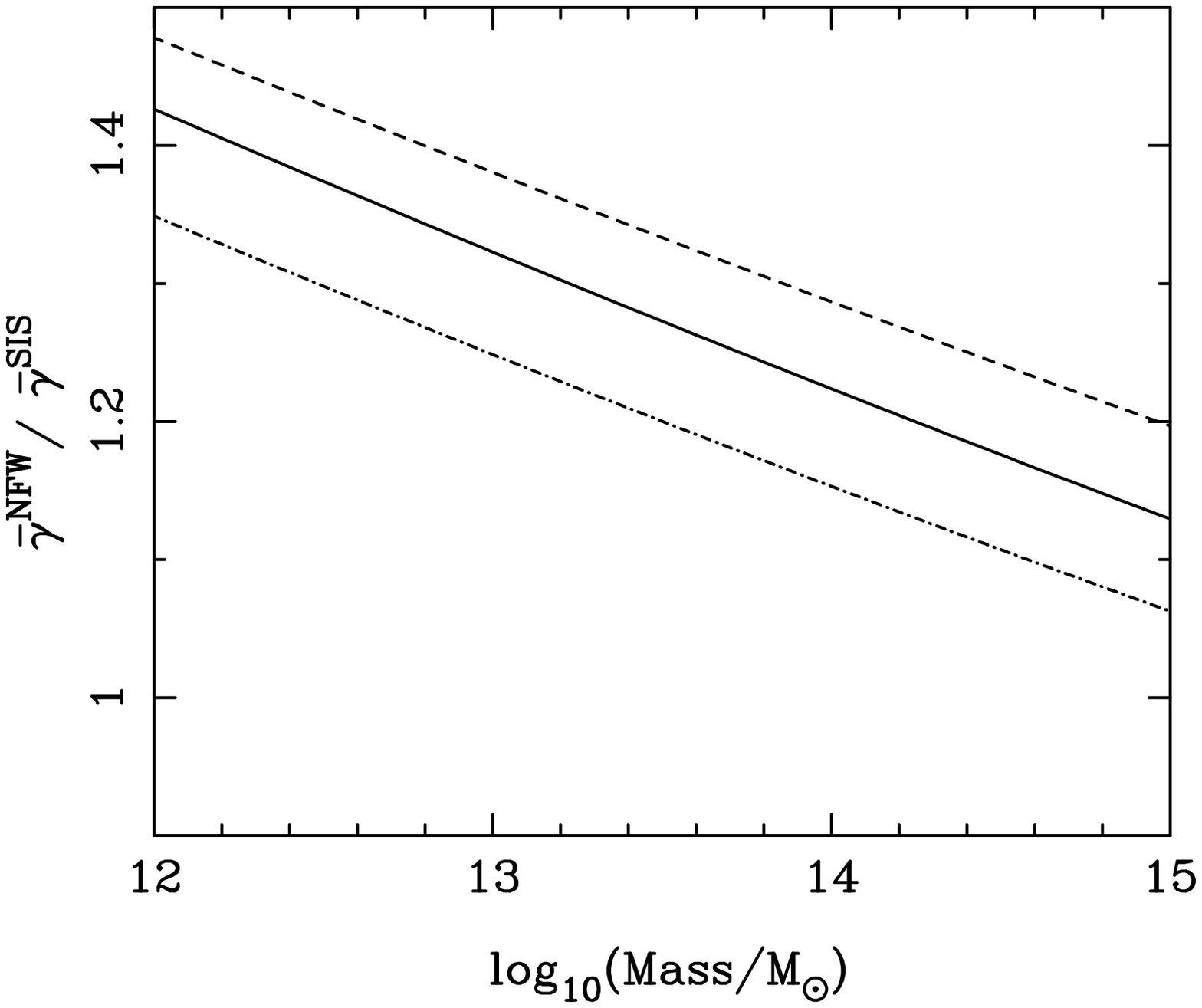,width=\columnwidth,angle=0}
 \caption{The ratio of mean shears for an SIS and NFW haloes of
 varying mass for haloes at a redshift of $z_c=0.1$. The solid line is
 for a $\Lambda$CDM fiducial cosmology, the dashed line is for a SUGRA
 fiducial model and the dot-dashed for a Phantom model, see Section
 \ref{The effect of changing the fiducial dark energy model} for details.}
 \label{testNFW}
\end{figure}

In Figure \ref{numberdensity_threshold} we show the expected
cumulative number count of dark matter haloes for a range of
median redshifts, $z_m$, $\N(>M,z_m)$ per square degree. The dotted
lines represent various upper redshift limits with no signal-to-noise
limit on cluster detection for $z_m=0.7$. The solid lines are for a
detection
threshold of clusters with
signal-to-noise of unity for various median redshifts, given by
equation (\ref{thresholdeq}).
In assuming a SIS model for the lensing clusters the average shear
around a cluster may be systematically underestimated. A more
reliable model is the Navarro-Frenk-White (NFW) profile, although
the density profile form would yield a more complex relation for
the shot noise term than the SIS profile. There is, however, an
approximate scaling relation which relates $\bar\gamma_{SIS}$ to
$\bar\gamma_{NFW}$ outlined in Wright \& Brainerd (2000) which, since
we take the average tangential shear in an apeture, should be
adequate. Adopting the 
techniques outlined in Wright \& Brainerd (2000), the concentration
parameter depends on the mass, redshift and fiducial cosmology. We use
the concentration parameter from Dolag et al. (2004), our
scaling from $\bar\gamma_{NFW}$ to $\bar\gamma_{SIS}$ depends on
the dark energy fiducial model, mass and redshift of the cluster. We
will use this scaling to correct the shear signal expected from
the halo model. Note
that this only affects the noise properties since the shear ratio only
depends on the redshift-distance relation. Figure \ref{testNFW} shows
how the scaling depends on mass and the fiducial dark energy models
(discussed in Section \ref{The effect of changing the fiducial dark
  energy model}) for clusters at a redshift of $z_c=0.1$. 

If the haloes are assumed to be randomly distributed over the sky,
and we take their physical size to be the virial radius, the
effect of overlapping haloes projected onto the sky is negligible.
For instance, a $M=10^{15}M_\odot$ halo has a virial radius of
$r=0.75$Mpc and a number density of $n \approx 10^{-2}$ per square
degree, while a $M=10^{13}M_\odot$ halo has a virial radius of
$r=0.15$Mpc and a number density of $n \approx 10^2$ per square
degree. At $z=0.2$, the physical distances $0.75$Mpc and $0.15$Mpc
subtend $0.12$ degrees and $0.025$ degrees respectively. Hence we
shall assume that halo overlaps are not important.

\subsection{Photometric Redshift Uncertainty}
\label{photozs}
In Section \ref{Photometric Redshift Errors} we introduced the effects
of including photometric 
redshifts on the lensing measurements. Here we detail our estimate of
the photometric redshift errors.

The uncertainty on the photometric redshift error on an individual
galaxy with redshift $z$ and magnitude $R$ for a multi-band survey
is well fitted by (Wolf et al. 2004);
 \be
 \sigma_z(z,R) =  A (1+z)
 \left[ 1 + 10^{B(R-R_*)}\right]^{1/2},
 \label{photo_z_err}
 \ee
where $A=0.035$, $B=0.8$ and $R_*=23.0$ for galaxies in a $5$-band
survey, and $A=0.007$, $B=0.8$ and $R_*=21.6$ in a $17$-band,
COMBO-17-type survey. This shows that the redshift errors are well
constrained at bright magnitudes but poorly constrained at faint
magnitudes. The first parameter, $A$, characterizes the best
performance achievable in the bright domain, where photon noise is
irrelevant and spectral resolution limits the redshift estimate.
The second parameter, $B$, describes how a decrease in photon
signal propagates into the redshift signal. This should be $0.8$
if we consider all galaxies, but can be made smaller by filtering
out galaxies with outlying redshift errors. The final parameter,
$R_*$, determines the magnitude where we see a sharp rise in the
redshift error function when we change from the
spectral-resolution limited r\'egime at bright magnitudes into the
r\'egime where photon noise drives the redshift noise by a factor of
$\approx 2.5$ per magnitude under the assumption of a locally
linear transformation from  colour-space into redshift-space.

The average redshift error in a bin at redshift $z$ is given by
averaging over all observable galaxies below a limiting absolute
magnitude in that bin,
 \be
 \overline{\sigma}_z(z) = \frac{\int^{{\cal M}_{\rm lim}(z)}_{-\infty}
  \!\!dM\,
  \Psi(M)
  \sigma_z(z,M)}{\int^{{\cal M}_{\rm lim}(z)}_{-\infty}\!\!dM\
  \Psi(M) }.
 \label{intdz}
 \ee
Here $\Psi(M)$ is a sum of Schechter functions $\Phi_{{\rm red}}$
and $\Phi_{{\rm blue}}$ (see Wolf et al, 2003 for details of the
COMBO-17 luminosity functions) for a red and blue sample of
galaxies. The luminosity functions are defined for a colour, $c$,
as
 \be
 \Phi_c(M)dM = 0.4 \ln \! 10 \,
 \phi^*_c X^\alpha_c(M) e^{-X(M)} dM,
 \ee
where
 \be
 X(M) =10^{-0.4(M-M^*_c)},
 \ee
and
 \ba
 \phi^*_{{\rm red}}(z) & = &  (2.0-z)\times 10^{-3} [h^{-1}  {\rm
    Mpc}]^{-3},\\
 \phi^*_{{\rm blue}}(z)& = & 3.0\times 10^{-3} [h^{-1}  {\rm
    Mpc}]^{-3},
 \ea
valid for $z<2$, are the characteristic space-densities of
galaxies. The slope of the luminosity functions are
 \ba
 \alpha_{{\rm red}}=-0.5,\\
 \alpha_{{\rm blue}}=-1.3,
 \ea
and
 \ba
 M^*_{{\rm red}}(z) = -20.18-1.04z,\\
 M^*_{{\rm blue}}(z)= -20.09-1.28z,
 \ea
are the characteristic magnitudes of reds and blue galaxies in the
COMBO-17 survey. $M_{\rm lim}(z_m)$ is the
limiting apparent magnitude of survey with median redshift $z_m$
given by (see Brown et al. 2003,  and equation \ref{zmR} in
Section 4.5)
 \be
 M_{\rm lim} = 20.8 + z_m/0.23
 \label{magzrel}
 \ee
for an optical survey,
which we then transform to the absolute limiting magnitude;
 \be
   {\cal M}_{\rm lim} = M_{\rm lim} - 5 \log_{10} \left\{(1+z) S_k[r(z)]
   \right\}+K(z).
 \ee
The K-correction, $K(z)$, is;
 \be
 K(z) = 2.5 (\nu -1) \log_{10} (1+z),
 \ee
where $\nu$ is the spectral slope of galaxies. We take this to be
$\nu=+1$, making the K-correction zero.
\begin{figure}
 \psfig{figure=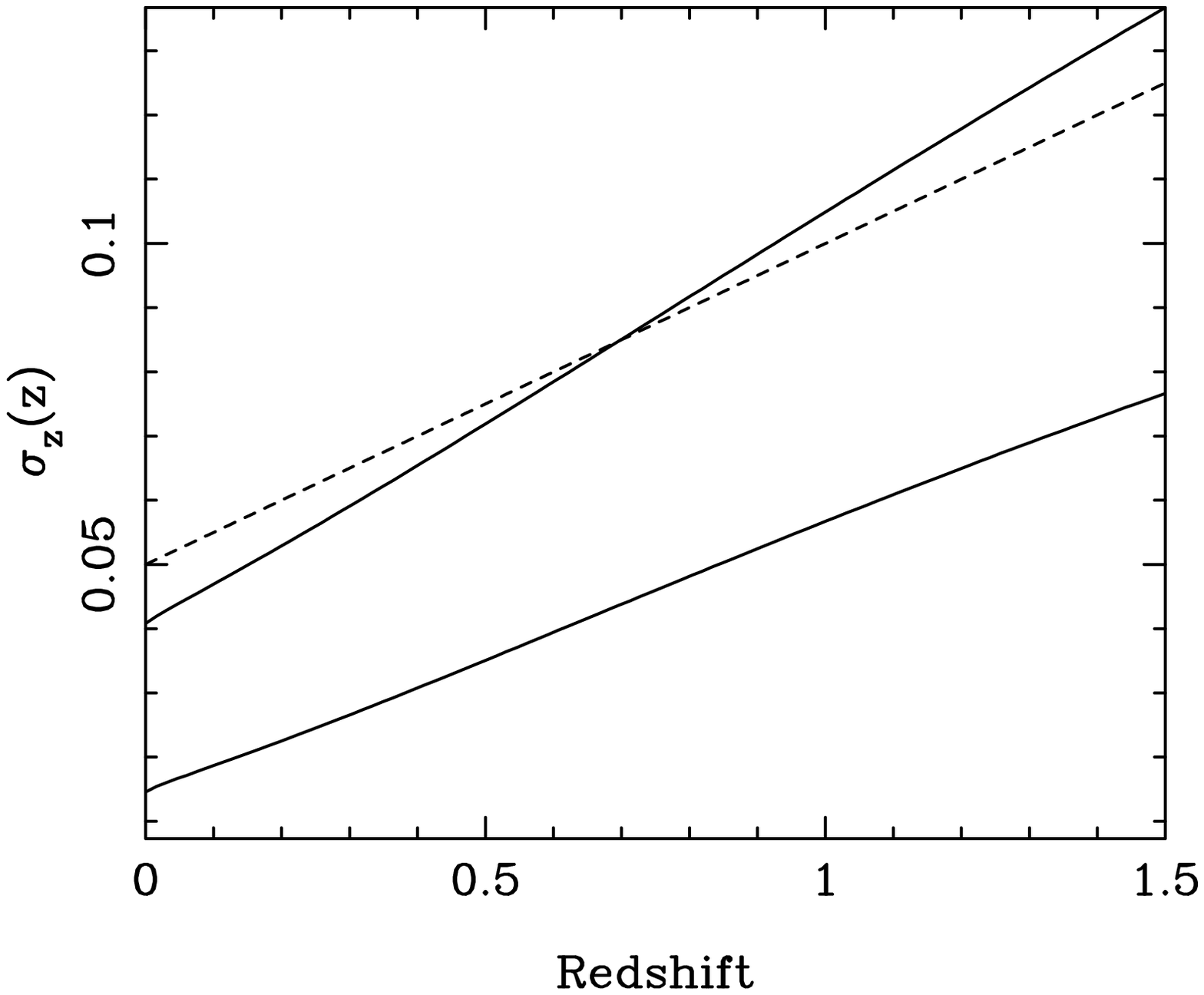,width=\columnwidth,angle=0}
 \caption{Variation of $\sigma_z(z)$ with redshift for a $5$-band
  (upper solid line) and a $17$-band (lower solid line) photometric
  redshift survey, 
  averaging over galaxy luminosities, for
  a survey with median redshift $z_m=0.7$ (solid line).
  Galaxy properties are from COMBO-17
  and described in the text.
  Also shown is a standard $5$-band photometric redshift model with
  $\sigma_z(z)=0.05(1+z)$ (dashed line).}
 \label{sigmaz}
\end{figure}
Figure \ref{sigmaz} shows the increase in mean photometric
redshift uncertainty for a $5$-band and a $17$-band survey with
median redshift $z_m=0.7$, based on the galaxy luminosity
functions. As the magnitude of a galaxy depends on its redshift,
the scaling of the photometric redshift noise is more complicated
than the simple $(1+z)$ scaling commonly used. Brodwin et al.
(2003) find $\sigma_z(z)=0.05(1+z)$ for a 5-band survey, which we
plot as the dashed line in Figure \ref{sigmaz}. Our estimate of
the redshift error for a 5-band survey predicts a higher error for
$z>0.7$, and a lower error for $z<0.7$. We have extrapolated these
formulae to $z=1.5$ though this extrapolation may be optimistic as
photometric redshift estimates can 
increases dramatically at $z\approx 1$ if IR data is not available.

For an intermediate $9$-band survey we linearly interpolate
between the $5$-band and $17$-band lines, assuming that at each
redshift the relationship between bands is linear. Over
all redshifts we find there is no simple linear scaling relation
with $(1+z)$. However we find approximate fitting formula for a 
$5$-band survey,
\be
\sigma_z(z)\approx 0.063(0.64+z),
\ee
and for a $17$-band survey,
\be
\sigma_z(z)\approx 0.041(0.37+z).
\ee
\begin{figure}
 \psfig{figure=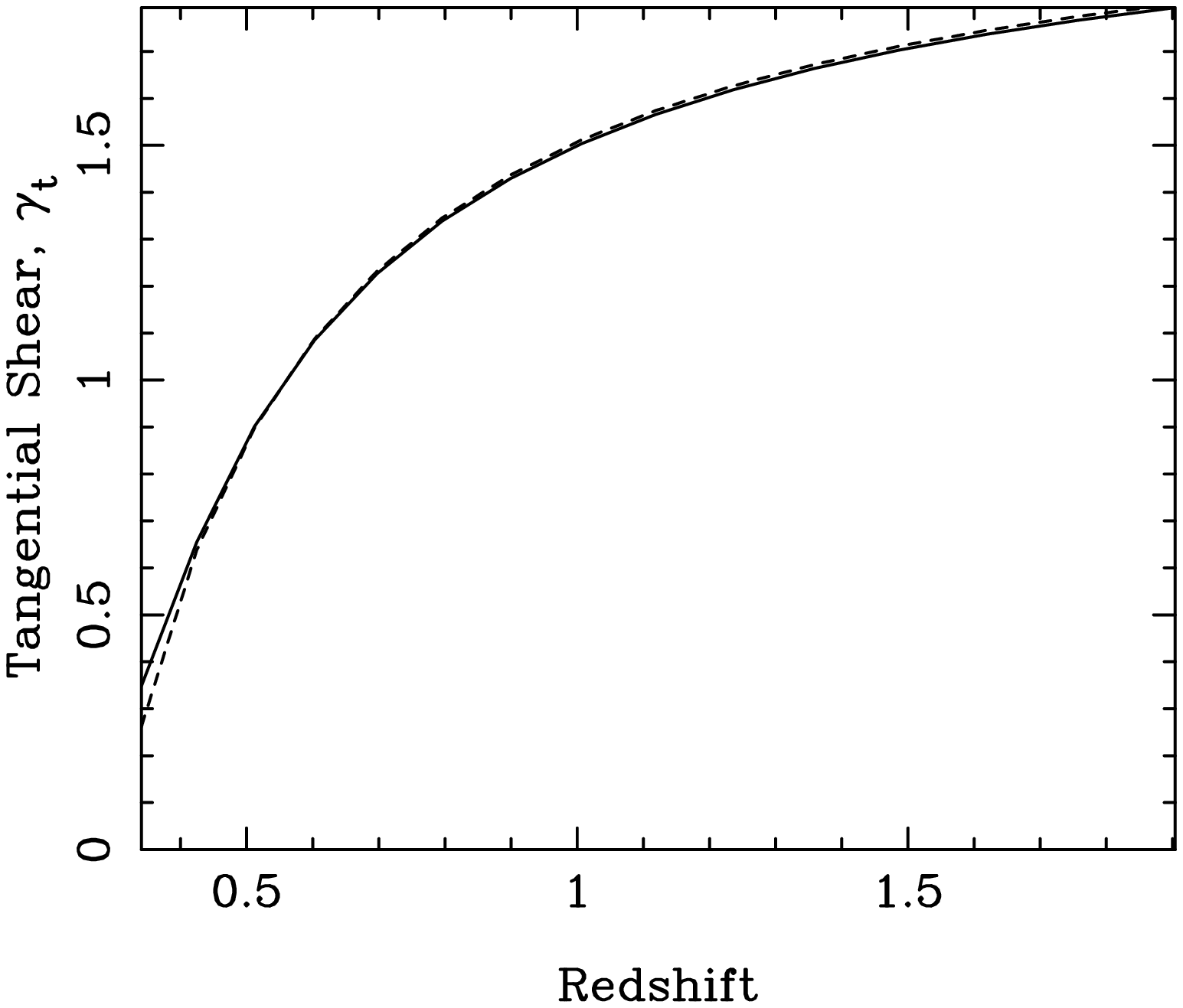,width=\columnwidth,angle=0}
 \caption{The effect of photometric redshift errors on the
 tangential shear behind a lensing cluster of mass
 $10^{15}M_{\odot}$ at a redshift of $0.2$, assuming $5$ bands. The dashed
 line is the true shear response, while the solid line is
 the shear with photometric redshift errors, using
 equation (\ref{mean_gamma}).}
 \label{shearresponse}
\end{figure}
Figure \ref{shearresponse} shows the effect of a 5-band
photometric redshift error, given by equation (\ref{mean_gamma}),
for a photometric galaxy survey parameterized the same as the
COMBO-17 survey, with median redshift $z_m=0.7$ and limiting
magnitude $R=24$, on the measured tangential shear distribution
behind a $M=10^{15}M_\odot$ halo at $z=0.2$. The main effect is a
suppression of the shear signal at low redshift, where the shear
is rapidly changing. This is due to the scattering of unlensed
galaxies in front of the lensing halo into bins just behind the
halo.

The photometric redshift error fit from COMBO-17, given by
equation (\ref{photo_z_err}), is per galaxy. In practice the
photometric redshifts produced by any multi-band analysis will
also provide an individual redshift error for every galaxy which
will also depend on redshift and magnitude. In this current
analysis photometric redshifts are averaged over all galaxy types
and magnitudes. In practice one would like to weight the 
data optimally to minimize the effect of both shear and photometric
redshift 
errors. Given the redshift dependence of the shear signal behind a 
lens, it is likely that both errors in the shear signal and
photometric redshift errors degrade the measurement of parameter,
while at redshifts far from the lens, shear errors will dominate.
This implies that there is an optimal weighting scheme which is a
function of galaxy redshift and magnitude for weak shear analysis
using photometric redshifts. We shall explore this elsewhere.

\subsection{Bias in the Photometric Redshifts}
\label{Bias in the Photometric Redshifts}

In addition to the uncertainty on photometric redshifts, we would
also like to know the effect of a bias in the photometric
redshifts, leading to an off-set in their calibration. We can
model the effects of this by considering the first-order effect of
such a bias on the measurable parameters. In Appendix B  we show
that for a Gaussian distributed likelihood function, the linear
bias in a parameter, which we shall call $\delta \theta_i$, due to
a bias in a fixed model parameter (i.e., one whose value we have
assumed and is not being measured), which we shall call $\delta
\psi_j$, is given by (see also Kim et al., 2004)
 \be
        \delta \theta_i = - [F^{\theta \theta}]^{-1}_{ik} F^{\theta
         \psi}_{kj} \delta \psi_j,
 \ee
where $\F^{\theta \theta}$ is the parameter Fisher matrix and
$\F^{\theta \psi}$ is a pseudo-Fisher matrix of derivatives with
respect to parameters which are assumed fixed ($\psi$) and those to be
determined ($\theta$).

Assuming there is a possible bias in the mean of the photometric
redshifts of the survey, $z_{{\rm bias}}$, (see Section
\ref{Photometric Redshift Errors}) due to poor calibration of the
photometric redshifts with spectroscopic redshifts, and
marginalizing over all other cosmological and dark energy
parameters, we find that the induced bias in $w_0$ due to the bias
in galaxy redshifts is
 \be
 \label{biaseqn}
 \delta w_0 = -C_{\rm bias}\,\delta z_{{\rm bias}},
 \ee
where $C_{\rm bias}$ is a constant.
If the bias in the mean of the photometric redshifts arises from
an overall bias in the photometric redshift calibration, the
calibration error will be
 \be
    \sigma(z_{{\rm bias}}) =
    \frac{\sigma(z)}{\sqrt{N_{\rm spec}}},
 \ee
where $N_{\rm spec}$ is
the number of galaxies with a spectroscopic redshift. If we set
$\delta z_{{\rm bias}} = \sigma(z_{{\rm bias}})$ and a requirement
that the bias in $w_0$ is half of the error, $\delta w_0 = 0.5
\Delta w_0$, then the number of galaxies with spectroscopic
redshifts we require is
 \be
    N_{\rm spec} = \left[  \frac{C_{\rm bias}
    \sigma(z)}{\delta w_0} \right]^2.
 \ee
We have found that $C_{\rm bias}\approx 9.0$ for the geometric test.
If we assume $\sigma(z) \approx 0.1$ and $\Delta w_0\approx 0.01$ then
 we require $N_{\rm spec}\approx 3 \times 10^4$.
The size of the required spectroscopic redshifts
required to calibrate the geometric test suggests that a large
spectroscopic survey, such as that proposed for the Wide-Field
Multi-Object Spectrometer (WFMOS; Bassett et al., 2005), would be
required and combined with a large-scale weak lensing survey.

We have also investigated the effect of an offset in the variance
of the photometric redshift errors
$\sigma_z(z)\rightarrow\sqrt{\sigma_z^2(z)+\Delta\sigma_z^2(z)}$.
We find that this effect is negligible for the geometric
test, so that the total bias due to photometric redshift errors is
only dependent on the bias in the offset of the mean. However in the
pseudo-Fisher analysis the variation about the mean of
$\Delta\sigma_z(z)$ is $\pm 0.05$, and we would expect there to be an
effect at some level if the variation was larger. We explore fully
marginalizing over nuisance parameters in a full Fisher analysis
elsewhere. 

\subsection{Limits on the Measurement of Galaxy Ellipticity }

\subsubsection{Ground-based Ellipticity Measurements}
\label{Ground-based Ellipticity Measurements}
For ground-based weak lensing observations estimates of galaxy
ellipticities are limited by atmospheric seeing. The angular sizes
of typical galaxies in the GOODS fields scale with redshift by (Ferguson
et al. 2002)
\be
\theta_g = 0.8 z^{-1} \,\,{\rm arcseconds}.
\ee
If $\theta_s$ is the typical
seeing during weak lensing observations, the post-seeing galaxy
image will be
\be
\theta_g' = \sqrt{\theta_g^2 + \theta_s^2}.
\ee
This will tend to decrease the ellipticity of galaxy images. Much
effort is put into weak lensing to correct this effect. However,
once the seeing disc exceeds the galaxy image and $\theta_g \ll
\theta_s$, this correction fails. Typically, galaxy sizes are
about $\theta_g =0.8$ arcseconds at redshift $z=1$. If the
groundbased seeing for weak lensing is typically $\theta_s \approx
0.7$ arcsecs, then by a redshift of $z=1.5$, the galaxy sizes have
dropped to $\theta=0.5$ arcseconds and galaxy ellipticities cannot
be recovered without the use of adaptive optics.

Another limitation which could potentially come into play is when
the galaxy image is too faint to properly measure the galaxy shape
against the sky background. However, Bacon et al. (2001) find that
the dispersion on the measured galaxy ellipticities is very
insensitive to the galaxy magnitude, and seems only limited by the
detection threshold for galaxy detection. For 5-$\sigma$ detected
galaxies, ellipticities can be measured down  to the limiting
magnitude of the survey, with $\sigma_e=0.3$.

Given these two results we shall assume that we cannot measure
redshifts beyond $z=1.5$ from the ground due to being unable to
recover the pre-seeing ellipticity.

\subsubsection{Space-based Ellipticity Measurements}

Rhodes et al. (2003) find no dependence on ellipticity dispersion
as a function of redshift for space-based data. Refregier et al.
(2003) and Massey et al. (2004) find that $\sigma_e=0.2$ is a
reasonable measure for the ellipticity dispersion for a
space-based weak lensing survey. They also find a maximum redshift
bound for space-based surveys can be set at $z=2.0$ corresponding
to a deep magnitude cut of $R=29.1$.

\subsection{Optical Surveys}
\label{optsurvey}
In this Section we outline how to parameterize a weak lensing and
photometric redshift survey, and how these will scale for
different telescopes. A reasonable way to compare between
potential survey designs is to consider equal-time observations.
Hence one can compare dark energy results both for a single
telescope class, and across telescope classes.
The time taken for an imaging survey on a given telescope scales
as (cf equation \ref{timelim})
 \be
 \label{surveyeq}
 T \propto  z_m^4 f_{\rm sky}
 D^{-2} \big({\rm fov}/1^\Box\big)^{-1},
 \ee
where $D$ is the diameter of the primary mirror of the telescope
and ${\rm fov}$ is its field of view. We normalize the timescale
of a survey to the 5-band (g', u, r', i', z') CFHT survey, where
 $T=162$
nights for $z_m=1.17$ ($r=25.9$), $f_{\rm sky}=4.25 \times
10^{-3}$, $D=3.6$m and ${\rm fov}=1$ square degree. The median
redshift for an R-band survey is (Brown et al., 2003)
 \be
 z_m = 0.23 (R-20.6),
 \label{zmR}
 \ee
while we find that the projected surface number count density of
galaxies in the COMBO-17 survey scales with the median redshift as
 \be
 n_2(z_m) = 30 \,z_m^{3.4} \,\,\,\,{\rm galaxies\,\, per\,\, square\,\,
   arcmin}.
 \ee
We also need to assume a functional form for the galaxy redshift
distribution which we take to be
 \be
 p(z|z_m)\propto z^2 \exp\big[ -\big(z/z^* \big)^{1.5}
  \big],
 \ee
where $z^*=z_m/1.412$ and
 \be
 \int_0^\infty dz \, p(z|z_m) =1.
 \ee
The space density of galaxies as a function of galaxy redshift,
$z$, and survey median redshift, $z_m$, is then
 \be
        n_3(z|z_m) = n_2(z_m) p(z|z_m).
 \ee
The 3-D galaxy redshift distribution, $n(z)=n_3(z|z_m)$, is used
in equations (\ref{mean_gamma}) and (\ref{mean_W}) when
calculating the effects of photometric redshift errors,  for
calculating the number of galaxies in redshift bin, $N_i$, for the
shot-noise, and for finding the cumulative surface density of
galaxies above a halo redshift, $n_2(>z)$, in equation
(\ref{shear_lim}).

The number of redshift bins used in the background to the lenses,
$N_B$, is determined by the photometric redshift uncertainty
(Section \ref{photozs}) by assigning a bin width at particular
redshift to be the average photometric uncertainty, $\sigma_z(z)$,
at that redshift. The bins exhaustively fill the available
redshift range.

\section{Survey Design Strategy}
Having formulated the basic method for estimating dark energy
parameters from shear ratios, we now consider the problem of what
type of survey would be optimal for measuring the properties of
the dark energy from the shear ratio geometric test. For
instance, should one construct a wide area, but shallow,
multi-band survey, or a narrow and deep multi-band survey with a
survey-class telescope, such as the VST (Belfiore et al., 2005), the
Dark Energy Survey on
the CTIO (Wester, 2005), darkCAM (Taylor, 2005) or Pan-STARRS (Kaiser,
2005)? Or 
should one instead take
snap-shots of galaxy clusters with a large but small field-of-view
telescope such as SUBARU, the VLT or the Keck Telescope? We shall
compare these different strategies by minimizing the marginalized
uncertainty on $w_0$ for fixed-time observations, assuming a prior
from the expected 14-month Planck
Surveyor experiment (Lamarre et al., 2003) to lift the main
degeneracies.

\label{ClusterDep} Broadly we have two observing strategies
available to us: targeted observations at individual clusters, and
a general wide-field survey. In the former, one would use a large
telescope with small field-of-view to take rapid observations of
each cluster, while in the latter a large telescope with a wide
field-of-view would make a general wide-field survey from which
one would extract haloes. With a halo decomposition analysis of
the matter distribution we can ask where most of the signal will
come from for a dark energy analysis with weak lensing, and so see
which strategy would be most effective in terms of telescope time.
We begin with targeted observations of clusters.

\subsection{Targeted Observation Mode}
\label{Targeted Observation Mode}
We shall assume we have a large telescope with a small
field-of-view which can target pre-selected galaxy clusters from a
pre-existing galaxy cluster catalogue. The survey would start by
imaging the largest clusters on the sky, and then move on to
subsequently smaller haloes. We shall assume that the telescope
has some fraction of the sky available to it, and we shall ignore
scheduling issues.

\begin{figure}
 \psfig{figure=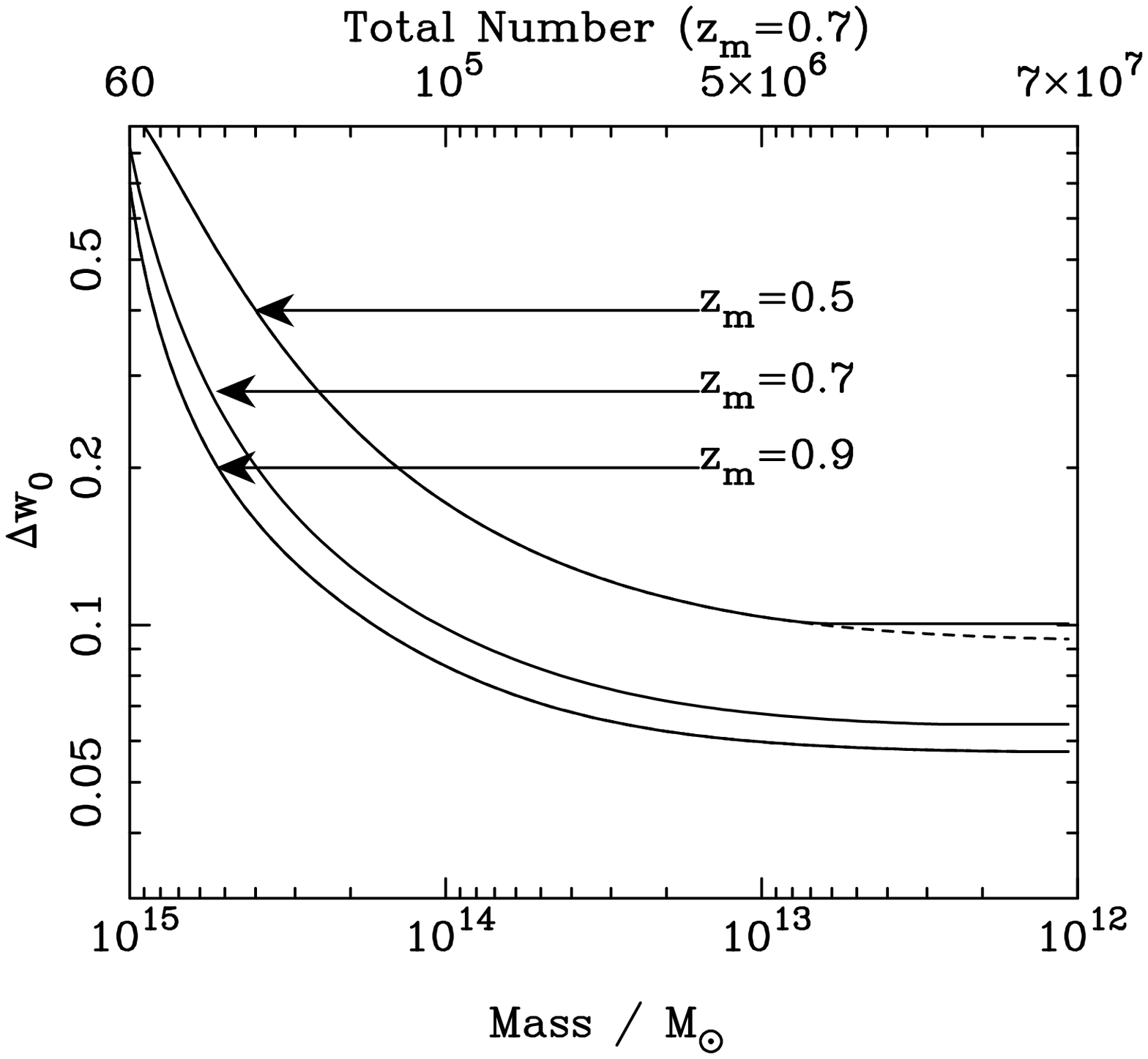,width=\columnwidth,angle=0}
 \caption{Variation of marginal error on $w_0$ with the mass of lensing
 cluster for a pointed survey with 20,000 square degrees accessible, and
 for $z_m=0.5$, $0.7$ and $0.9$.
 The dashed line has no S/N threshold,
 the solid line has a threshold condition set by equation
 (\ref{thresholdeq}). We assume a WMAP 4-year prior. Note that the
 cumulative total number of galaxies depends on the median
 redshift, $z_m$, (see Figure 5). Here it is calculated for $z_m=0.7$, but
 has little effect.}
 \label{signalmasstarg}
\end{figure}

Figure \ref{signalmasstarg} shows the accuracy on $w_0$,
marginalized over $\Omega_m$, $\Omega_v$ and $w_a$, which can be
achieved by a targeted survey as a function of the number of
clusters in decreasing mass. We have assumed half the sky
($20,000$ square degrees) is available, and combined the lensing
result with a $4$-year WMAP prior (see Section \ref{Priors}). The
time taken for such a survey is just the time taken to image down
to the median redshift for a given telescope, multiplied by the
number of clusters. Note that the cumulative total number of
haloes, ${\cal N}(>M,z_m)$, depends on the median redshift of the
survey, $z_m$; the upper scale on Figure \ref{signalmasstarg}
assumes $z_m=0.7$ and $z_{{\rm max}}=1.5$. However, comparing with
Figure
5 we see that varying $z_m$ has only a small effect in the number of
haloes
above $M=10^{13}M_\odot$, but does change the total number for
masses below this.

Figure \ref{signalmasstarg} implies that by imaging only $60$ of
the most massive clusters ($10^{15}M_{\odot}$) in a hemisphere to
$z_m=0.9$ ($R=24.7$) in five bands and combining with the 4-year
WMAP, one could reach an accuracy of $\Delta w_0=0.50$, after
marginalizing over all other parameters, including $w_a$. This
seems a viable strategy, a factor of $2$ improvement on 4-year WMAP
given a marginalization over $w_a$. If $w_a$ is fixed at $w_a=0$ then
the marginal error on $w_0$
reduces to $\Delta w_0\approx 0.25$, a factor of $2.5$ improvement on
4-year WMAP, marginalizing over other parameters.
To rapidly image each halo in five bands 
with an 8-metre class telescope with a $0.025$ square degree
field-of-view, such as with SuprimeCam on the Subaru telescope
(see Broadhurst et al., 2005, for the use of Subaru in a lensing
analysis) would take $10$ to $20$ nights.

Beyond this accuracy, there are diminishing returns for a pointed
survey from the geometric test. To reach an accuracy of $\Delta
w_0\approx 0.05$,
one would have to image around $10^7$ haloes, with the number of
galaxies scaling roughly as
 \be
    N\approx 10^{0.35/\Delta w_0}.
 \ee
For a targeted survey, this seems an unfeasible task. The weakness
in this relation is due to the fact that we have ranked haloes by
mass, and while the number of haloes is increasing the mass per
cluster, and hence lensing signal, is falling. By the time we are
imaging the $<10^{13}M_\odot$ haloes, the shear signal is so weak
as to no longer contribute to a significant measurement of dark
energy.

These curves scale with the survey median redshift roughly as
 \be
    \Delta w_0(>M,z_m) \approx \Delta w_0 (>M,z_m=0.7)
    \left(\frac{z_m}{0.7}\right)^{-1},
 \ee
where the increase in accuracy arises due to the increase in
number of background galaxies reducing the shot-noise, and the
increase in available clusters reducing clustering variance. This
approximation fails for the most massive clusters, where imaging
deeper does not help as we are clustering-limited.

\subsection{Time-Limited Survey Mode}
\label{Time-limited Survey Mode}
In contrast to a targeted observation mode, one could also use a
large survey telescope with a wide field-of-view to construct a
general wide-field survey, and extract haloes from this for the
shear ratio analysis. In this case it makes sense to restrict the
amount of telescope time one can allocate to such a survey. In the
next Section we discuss the optimization of such a survey. Here we
shall assume the optimum survey parameters and investigate how the
signal is distributed across the mass spectrum of haloes.

\begin{figure}
 \psfig{figure=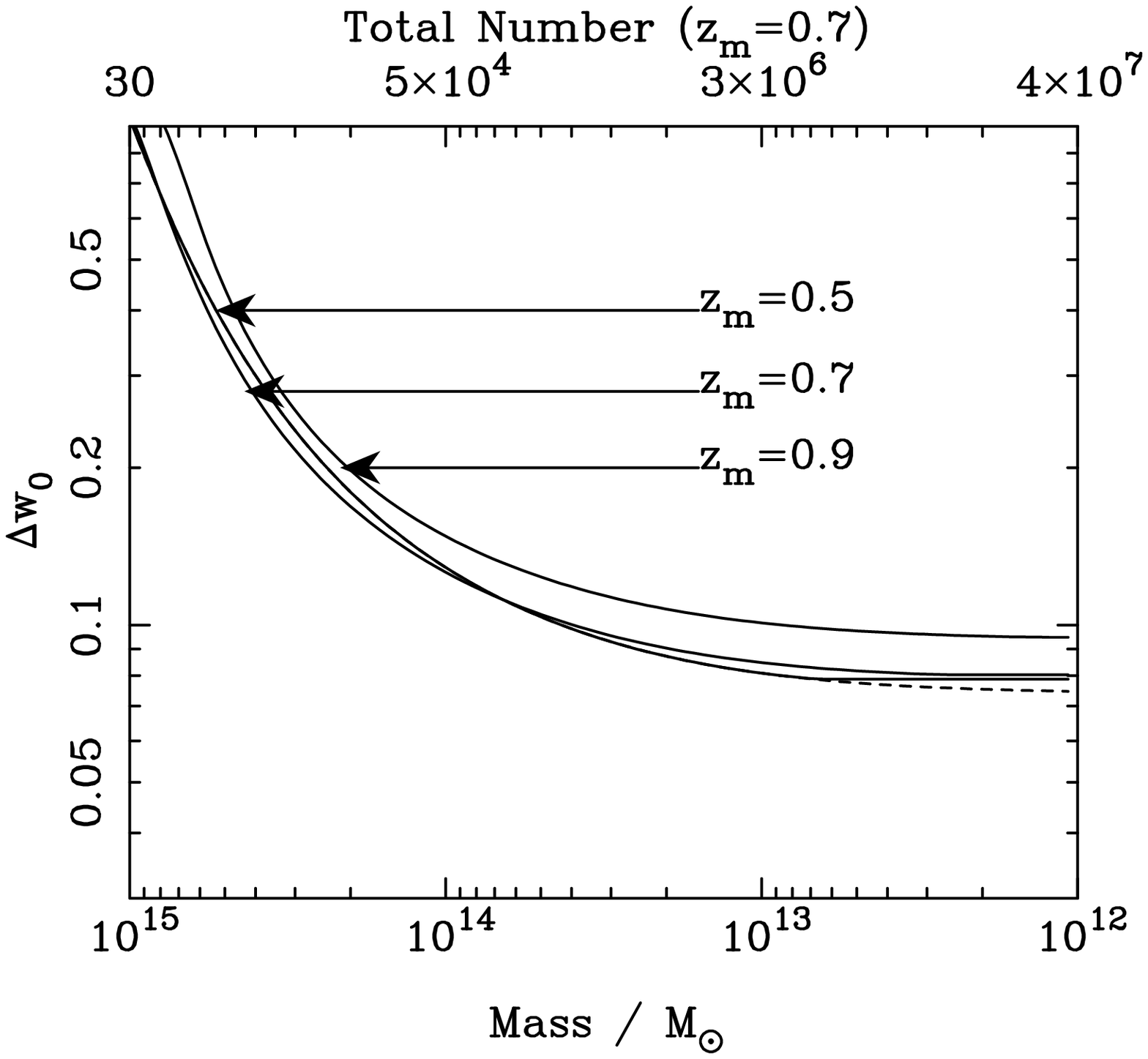,width=\columnwidth,angle=0}
 \caption{Variation of marginal error on $w_0$ with the mass of lensing
 cluster for a 10,000 square degree survey to $z_m=0.7$, and scaled to
 $z_m=0.5$ and $z_m=0.9$ with area of $10,000(0.7/z_m)^{4}$.
 The dashed line has no S/N threshold,
 the solid line has a threshold condition set by equation
 (\ref{thresholdeq}). We assume a WMAP 4-year prior.}
 \label{signalmass}
\end{figure}

Figure \ref{signalmass} shows the cumulative gain in accuracy on
$w_0$ as we add haloes of decreasing mass. We have marginalized
over the remaining parameters $(\Omega_m,\Omega_v,w_a)$, and
calculated the Fisher matrix using the analysis of Section
\ref{sectionpara}. We have assumed a fixed-time survey with median
redshift of $z_m=0.5$, $z_m=0.7$ and $z_m=0.9$ (limiting
magnitudes of $r=23$, $r=23.8$ and $r=24.7$, respectively),
combined with a $4$-year WMAP prior (see Section \ref{Priors}). The
lines for $z_m=0.5$ and $z_m=0.7$ cross at approximately $7\times
10^{14} M_{\odot}$ this is interpreted as for a fixed time survey the
optimal median redshift varies slightly with the mass range of clusters
used. As clusters of lower mass are included the optimal median
redshift behaviour converges so that $z_m=0.7$ yields the lowest
error, note Figure \ref{signalmass} includes a $4$-year WMAP prior. 
The area of each survey is of 38,400 square degrees, 10,000 square
degrees and 3,660 square degrees, respectively, appropriate for a
survey with one, or more, 4-metre telescopes with a 2 square degree
field-of-view (more than one would be needed for a $z_m=0.5$, 38,400
square degree survey). Note again that the upper scale (Total Number) for
number of
haloes depends on median redshift which is here assumed to be
$z_m=0.7$. The cumulative number of haloes is half of that for a
given mass than for Figure \ref{signalmasstarg} as the total area
probed is half.

Again we find that the largest haloes provide the largest
contribution to the measurement of $w_0$, with an error of $\Delta
w_0 = 0.6$ from the largest 30 haloes. The error has flattened off
from 60 to 30 haloes. As with the targeted survey mode, the
increase in accuracy for including smaller haloes has diminishing
returns. However, given these haloes will already be in the
survey, the limitation here is processing time, rather than
telescope time. For a 10,000 square degree survey to $z_m=0.7$
($r=23.8$) we can reach an accuracy of $\Delta w_0=0.08$ from the
analysis of $N=3 \times 10^6$ haloes, down to haloes with
$M>10^{13}M_\odot$. The majority of the signal (the steepest
gradient in Figure \ref{signalmass}) comes from the relatively
numerous intermediate mass haloes with $M\sim 5\times 10^{14}M_{\odot}$.
Beyond this the signal-to-noise per cluster is too small to
contribute to a measurement of $w_0$.

For a time-limited survey, it is useful to parameterize how the
uncertainty on $w_0$ scales with different telescopes and surveys
by scaling the error with the fractional survey sky coverage,
 \be
    f_{\rm sky}=\frac{A}{40,000\, {\rm sq.\,\, deg.}},
 \ee
 where $A$ is the survey area, so that
 \be
    \Delta w_0 = \Delta w_0 (f_{\rm sky} = 0.25)\left(\frac{f_{\rm
    sky}}{0.25}\right)^{-1/2},
 \ee
where
 \be
 \label{timelim}
 f_{\rm sky} = 0.25 \left(\frac{T}{600\,{\rm nights}} \right)\left(\frac{z_m}{0.7}
 \right)^{-4}\left(\frac{{\rm fov}}{1^\Box}\right)\left(\frac{D^2}{4 {\rm m}^2} \right).
 \ee
Hence one can trade off telescope size and field-of-view (fov) with
the survey time-limit, $T$, and the median depth, $z_m$.

To summarise Sections \ref{Targeted Observation Mode} and
\ref{Time-limited Survey Mode}, while a reasonable sized pointed
survey of around 
$60$ of the largest clusters in a hemisphere combined with the
4-year WMAP results could rapidly measure $w_0$ to around $\Delta
w_0=0.50$ in a short space of time, to improve the accuracy to a
few percent would require an unfeasible amount of telescope time.
However, a time-limited wide-field lensing and 5-band photometric
redshift survey could push the accuracy down to a few percent
accuracy, for example $\Delta w_0=0.08$ for a 10,000 square degree
survey to $z_m=0.7$, with the analysis of the millions of medium
sized clusters and groups ($M>10^{13}M_\odot$).

Time-limited survey designs and their optimization in measuring
$w(z)$ are considered in further detail in Section \ref{Optimization
for a Wide-Field Cluster Lensing Survey}.

\subsection{Area-limited Survey Mode}
A further distinct class of experiments, such as the LSST (see
Tyson et al., 2002) and Pan-STARRS (PS4; Kaiser, 2005), will
repeatedly image an entire hemisphere ($20,000$ square degrees) to
a given median redshift; this is proposed to be done by stacking
multiple images. In these cases the limiting factor is the amount
of sky available to a given telescope, and time allowing for a
given median redshift to be reached. Figure \ref{signalmasstarg}
shows that the marginal error on $w_0$ will vary as the median
redshift of the survey as
 \be
  \Delta
  w_0=0.07\left(\frac{z_m}{0.7}\right)^{-1}
  \left(\frac{f_{\rm sky}}{0.5}\right)^{-1/2}
 \ee
so that a $z_m=0.9$, $A=20,000$ square degree survey could image
approximately $7\times 10^7$ clusters between $10^{12}$ and
$10^{15}$ $M_{\odot}$, and achieve a marginal error of $\Delta
w_0=0.05$. A survey of this type is a viable
alternative to the time-limited wide-field survey.

\section{Optimization for a Wide-Field Cluster Lensing Survey}
\label{Optimization for a Wide-Field Cluster Lensing Survey}
Having investigated the source of the lensing signal which
contributes to the measurement of $w_0$, and shown that a
time-limited, wide-field survey can reach high-accuracy
measurements of $w_0$, we now proceed to optimize such a weak
lensing and photometric redshift survey for a fixed time to
measure the properties of dark energy from the geometric test.

\subsection{Combining lensing with other dark energy experiments}
\label{Priors}

As well as gravitational lensing, there are other experiments which
can probe dark energy, notably the CMB, Baryon Acoustic Oscillations
(BAO) in the galaxy power spectrum, and the supernova Type Ia Hubble
diagram. Individually each of these probes can probe dark energy,
but suffer from degeneracies between $w_0$ and $w_a$, and with other
parameters. These degeneracies can be lifted by combining methods.
Since there are a number of different probes, these experiments can
generate a number of combinations which can be compared for
consistency and as a test for systematics. In addition, dark energy
probes can be divided into methods that probe just the geometric
properties of the Universe, and those that combine the evolution of
mass clustering and geometry. These may respond differently
depending on whether the apparent dark energy is vacuum energy,
modelled as a fluid with negative equation of state, or a change in
gravity on large scales. Again, with a combination of methods these
possibilities can be explored. In this paper we shall only address
the combination of methods under the assumption that the dark energy
can be modelled by a negative-pressure equation of state. Finally, in
this paper we do not consider the Integrated Sachs-Wolfe (ISW)
effect directly, via cross-correlating galaxy surveys with the CMB,
although this too can probe dark energy.

The error analysis of a combination of independent experiments can
simply be accounted for by summing over each Fisher matrix for a
CMB, Type-Ia supernovae (SNIa) or a Baryon Acoustic Oscillation
experiment (BAO)
 \be
        F^{\rm TOT}_{ij} = F^{\rm GL}_{ij} + \sum_{P}F^{P}_{ij},
  \ee
where $F^P$ are the predicted Fisher matrices for each type of
data. We examine three different dark energy probes, motivated by
experiments which will be contemporary with any experiment that
could use the geometric test. The fiducial cosmological model used
in the Fisher calculations for these CMB, BAO and SNIa experiments
are: $\Omega_m=0.27$, $\Omega_v=0.73$, $h=0.71$, $\sigma_8=0.80$,
$\Omega_b=0.04$, $w_0=-1.0$, $w_a=0.0$, the scalar spectral index
$n_s=1.0$, optical depth to the surface of last scattering
$\tau=0.09$, the running of the spectral index,\
 \be
    \alpha_n= \frac{d  n(k)}{d\ln k},
 \ee
with $\alpha_n=0.0$, the tensor to scalar ratio $r=T/S$ with
$r=0.01$ and the galaxy bias factor, $b$, which we set to $b=1.2$.

\subsubsection{WMAP and Planck Surveyor CMB experiments}

Here we consider both a $4$-year WMAP experiment and a
$14$-month Planck experiment, with predictions calculated using
CMBfast (version 
4.5.1, Seljak \& Zaldarriaga, 1996). We have used a similar
procedure to that outlined in Hu (2002) and Eisenstein et al.
(1998). The Fisher matrix for a CMB experiment is:
\begin{equation}
  F^{CMB}_{ij}=\sum^{\ell_{max}}_{\ell_{min}}\sum_{X,Y}
  \frac{\partial C_{X\ell}}{\partial\theta_i}
  ({\rm Cov}_{\ell})_{XY}^{-1}
  \frac{\partial C_{Y\ell}}{\partial\theta_j}
\end{equation}
where $C_{X\ell}$ is the power for $X=T,E,TE$ or $B$ (Temperature,
E channel polarization, Temperature-E channel cross correlation
and B channel polarization) in the $\ell^{th}$ multipole.

The elements of the symmetric covariance matrix are given in
Eisenstein et al. (1998). For example the $TT$ element of the
covariance matrix is given by:
\begin{equation}
  {\rm Cov_{\ell}}_{TT}=\frac{2}{(2\ell+1)f_{sky}}\left(
  C_{T\ell}+w_T^{-1}B_{\ell}^{-2}\right)
\end{equation}
where $B^2_{\ell}$ is a Gaussian beam window function
$B^2_{\ell}={\rm exp}(-\ell(\ell+1)\theta^2_{{\rm beam}}/8{\rm
ln}2)$ and $\theta_{{\rm beam}}$ is the full-width, half-maximum
(FWHM) of the beam in radians. The inverse square of the detector
noise level on a steradian patch for temperature and polarization
is given by $w_i=(\theta_{{\rm beam}}\sigma_i)^{-2}$ where
$i=T,P$. The sensitivity in $\mu{\rm K}$ per FWHM beam ($\Delta
T/T$ or $\Delta P/T$) is $\sigma_i=\sigma^i_{pix}$.

For multiple channels the quantity $wB_{\ell}^2$ is replaced by
the sum of this quantity for each channel. The values for
$\theta_{{\rm beam}}$ and $\sigma_i$ for the various
experiments were taken from Hu (2002) (Table I), the Planck parameters
are shown in Table \ref{planckparam}. We have used a
maximum  $\ell_{max}=2000$  and minimum $\ell_{min}=10$ in the summation 
over wavenumber. $f_{sky}$ is set to $0.66$ to simulate a
typical galactic cut.

The 11-parameter CMB cosmological parameter set is
($\Omega_m$, $\Omega_v$, $h$, $\sigma_8$, $\Omega_b$, $w_0$,
$w_a$,
 $n_s$, $\tau$, $\alpha_n$, $r=T/S$). We do not include a
marginalization over calibration of the CMB instrument.

\begin{table}
\begin{center}
\begin{tabular}{|l|c|c|c}
 {\bf darkCAM}&&&\\
 \hline
 Area/sq degrees&$z_m$&$z_{\rm max}$&$N_{\rm Bands}$\\
 \hline
 $10$,$000$&$0.70$&$1.5$&5\\
 \hline
 \\
 \\
 \hline
 {\bf Planck}&&&\\
 \hline
 Band/GHz & $\theta_{{\rm beam}}$ &
 $\sigma_T$/$10^{-6}$ & $\sigma_P$/$10^{-6}$\\ 
 \hline
 $44$& $23'$&$2.4$&$3.4$\\
 $70$& $14'$&$3.6$&$5.1$\\
 $143$&$8.0'$&$2.0$&$3.7$\\
 $217$&$5.5'$&$4.3$&$8.9$\\
 \hline
 \\
 \\
 \hline
 {\bf WFMOS}&&&\\
 \hline
 Area/sq degrees&$z_{\rm bin}$&$k_{\rm max}$/$hMpc^{-1}$&Bias\\
 \hline
 $2000$&$1.0$&$0.15$&$1.25$\\
 $300$&$1.0$&$0.15$&$1.25$\\
 \hline
 \\
 \\
 \hline
 {\bf SNAP}&&&\\
 \hline
 $z_{\rm max}$&$N_{\rm bin}$&$N_{SNIa}$&$\sigma_m$\\
 \hline
 $1.5$&$17$&$2000$&$0.15$\\
 \hline
\end{tabular}
\caption{The main default values parameterising the Lensing, CMB,
  BAO and SNIa experiments considered in this paper. For further
  details of the 
  surveys see Section \ref{Priors} and Table \ref{fullresults}.}
\label{planckparam}
\end{center}
\end{table}

\subsubsection{Combining with SNIa experiments}
We have calculated errors on parameters for SNIa experiments for
the proposed SuperNova Acceleration Probe (SNAP; Aldering, 2005)
supernovae experiment using a prescription similar to that outline
in Ishak (2005) and Yeche et al. (2006). The Fisher matrix,
defined by Tegmark et al. (1998) and Huterer \& Turner (2001), is:
 \be
   F^{SNIa}_{ij}=\sum_z^{N_z}\frac{1}{[\Delta {m(z)}]^2}\frac{\partial
  m(z)}{\partial\theta_i}\frac{\partial
  m(z)}{\partial\theta_j}
 \ee
 where $m(z)$ is the apparent magnitude of a supernova at a
given redshift and $N_z$ is the number of supernova bins
in redshift. The apparent magnitude is related to the luminosity
distance by $m(z)={\mathcal M}+5{\rm log_{10}}D_L(z)$ 
where $D_L(z)\equiv (H_0/c)(1+z)r(z)$ is the $H_0$-independent
luminosity distance. The normalization parameter is ${\mathcal
  M}\equiv M-5{\rm log_{10}}(H_0/c)+{\rm constant}$, where $M$ is the
absolute magnitude of a SNIa. 

The effective magnitude uncertainty in a given bin at a particular
redshift, taking into account luminosity evolution,
gravitational lensing and dust and the effect of peculiar velocity
uncertainty is given by (Kim et al., 2003)
 \be
 \Delta m(z)=\sqrt{\sigma_m^2+\left(\frac{5\sigma_{\nu}}{
      cz \ln 10}\right)^2+N_{\rm bin}\delta_m^2}
 \ee
where the scatter in peculiar velocities of $\sigma_{\nu}=500 \
{\rm kms^{-1}}$ is assumed, and the systematic limit
$\delta_m=0.02$ (for a space based experiment). We use the
standard set of $2000$ simulated SNAP supernova distributed in
$16$ redshift bins
 of width $\Delta z=0.2$ between redshifts $0.0\leq z \leq 1.8$ the
 number per bin taken to be the simulated sample from Yeche et
 al. (2006) and Virey et al. (2004).
The full SNIa parameter set is ($\Omega_m$, $\Omega_v$, $w_0$,
$w_a$, $h$).

\subsubsection{Combining with Baryon Acoustic Oscillations experiments}
We have modelled the errors on cosmological parameters for a BAO
experiment, taking a WFMOS-type experiment, following Seo \&
Eisenstein (2003), Blake and Glazebrook (2003) and  Wang (2006). The
Fisher matrix for a BAO experiment can be approximated by
 \be
    F^{BAO}_{ij}=\sum_{k,z}\left[\Delta \ln P(k,z)\right]^{-2}\frac{\partial
    \ln P(k_{\rm eff},z)}{\partial\theta_i}\frac{\partial
    \ln P(k_{\rm eff},z)}{\partial\theta_j}
 \ee
 where $P(k_{\rm eff},z)$ is the linear matter power spectrum
(see Eisenstein \& Hu, 1998) at a redshift $z$ including growth
factors for an arbitrary dark energy cosmology (see Linder, 2003).
The summation is over redshift bins, $z$, and wavenumber $k$.
$k_{\rm eff}$ is an approximation to the observable wavenumber
averaged over both radial and angular direction and is given by
 \be
    k_{\rm eff}= k \left[\frac{r(z)H_{\rm fid}(z)}{
                r_{\rm fid}(z)H(z)}\right]^{1/3}
 \ee
 where the subscript ${\rm fid}$ refers to the comoving distance
$r(z)$ and Hubble parameter $H(z)$ at the fiducial $\Lambda$CDM
cosmology. The fractional uncertainty on the measurement of the power
spectrum is given by
 \be
        \Delta \ln P(k,z)=2\pi\sqrt{\frac{1}{Vk^2\Delta k}}
        \left[1+\frac{1}{nP(k,z)}\right]
 \ee
where $V$ is the volume of the survey. We assume $nP=1$ for all
surveys (see Seo \& Eisenstein, 2003).

The BAO survey assumed has two
redshift slices centred on $z=1.0$ ($0.5<z<1.3$) covering $2000$
square degrees and $z=3.0$ ($2.3<z<3.5$) covering $300$
square degrees. The volume is calculated assuming the area and
redshift ranges at the fiducial cosmology.

We have also calculated the BAO prediction for a survey with an
area of $10000$ square degrees with a median redshift of
$z_m=0.7$, using five redshifts bins with ranges centred upon
$z=0.4$ ($0.3<z<0.5$), $z=0.6$ ($0.5<z<0.7$), $z=0.8$
($0.7<z<0.9$), $z=1.0$ ($0.9<z<1.1$) and $z=1.2$ ($1.1<z<1.3$). To
include the effect of photometric redshift uncertainty 
we add a radial damping term (see Zhan et al., 2005)
 \be
        P(k_{\rm eff},z)\rightarrow P(k_{\rm eff},z)
        e^{-c^2\sigma^2_z(z)k_{\rm eff}^2/H_{\rm fid}^2(z)}
 \ee
where $\sigma_z(z)$ is given by equation (\ref{intdz}).

Alternatively, in an effort to reduce the photometric redshift
error, the matter distribution could be estimated by grouping
galaxies into clusters each containing $n_{{\rm per cluster}}$
galaxies (Angulo et al., 2005). This would have the combined
effects of decreasing the effective number density $n\rightarrow
n/n_{{\rm per cluster}}$ and decreasing the redshift error by
averaging the error over the group
$\sigma_z(z)\rightarrow\sigma_z(z)/\sqrt{n_{{\rm per cluster}}}$. We
found for $n_{{\rm per cluster}}>1$
the marginal errors on $w_0$ and $w_a$ increase, since the effect of 
decreasing number density increases the fraction error on the
power spectrum by more than the decrease in the photometric
redshift error can compensate. Hence we find that using clusters
for the BAO experiment here does not add to the results of the
Planck CMB experiment.

To ensure we are in the linear r\'egime the maximum wavenumber used in all
the surveys is $k=0.15 \ h{\rm Mpc^{-1}}$, and we use $\Delta
k=5\times 10^{-3}\ h{\rm Mpc^{-1}}$. The full parameter set used is ($\Omega_m$,
$\Omega_v$, $h$, $b\sigma_8$, $\Omega_b$, $w_0$, $w_a$,
 $n_s$, $\alpha_n$) where $b$ is a bias factor parameterizing the mapping
of the dark matter distribution to the galaxy distribution. An
important assumption is that the bias is a constant on the scales probed.

\subsection{A Simplified Error Model}
Before considering the full problem of optimizing a weak lensing
survey for the geometric dark energy test, it is useful to
consider a simplified estimate of the parameter uncertainty, so that
the more complex results can be understood in terms of simple relations
between competing effects. The
uncertainty on $w_0$ is roughly given by
 \be
 \label{simp}
 \frac{\Delta w_0}{w_0} \approx \frac{2}{\gamma  \sqrt{N_{\rm B}N_{\rm
      cl}}} \left( \frac{\de \ln R}{\de \ln w_0} \right)^{-1}
 \left( \frac{\sigma_e^2}{N_i} + C^{\gamma \gamma} \right)^{1/2},
 \ee where
 \be
        N_{\rm cl}=A/{\rm fov}
 \ee
is the number of independent clusters or fields in the  analysis,
 \be
    N_{\rm B}\approx z_m/\Delta z
 \ee
is the number of redshift bins behind the lens, where $z_m$ is
here the median redshift of the survey and $\Delta z$ is the
typical redshift error at that depth. The typical number of
galaxies per bin is
 \be
        N_i \approx f_l N_{\rm tot}/N_B N_{\rm cl},
 \ee
where $f_l$ is the fraction of galaxies in the field behind the
cluster, and $N_{\rm tot}$ is the total number of galaxies in the
survey. The terms in this expression arise from two sources. The
first, proportional to $\sigma_e$, is the intrinsic uncertainty
per shear mode due to galaxy ellipticities, and can be beaten down
by increasing the number of galaxies per redshift bin, or by
averaging over more bins, or more clusters. The second term,
proportional to $C^{\gamma \gamma}$ is due to lensing by
large-scale structure in between the lens and the source bins, and
can be reduced by increasing the number of redshift bins (with the
approximation that each lensing bin is independent) and by
averaging over independent clusters. The number of clusters in the
sample scales with median survey redshift as
 \be
    N_{cl}(M\ge 10^{14}M_\odot) = 10 z_m^{3.4}
 \ee
clusters per square degree, where we have cut the cluster sample
off at $10^{14}M_\odot$, where we find the signal contributing to
the measurement of $w_0$ vanishes (see Section 5).

In general we will be interested in fixed-time surveys, where the
survey time scales roughly as \be T = T_0 z_m^{4}f_{\rm sky}, \ee
where $f_{\rm sky}$ is the fraction of the sky covered by the
survey, $z_m$ is the median redshift of the survey, and $T_0$ is a
time constant, the time to observe the whole sky to a median
redshift $z_m=1$ (i.e. to a limiting magnitude of $25$ in the
r-band; see equation \ref{magzrel}), set by the telescope
specifications and number of observed bands. The time scales as
the fourth power of the median redshift due to cosmological
dimming effects and the need to detect the object against the sky
background. As a concrete example we shall use the
Canada-France-Hawaii Telescope (CFHT; Semboloni et al., 2006; Tereno
et al., 2004), which is a 3.6m
telescope with a 1 square degree field of view, integrating over 5
bands, for which $T_0=2 \times 10^4$ nights. We shall also assume
a projected number density on the sky which scales with the median
redshift of the sample as
 \be
    n_2(<z)=30 z_m^{3.4}\,\,\,\, {\rm galaxies\,\, per\,\,
    square\,\,arcmin},
 \ee
as measured from the COMBO-17 survey, an angle averaged shear-shear
correlation function,
 \be
        C^{\gamma\gamma} = 10^{-5} z_m^{1.6},
 \ee
and an intrinsic ellipticity dispersion
 \be
        \sigma_e =0.3.
 \ee
With this simplified error model, we find the fractional error on
$w_0$ scales as
 \be
        \frac{\Delta w_0}{w_0}= 0.062 z_m^{-1.35}(1+24.1z_m^{4}\Delta z)^{1/2}.
 \label{simperror}
 \ee
The leading term here is due to shot-noise, while the second term
in quadrature is due to large-scale sampling variance. Assuming we
have ten redshift bins, so that $\Delta z=0.07$ is typical of the
photometric redshift error, equation (\ref{simperror}) minimizes
at $z_m \approx 1.0$. For a fixed-time survey we find that for a
shallow, low-$z$, wide area survey, the error on $w_0$ is
dominated by shot-noise. Here the signal is not very large, and
the number of background galaxies (and therefore combinations of
background source planes) is too low. For a deep survey this
becomes dominated by large-scale structure clustering. This occurs
because we have to make the survey area smaller to compensate for
the depth. Hence we have fewer clusters to average over and reduce
the clustering noise. Both sources of noise increase with the size
of the redshift error, $\Delta z$. In the case of shot-noise this
is again because we have fewer combinations of source planes to
sum over. In the case of clustering noise-dominated there is a
stronger effect because we have fewer source planes to average out
the effects of clustering.

\subsection{Survey Optimization}
The optimizations discussed in the following Sections only include a
CMB Planck experiment, the combination with further experiments is discussed
in Section \ref{COMSBS}. For a weak lensing and photometric redshift
survey on a given telescope for a set amount of observing time, the
survey itself is characterized by the area, parameterized here by
$f_{\rm sky}$, the median redshift, $z_m$, of the survey in the band
used for weak lensing (usually the $r$- or $i$-band) and the number
of bands used for photometric redshift accuracy, $N_{\rm bands}$.
For a given number of bands we only
have one free parameter, which we shall assume is the median
redshift, $z_m$.

\begin{figure}
 \psfig{figure=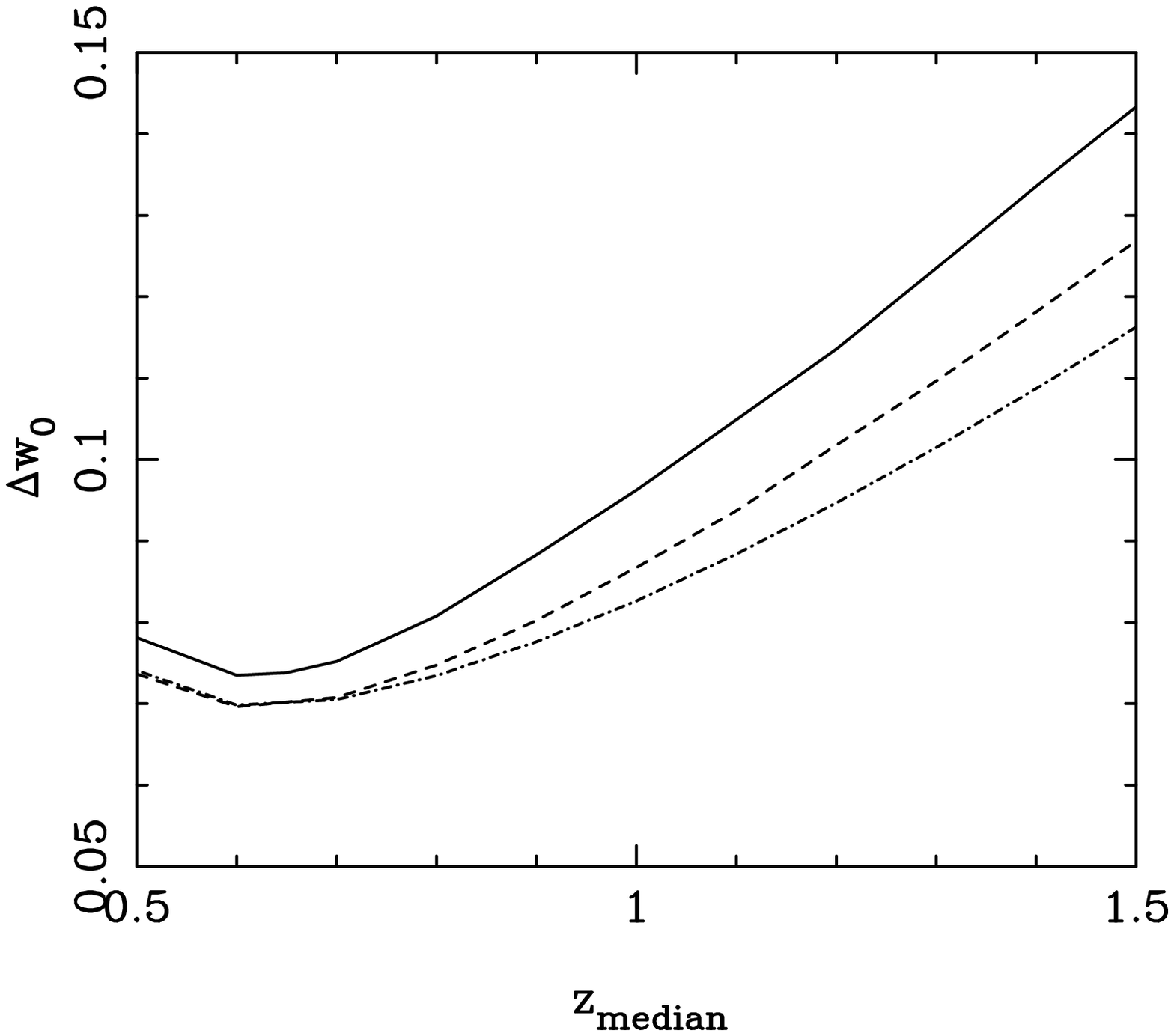,width=\columnwidth,angle=0}
\caption{The uncertainty on $w_0$, marginalized over all other
parameters, as a function of median redshift, $z_m$, for a
time-limited survey, assuming a prior from a 14-month Planck
experiment. The survey 
area is $A=10,000(z_m/0.7)^{-4}$ square degrees. We set a lower
limit of $z=0.5$, which would correspond to a hemisphere. The
solid line is for $5$-band photometric redshift survey, the dashed
line for $9$-band and the dot-dashed line for $17$-band. Note that
the time constraint is only on the 5-bands, assuming that the
other bands will come from other surveys. Note, see Section
\ref{Ground-based Ellipticity Measurements},
that we take an upper redshift limit of $z_{\rm max}=1.5$.}
 \label{VISTAdwzmed}
\end{figure}

Our procedure is to vary $z_m$, calculating the survey area by
equation (\ref{surveyeq}). With the galaxy number distribution and
number counts, we can calculate the Fisher matrix and hence the
marginalized uncertainty on a measurement of $w_0$. Figure
\ref{VISTAdwzmed} shows the  marginalized error on $w_0$ (assuming
a 14-month Planck experiment) for a $D=4$m class telescope with a 2
square degree field of view for a variety of numbers of
photometric bands. For example a $5$-band survey would be the case
for, e.g., the Dark Energy Survey on the CTIO Blanco telescope or
the darkCAM survey. The results reflect our analysis of the simple
analytic model. For a shallow, wide survey the lensing signal is
not strong, the number of background galaxies is low and so the
error on $w_0$ is shot-noise dominated. The error on $w_0$ is poor
beyond $z_m\sim 0.7$, indicating that clustering noise is a strong
effect. The small variation with the number of optical bands is due to
the 
effect that, despite the marginal error of the geometric test
decreasing, the intersection with the Planck experiment does not
substantially change. This is investigated further in Section
\ref{Optical and Infrared surveys}.

The optimal survey is a, 5-band, 18,500 square degree survey with median
redshift $z_m=0.6$, combined with a 14-month Planck survey. However
note that the dependence on
median redshift is shallow about the minimum and that the optimal survey when
considering a figure of merit (see Section \ref{Figure of Merit}) is a
5-band, 10,000 square degree survey with median redshift $z_m=0.7$, so
that from hereon, and in Section \ref{paraforecast}, we will use this
as our fiducial survey design.

\subsection{Optical and Infrared surveys}
\label{Optical and Infrared surveys}
In the last few years multi-band surveys have started to
open up the high redshift Universe. Hence it is now possible to
combine 5-band optical surveys with 4-band infrared surveys for
9-band photometric redshifts. We can study the effect of varying
the number of assumed additional bands available on the
measurement of dark energy parameters by varying the photometric
redshift error.
\begin{figure}
 \psfig{figure=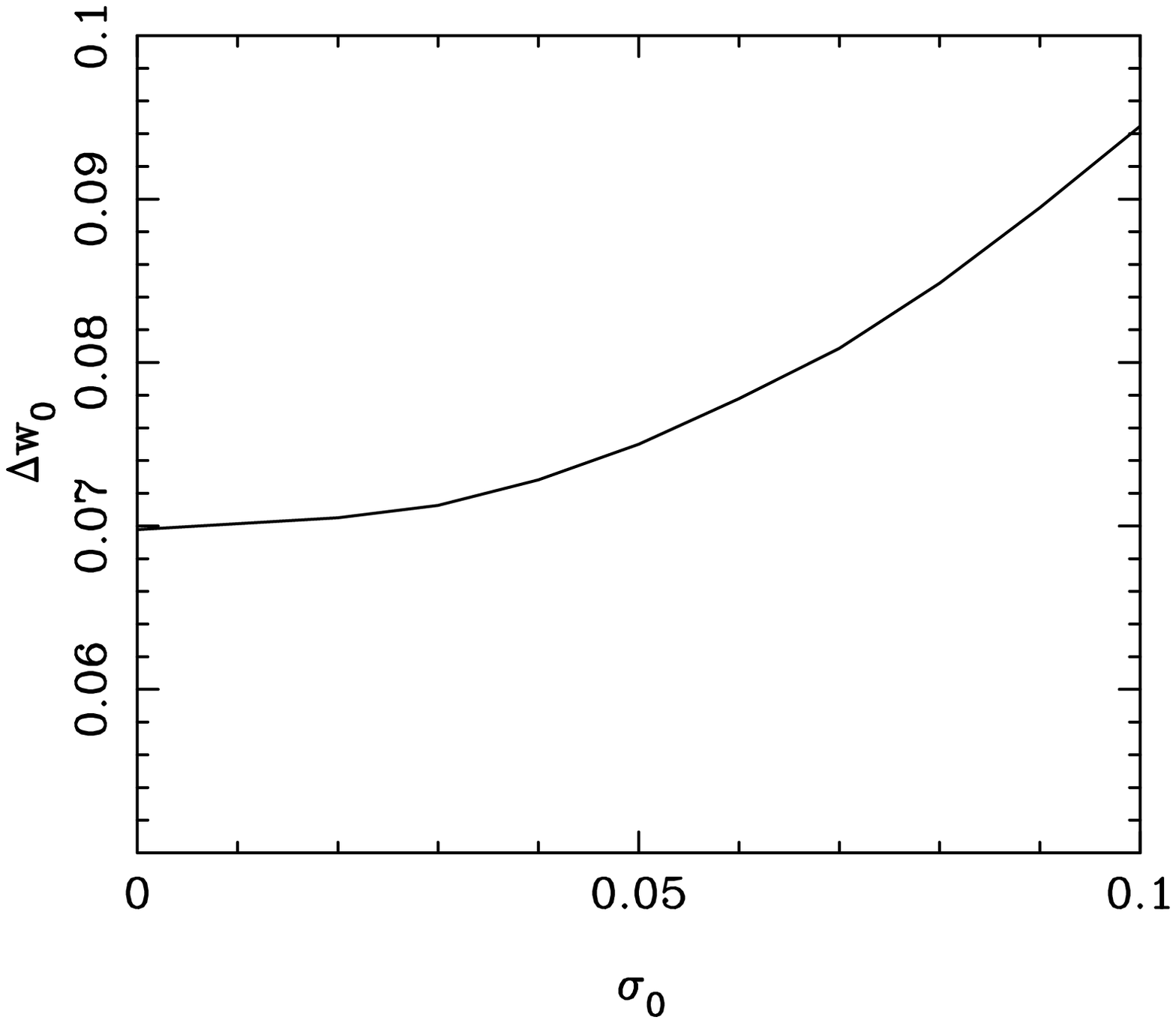,width=\columnwidth,angle=0}
\caption{The uncertainty on $w_0$, marginalized over all other
parameters with a 14-month Planck experiment, as a function of
photometric redshift accuracy, parameterized by
$\sigma(z)=\sigma_0(1+z)$. The normalization, $\sigma_0$ scales
roughly as the number of photometric bands as $\sigma_0\propto
N_{\rm bands}^{-1}$, where we find $\sigma_0=0.05$ for a 5-band
photometric redshift survey and $\sigma_0=0.01$ for a 9-band (4-band
infrared and 5-band optical) 
infrared and optical photometric redshift survey. }
 \label{w_vs_sigz}
\end{figure}
Figure \ref{w_vs_sigz} shows the variation of the accuracy on
$w_0$, marginalized over all the other parameters with a 14-month
Planck experiment, as a function of varying the accuracy of the
photometric redshifts. We parameterize this by defining
 \be
    \sigma_z(z)= \sigma_0(1+z).
 \ee
A value of $\sigma_0=0.05$ is
approximately appropriate for a 5-band photometric redshift
survey, while $\sigma_0=0.01$ corresponds to a 9-band
(4-band infrared and 5-band optical) photometric redshift survey.
For a 5-band survey ($\sigma_0=0.05$) we find $\Delta w_0=0.075$,
while for a 9-band (4-band infrared and 5-band optical) photometric
redshift survey ($\sigma_0=0.01$) 
we find $\Delta w_0=0.071$. Note this is distinct from a $9$ band
optical survey considered up until this point.

If the photometric redshifts are degraded, for instance if fewer
than five bands are available, the accuracy of $w_0$ is also
degraded. By the time $\sigma_0=0.1$ (for, say, 3-bands), the
error has increased to $\Delta w_0=0.094$. Note we have not
included the effect of outliers here (see Section 8.4), which
will degrade the signal further.

\begin{figure}
 \psfig{figure=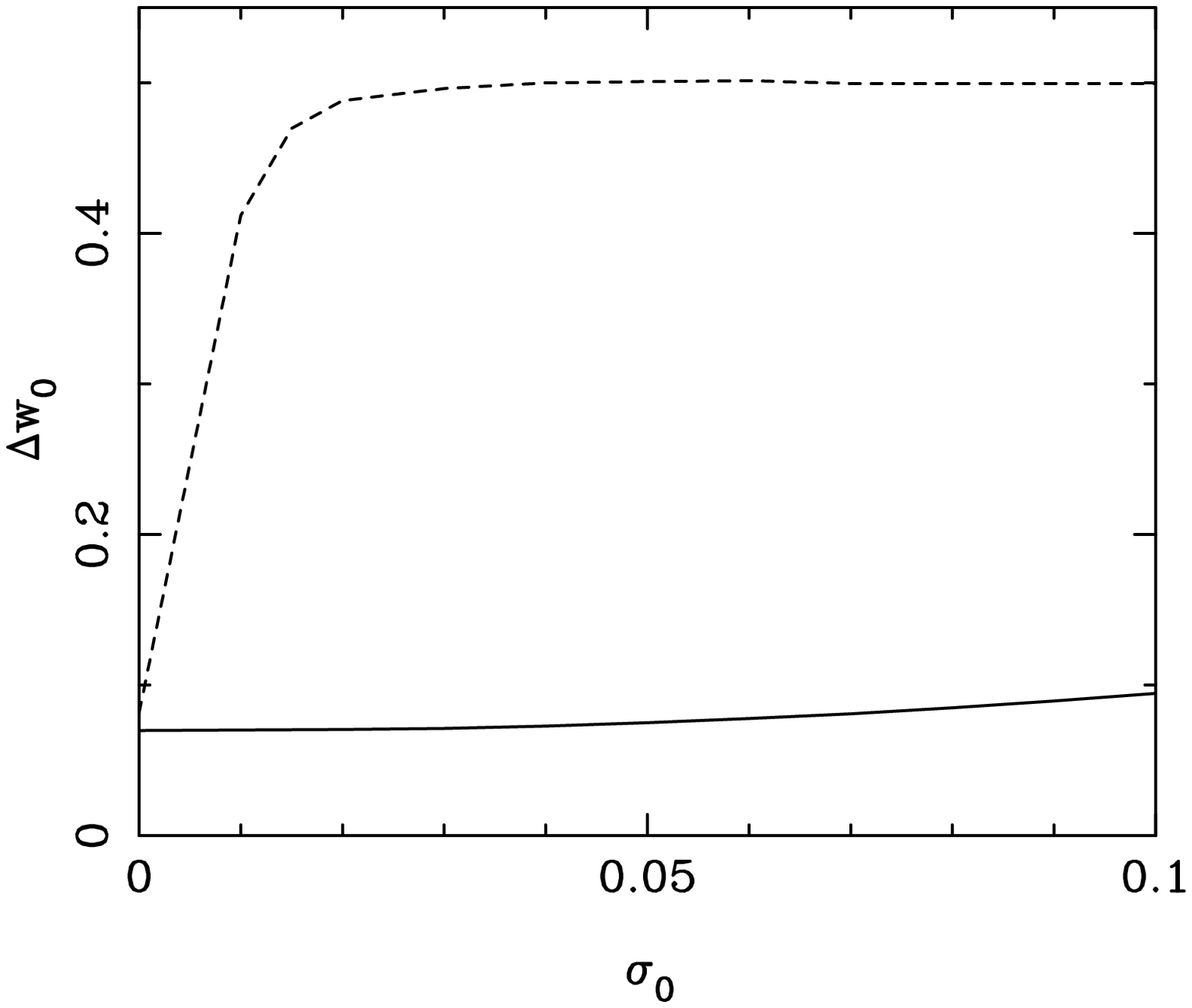,width=\columnwidth,angle=0}
\caption{The uncertainty on $w_0$, marginalized over all other
parameters with a 14-month Planck experiment, as a function of
photometric redshift accuracy, parameterized by $\sigma_0$. The solid
line are the geometric test constraints, the dashed line are the
constraints using BAO from a $10,000$ square degree survey with $z_m=0.7$.}
 \label{sigma0_BAO}
\end{figure}

We have found that using BAO to measure dark energy from a
photometric redshift survey is difficult as the damping term due to
the photometric 
redshifts, effectively constraining the range of Fourier modes
available to analyze, quickly reduces the amount of cosmological
information that can be extracted. Figure \ref{sigma0_BAO} shows the
variation of the error achievable using BAO from a photometric redshift
survey, the error is simply the CMB error until $\sigma_0\approx
0.02$ where the BAO constraint begins to improve the a 14-month Planck
CMB error. To constrain dark energy using a photometric redshift
survey many 
bands (possibly infrared) would be vital over the whole redshift range
to decrease the 
photometric redshift error. As the redshift error becomes
$\sigma_z(z)\rightarrow 0$, as would effectively 
be the case for a spectroscopic survey, the geometric test constraints
and the BAO constraints are comparable. 

\subsection{Scaling results to other surveys}
To scale these results to other weak lensing surveys, equation
(\ref{timelim}) should be used with a time calibration i.e.
 \begin{equation}
  \label{TT0cal}
  \frac{T}{T_0}= \left(\frac{z_m}{z_{m0}}\right)^{4}
  \left(\frac{A}{A_{0}}\right)
  \left(\frac{D}{D_{0}}\right)^{-2}
  \left(\frac{fov}{fov_{0}}\right)^{-1}.
 \end{equation}
The subscript $0$ refers to parameters time, median redshift and area of
a survey on a telescope with certain diameter and field of view.
The scaling applies between surveys with equal number of bands;
for $5$ bands the Canada-France-Hawaii Telescope Legacy Survey
(CFHTLS) can be used, while for $17$ bands COMBO-17 can be used.
Although it can be naively assumed that the time for a given
survey scales proportionally with the number of bands so that $T_0
\rightarrow T_0 N_{b0}/ N_{b}$ where $N_{b}$ is the number of
bands in the survey.

\begin{table}
\begin{center}
\begin{tabular}{|l|c|c|}
 \hline
      &   CFHT    &  COMBO-17 \\
 \hline
 D(m) &   3.6      &    2.2 \\
 fov (sq deg.) & 3 &    1    \\
 N (bands) &   5   &    17   \\
 $z_m$  &   1.17   &  0.7    \\
 Area (sq. deg.)& 170 &  1  \\
 T (nights)& 500   &   6   \\
 \hline
\end{tabular}
\caption{Default survey parameters for the 5-band CFHT Legacy
Survey and the 17-band COMBO-17 survey. }
\label{surveytable}
\end{center}
\end{table}

One of two questions may arise. What is the error on $w_0$ (or
$w_a$) that can be achieved given $T$ nights on a given telescope,
and freedom to choose the survey design? Or, given a
survey of area $A$ and median redshift $z_{m}$ what is the
constraint on $w_0$ (or $w_a$) that can be achieved? Both of these
questions can be answered using the information given here.

If the field of view of the telescope is small enough so that only
approximately one cluster will be observable per pointing then a
targeting strategy should be used. In this case
Figure 6 should be used so that given $P$ pointings on a given
telescope the appropriate marginal error can be predicted. For a
targeting strategy the time trade-off is determined not by the
total area covered but by the number of pointings. The number of
pointings achievable given $T$ nights to a redshift $z_{m}$ can
be expressed, as
\begin{equation}
  P=\left(\frac{T}{T_0}\right)
            \left(\frac{z_{m0}}{z_{m}}\right)^4
        \left(\frac{D}{D_0}\right)^2\frac{fov_0}{A_0}.
\end{equation}
The achievable marginal errors from a targeting strategy are
however limited due to the large amount of clusters which need to
be observed for a tight dark energy constraint.

Given the freedom to choose any wide-field surveys median redshift,
the optimal median redshift of $z_m\approx 0.7$ is insensitive to the
number of bands, when combined with a Planck
prior (see Figure 10). Equation (\ref{TT0cal}) should then be used, with the
appropriate calibration, to calculate the area achievable given $T$
nights. If the number of bands is $5$, $9$ or $17$ the appropriate
line in Figure 10 then scales proportionally up (and down) with
decreased (or increased) arial coverage from $10,000$ square
degrees, for a $5$ band survey i.e. $\Delta
w_0(A)=(0.075)(A/10,000)^{-1}$. If the number of bands is not shown in
Figure 10 then Figure 11 can be used to find the minimum of the
appropriate $\Delta w_0$ vs. $z_m$ line (at $z_m=0.7$). This can
then be scaled for a differing arial coverage as before.

Given a fixed survey of area $A$ and median redshift $z_{m}$
Figures 10 and 11 can be used in a similar way. Given the error in
Figure 10 for a given median redshift $\Delta w_0(z_{m})$ the
achievable error can be calculated using $\Delta w_0(A)=\Delta
w_0(z_{m})(10,000/A)$. In scaling between bands a similar
interpolation between Figure 10 and Figure 11 can be performed.

\subsection{Constraining $w(z)$ at higher redshifts}
\subsubsection{Pivot redshifts}
As well as constraining the marginalized dark energy equation of
state, $w(z)$, at $z=0$ ($w_0$), we can combine the measured
accuracy of $w_0$ and $w_a$ to estimate the measured accuracy of
$w(z)$ at higher redshift. Here we can gain some information by
using the degeneracy between $w_0$ and $w_a$ (see Section
\ref{paraforecast}), to find a redshift where the anti-correlation
combines to minimize the error.
\begin{figure}
 \psfig{figure=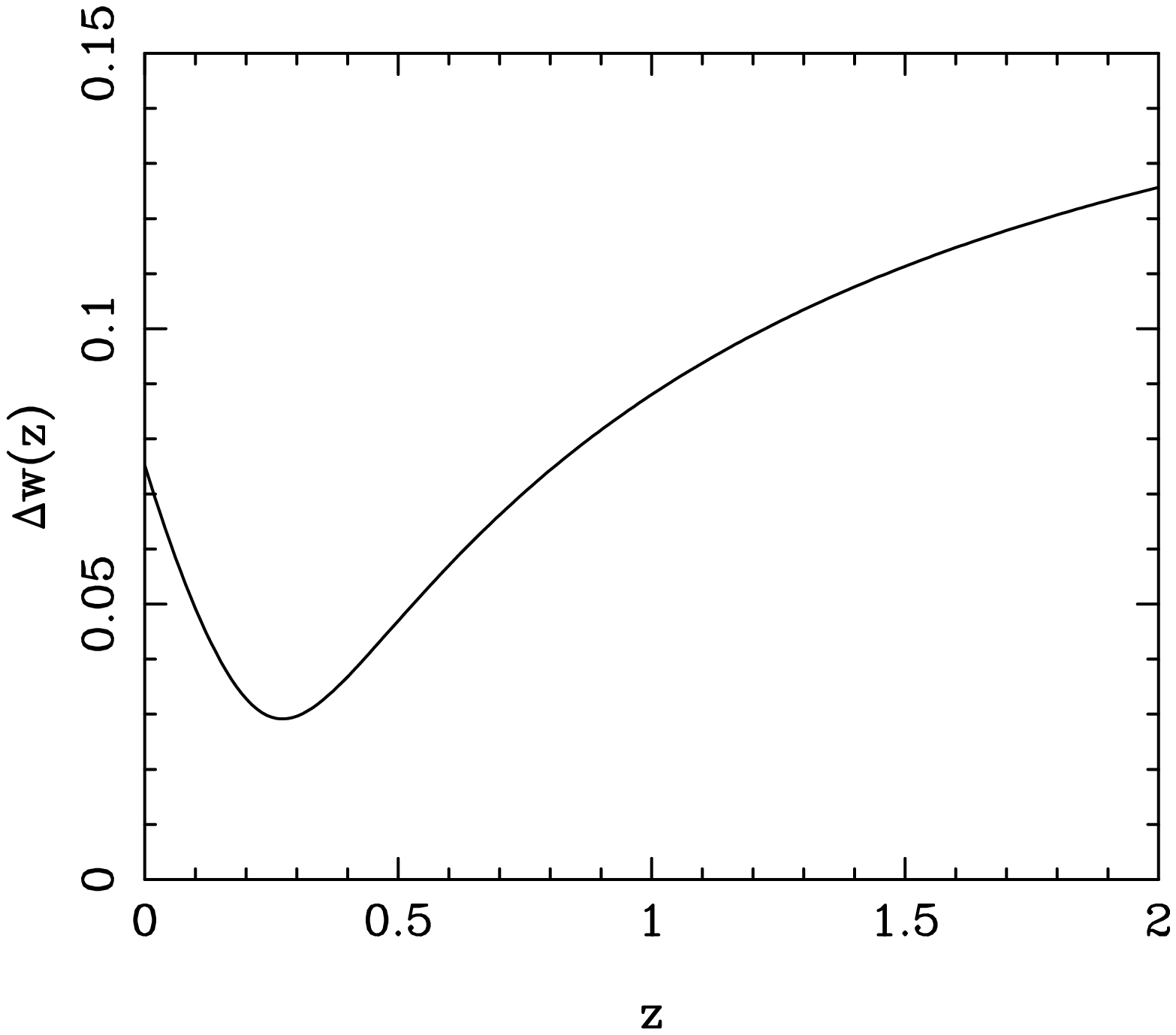,width=\columnwidth,angle=0}
\caption{The uncertainty on $w(z)$, the dark energy equation of
state measured at different redshifts, marginalized over all other
parameters. For gravitational lensing combined with 14-month Planck
experiment.  This 
shows that the highest accuracy constraint on $w(z)$ occurs at
$z=0.27$ with $\Delta w(z=0.27)=0.0298$.}
 \label{zerror}
\end{figure}
\begin{figure}
 \psfig{figure=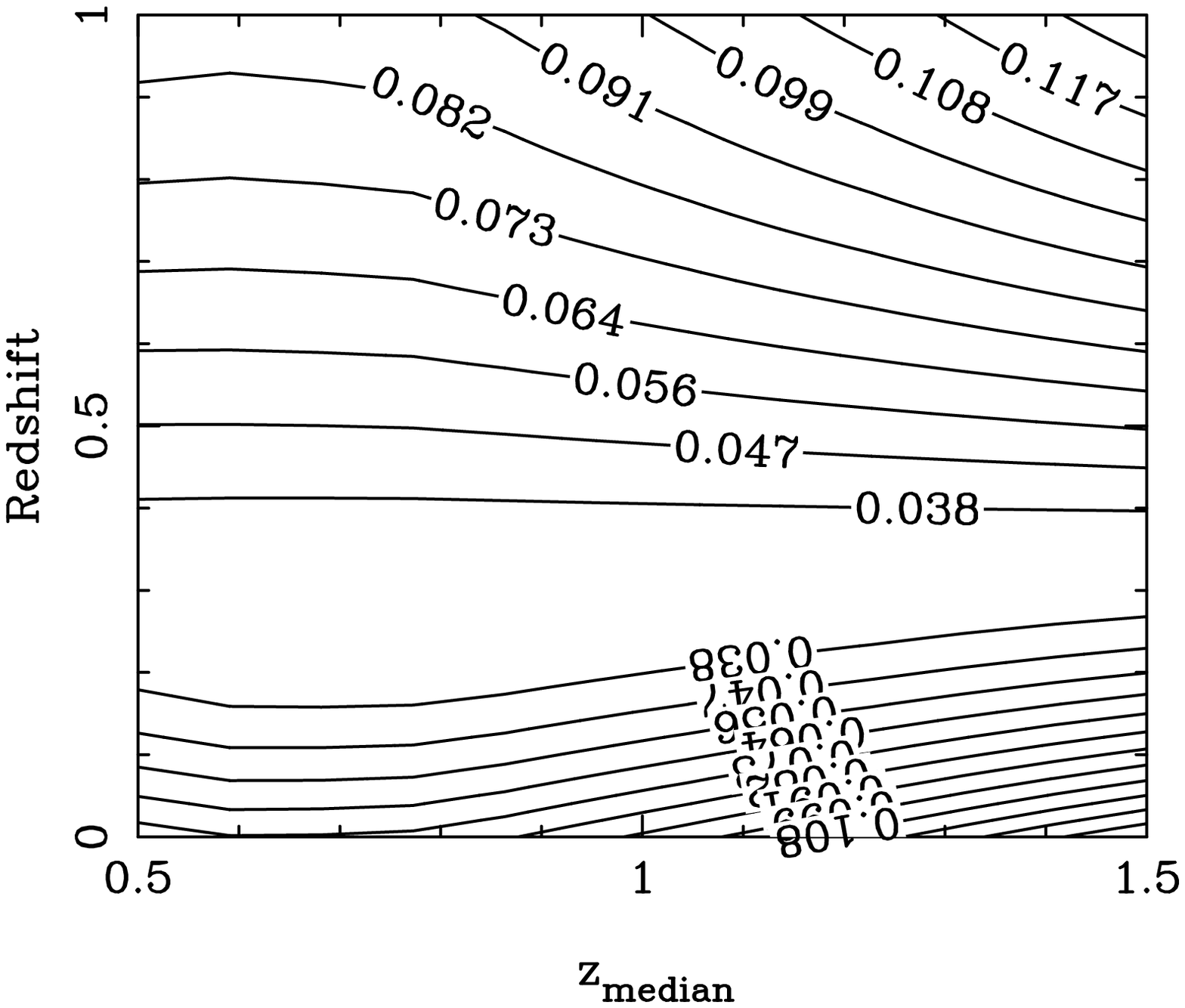,width=\columnwidth,angle=0}
\caption{The uncertainty on $w(z)$, the dark energy equation of
state measured at different redshifts, marginalized over all other
parameters for gravitational lensing combined with 14-month Planck
experiment, 
and its dependence on median redshift. The contours are
lines of equal marginalized $w(z)$ error, the numbers on the lines
being the marginal error on that line.}
 \label{zerrorcont}
\end{figure}
Figure \ref{zerror} shows the expected accuracy of
$w(z)=w_0+w_az/(1+z)$ as a function of redshift for a
5-band, 10,000 square degree survey with median redshift $z_m=0.7$,
combined with a 14-month Planck survey. The
highest accuracy
measurement occurs at $z=0.27$, where $w(z=0.27)=0.0298$. This
low-redshift pivot redshift for the geometric test is due to its
insensitivity to $w_a$.

Figure \ref{zerrorcont} shows how the error on $w(z)$ varies with
both redshift, $z$, and with median redshift of the survey, $z_m$,
for the same time-limited survey. It can be seen that the
minimization in the error at $z\approx 0.7$ in Figure
\ref{VISTAdwzmed} is reproduced at the $z=0$ line (along the
x-axis) of the plot, and Figure \ref{zerror} is reproduced by
considering the variation in the error along the $z_m=0.7$ line. It is
clear that if one is concerned with
optimizing a survey design to constrain the error on $w(z)$ at an
optimal redshift then there is little sensitivity to the survey
design. This is due to the effect of intersection, that is even though
the lensing only error may be varying the intersection of the
lensing ellipse with the Planck experiment ellipse does not vary
considerably in width (characterized by the width of the inner
contour) or orientation (characterized by the value of $z$ at
which the error on $w(z)$ minimizes).

\subsubsection{Figure of Merit}
\label{Figure of Merit}
\begin{figure}
\psfig{figure=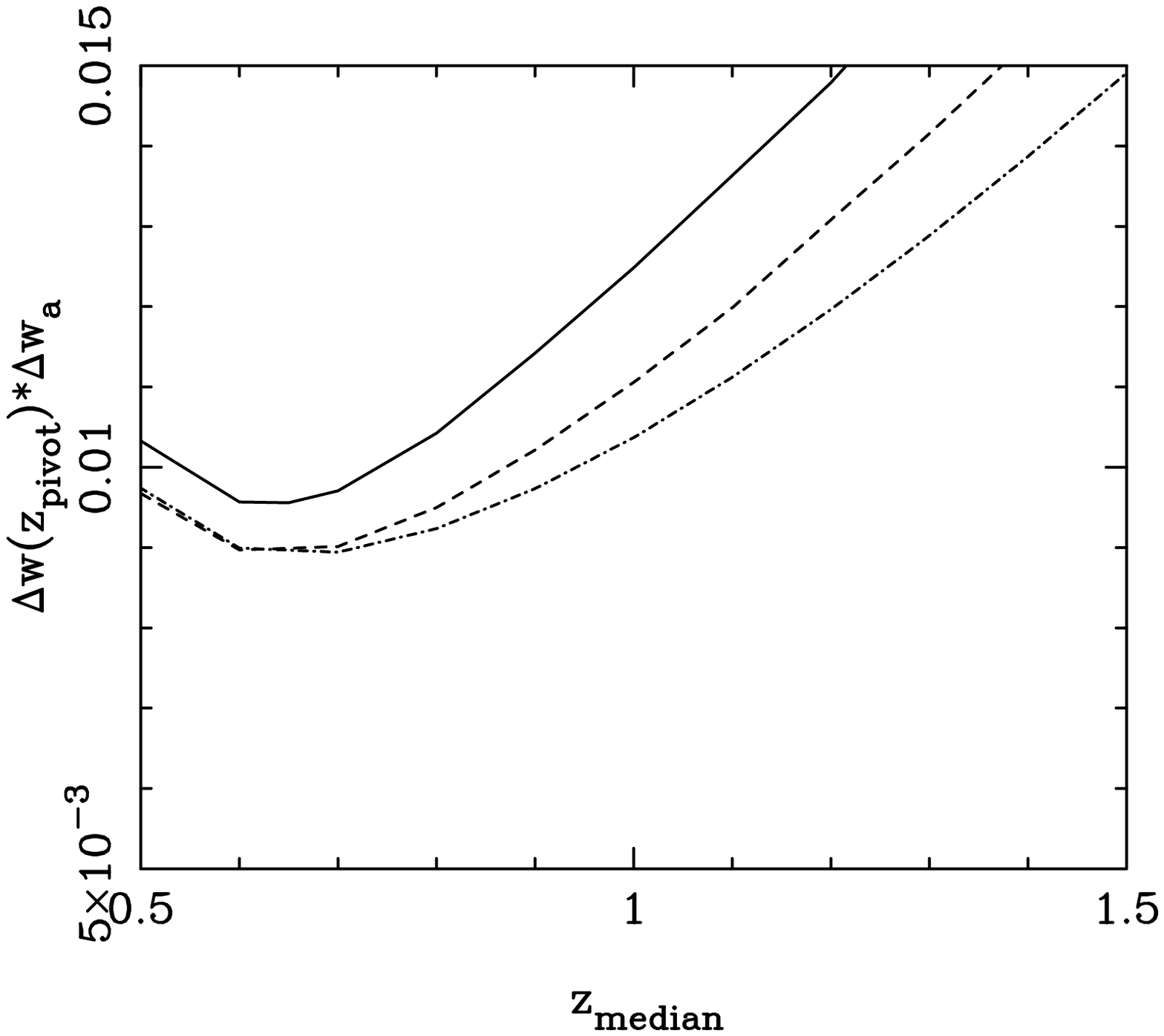,width=\columnwidth,angle=0}
\caption{The figure of merit as a function of median redshift, $z_m$,
for a  time-limited survey, assuming a 14-month Planck prior. The survey
area is $A=10,000(z_m/0.7)^{-4}$ square degrees. The
solid line is for $5$-band photometric redshift survey, the dashed
line for $9$-band and the dot-dashed line for $17$-band.}
 \label{figomedian}
\end{figure}

A useful `figure of merit' (Linder, 2003; Linder, 2006; Dark Energy
Task Force, DETF 2006) in dark
energy predictions can be constrained
by considering the smallest area of parameter space constrained by a
given experiment. The dark energy equation of state can be written as:
\be
w(a)=w_i+w_a(a_i-a)
\ee
where $w_i\equiv w(a_i)$ and we have expanded around scale factor
$a_i$. The error on $w(a)$ is:
\be
\Delta w(a)^2=\Delta w_i^2+(a_i-a)^2\Delta w_a^2+2(a_i-a){\rm Cov}(w_i,w_a)
\ee where ${\rm Cov}(w_i,w_a)$ is the covariance between $w_i$ an
$w_a$ (equal to the corresponding inverse Fisher matrix element). By
taking the derivative of this quantity the scale factor at which the
error minimizes can be found 
\be 
a_{{\rm min}}=a_i+\frac{{\rm Cov}(w_i,w_a)}{\Delta w_a^2}. 
\ee 
In the standard expansion in
equation (6) $a_1=1$ and the above expression reduces to the
equation for the pivot redshift. In this formalism the pivot
redshift occurs when the covariance between the $w_i$ and $w_a$ is
zero. This is equivalent to the pivot redshift in
the formalism of equation (6). The ellipse at the pivot redshift is
then the smallest ellipse constrained by a given experiment. Since
this ellipse is de-correlated its area can be simply approximated by
\be 
\Delta w(a_{\rm pivot})*\Delta w_a. 
\ee 
This is the figure of
merit used to quantify the performance of any given experiment: the
smaller the figure of merit the tighter the constraints on the
equation of state of dark energy will be over a larger redshift
range. Broadly it can visualized by comparing Figure
\ref{zerrorcont} and Figure \ref{figomedian}, the figure of merit is
minimized where the lowest contour in Figure \ref{zerrorcont} is
widest, this can be seen in Figure \ref{figomedian}. It can be seen
that the optimal experiment when considering the figure of merit is
at a median redshift of $z_m=0.7$ for $5$ bands. The figure of merit
is shown for all considered experiments in Table \ref{fullresults}.

\section{Parameter Forecasts}
\label{paraforecast}
Having found the optimal survey strategy to measure the dark
energy equation of state for a given experiment, we can now
investigate the constraints on the full parameter space.
Throughout we shall assume a 10,000 square degree 5-band
photometric redshift weak lensing survey with a median redshift of
$z_m=0.7$ ($r=23.8$).

In this Section we shall discuss dark energy parameter constraints
from geometric lensing alone (Section 7.1), combined with the WMAP
4-year and 14-month Planck experiments (Section 7.2), and combined with
a WFMOS BAO experiment and SNAP SNIa experiment in Section 7.2 and 7.3. A 
table of the different surveys we have considered, and the
predicted marginal errors on the dark energy parameters, is
presented in Table \ref{fullresults}.

Using the full Fisher matrix formalism for parameters in a
consistent cosmological model we can estimate 
the accuracy on a set of cosmological parameters for a given
experiment, taking into account marginalization over all other
parameters. The details of the Fisher analysis
are discussed in Section 6.1. The
11-parameter cosmological parameter set we shall use is
($\Omega_m$, $\Omega_v$, $h$, $\sigma_8$, $\Omega_b$, $w_0$,
$w_a$, $n_s$, $\tau$, $\alpha_n$, $r=T/S$), with default values
(0.27, 0.73, 0.71, 0.8, 0.04, -1.0,0.0, 1.0, 0.09, 0.0, 0.01). We
shall compare and combine analysis with the results from a weak
shear spectral analysis (e.g. Heavens et al., 2006) elsewhere.

\subsection{Parameter forecasts for the geometric lensing test alone}
\begin{figure}
 \psfig{figure=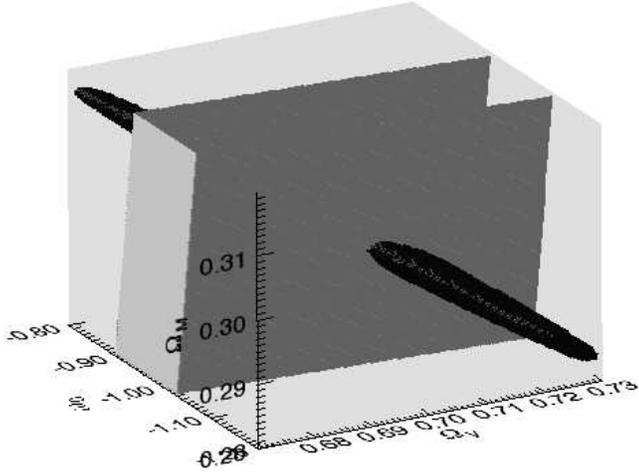,width=\columnwidth,angle=0}
 \caption{3D parameter space for a 10,000 square degree lensing survey
 to a median redshift of $z_m=0.7$ with 14-month Planck
 experiment, with no dark energy evolution. The volumes bounded by
 light and dark grey which represent the 1-$\sigma$ parameter
 estimations for weak lensing 
 and a  14-month Planck experiment respectively.}
 \label{VISTA-A3D}
\end{figure}

On its own, the geometric test constrains a sheet in the
likelihood space of ($w_0$, $w_a$, $\Omega_v$, $\Omega_m$). Figure
\ref{VISTA-A3D} shows this plane in the 3-space of ($w_0$,
$\Omega_v$, $\Omega_m$), having marginalized over $w_a$ (light grey
plane). The surface here encloses the 3-parameter, 
1-$\sigma$ likelihood surface. The equation of
this plane in the full 4-parameter space is
 \be
        X= 0.64 w_0 -0.31 w_a -0.35 \Omega_v - 0.67 \Omega_m.
 \ee
For model
parameters of $w_0=-1$, $w_a=0$, $\Omega_m=0.27$, and $\Omega_v=0.73$ this
can be evaluated to give
 \be
    X = -1.08,
 \ee
which can be measured with an expected accuracy of
 \be
    \Delta X = 0.031.
 \ee
If we fix $w_0=-1$ and $w_a=0$, we can see that the geometric test constrains
the degenerate line $\Omega_v + 1.91 \Omega_m=1.26$. This can be
compared with the CMB constraint on the density parameter plane of
$\Omega_v + \Omega_m =1$.

We can project this onto a 2-parameter space, marginalizing over
all other parameters. Figure \ref{WMAPLENS} show the 2-parameter,
1-$\sigma$ (68.3\% confidence) likelihood contours for the
parameter space of $\Omega_v$, $\Omega_m$, $w_0$ and $w_a$. The
lightest grey solid block is the constraint on parameters from
the lensing geometric method only. Here again we see the large
degeneracies between the geometric parameters. In particular it is
again clear that the geometric test is very insensitive to $w_a$
(see Section 2.3). The 1-parameter, 1-$\sigma$
marginalized parameter 
uncertainties can be found by projecting these contours onto each
axis and dividing by $2.3$. These are presented in Table
\ref{fullresults}.

\subsection{Comparing and combining the geometric lensing and the
CMB}
To lift the degeneracies in the geometric test we can combine our
predictions with 
results expected from the CMB. Here we consider combining with the expected
results from the 4-year WMAP experiment. Below we shall compare
with the results expected from a 14-month experiment with the
Planck Surveyor.

\subsubsection{Combining with WMAP}
\begin{figure}
\psfig{figure=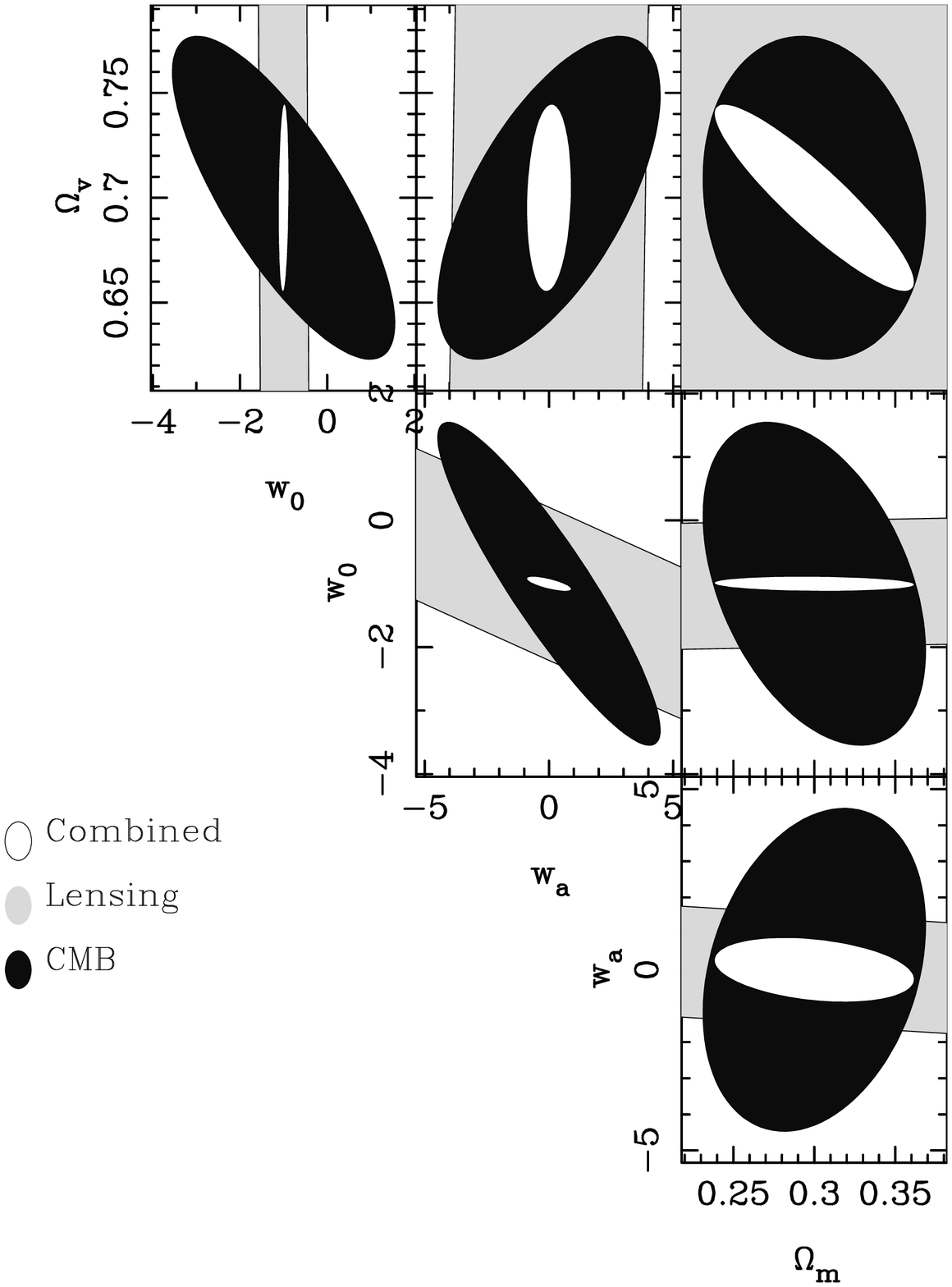,width=\columnwidth,angle=0,clip=}
 \caption{Two-parameter, 1-$\sigma$ (68.3\% confidence) likelihood contours
 for geometric parameters for a 10,000 square degree
 lensing survey geometric analysis to a median depth $z_m=0.7$,
 compared and combined with the expected 4-year WMAP results.}
 \label{WMAPLENS}
\end{figure}

The parameter forecasts for a $4$-year WMAP survey are compared
and combined with the geometric test, allowing for spatial
curvature, in Figure \ref{WMAPLENS}. The lightest grey ellipses
are the geometric test alone, the darkest ellipses are the
marginalized parameter forecasts for WMAP, while the central
white ellipses, show the combined likelihood contours for the
combined CMB and geometric methods. We have suppressed the
amplitude of density perturbations parameterized by $\sigma_8$,
the Hubble parameter $h$, the optical depth $\tau$, and the
tensor-to-scalar ratio $r$, which are also estimated by the CMB.
We shall consider these parameters in Section 7.2.2.

Figure \ref{WMAPLENS} illustrates the poor sensitivity of the CMB
to $w_0$ and $w_a$, but constrains the curvature of the model by
the combination $\Omega_m+\Omega_v$. The response of the CMB to
dark energy comes mainly from the Integrated Sachs-Wolf (ISW)
effect. Combining the geometric lensing test and the CMB, we find
the orthogonality of the two methods reduces the error on the dark
energy parameters from $\Delta w_0({\rm WMAP})=1.268$, $\Delta
w_a({\rm WMAP)}=2.225$ to $\Delta w_0({\rm
WMAP+GL})=0.089$ and $\Delta w_a({\rm WMAP+GL})=0.714$. There is also
marginal improvement in $\Delta \Omega_m$ and $\Delta \Omega_v$.
The main improvement to the lensing analysis is the WMAP
constraint on the curvature of the Universe in the $\Omega_m$,
$\Omega_v$ parameter plane. To get a clearer picture of the
orthogonality of the CMB 4-year WMAP and lensing geometric test
results, we plot a 3-D view of the one-parameter, 1-$\sigma$
parameter surfaces in Figure \ref{VISTA-A3D}. This shows the
$w_0$, $\Omega_m$, $\Omega_v$ parameter surfaces, marginalized
over all other parameters, including $w_a$.

\subsubsection{Combining with Planck Surveyor}
\begin{figure}
 \psfig{figure=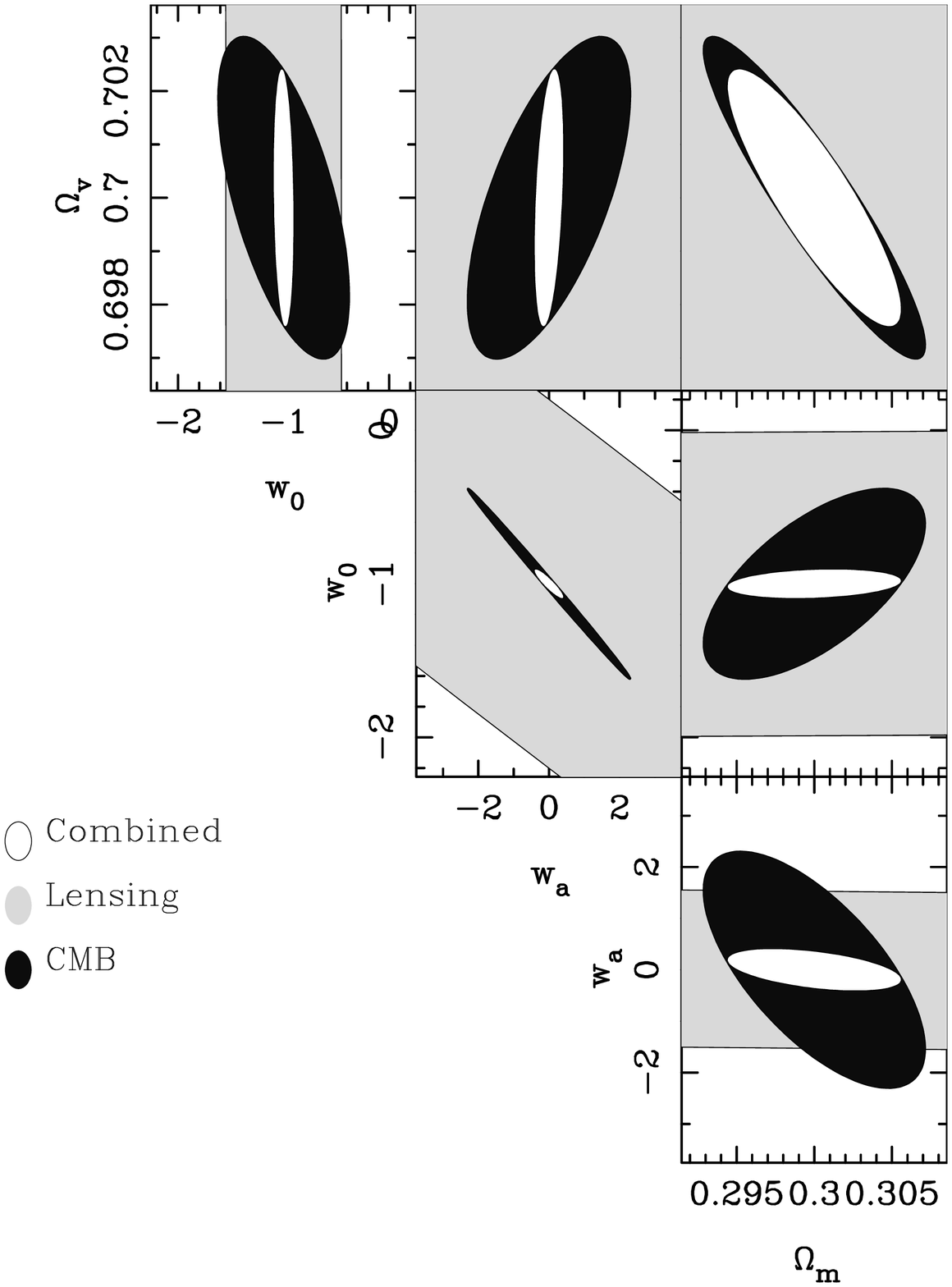,width=\columnwidth,angle=0,clip=}
 \caption{Two-parameter, 1-$\sigma$ (68.3\% confidence) likelihood contours
 for geometric parameters  for a 10,000 square degree
 lensing survey to a median depth $z_m=0.7$, with a 14-month Planck
 experiment. Note the change in the scale of the axes from Figure
 \ref{WMAPLENS}, from hereon the remaining Figures will uses the scale
 introduced in this Figure.} 
 \label{PlanckLENS}
\end{figure}

We can compare the information in Figure \ref{WMAPLENS} from a
4-year WMAP experiment with that of a $14$-month Planck Surveyor
experiment, shown in Figure \ref{PlanckLENS}. While the Planck
error ellipses (darkest grey) are considerably smaller than those of
the 4-year WMAP, the degeneracy between $w_0$ and $w_a$ remains.
On its own Planck can measure $w_0$ to an accuracy of $\Delta
w_0=0.502$ and on $w_a$ to an accuracy of $\Delta w_a=1.86$, with
the main source of information from the Integrated Sachs-Wolfe
(ISW) effect. Again the curvature of the model is well constrained
by the CMB. Combining Planck with the geometric lensing test
reduced the dark energy parameter uncertainties to $\Delta
w_0=0.075$ and $\Delta w_a=0.326$, a factor of $\sim 7$ improvement in
the measurement of $w_0$ over Planck alone.

\begin{figure*}
 \psfig{figure=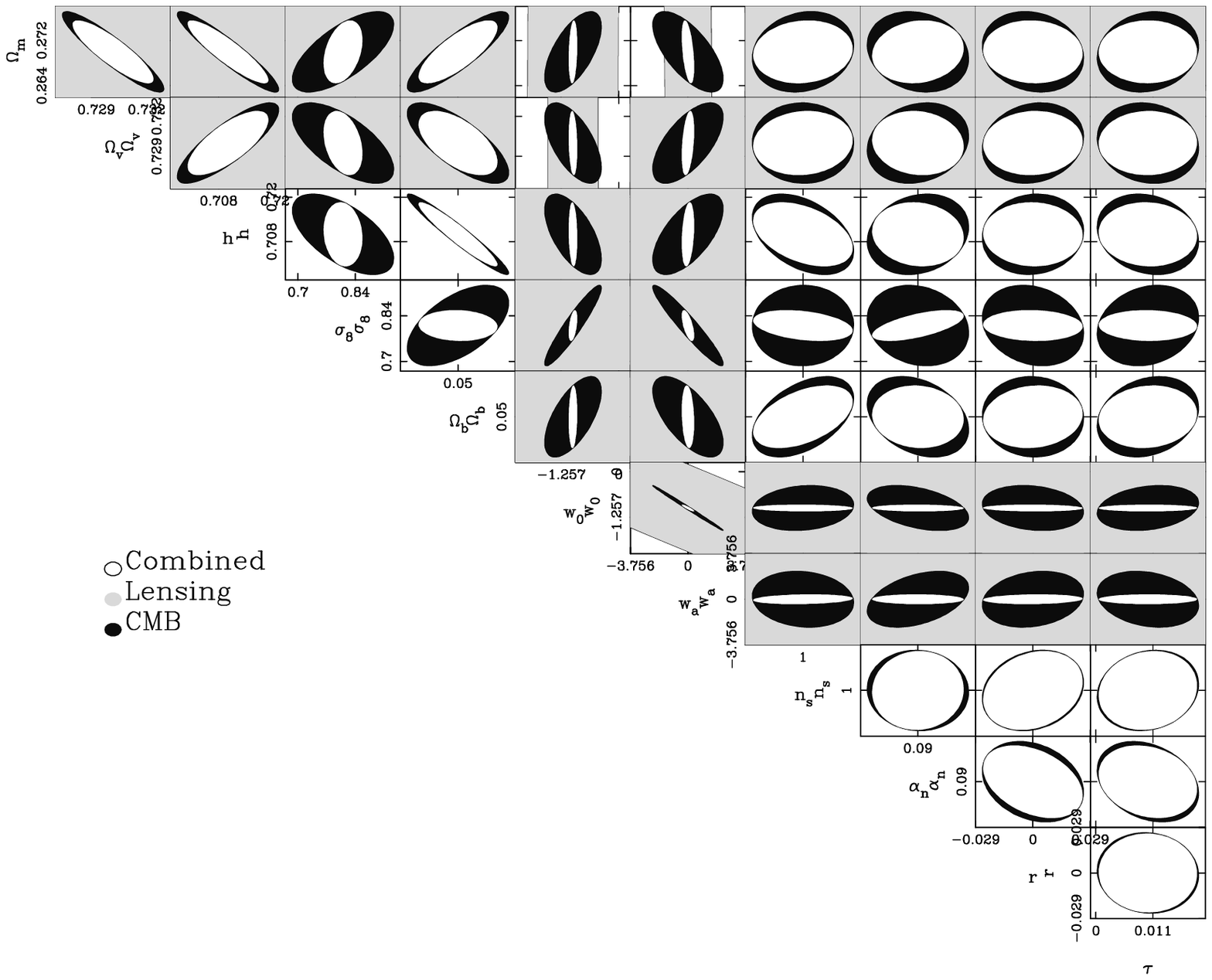,width=2.0\columnwidth,angle=0,clip=}
 \caption{The two parameter 1-$\sigma$ (68.3\% confidence) geometric
 constraints for a 10,000 square degree 
 lensing survey to a median depth $z_m=0.7$, with a 14-month Planck
 experiment in the 11-dimensional parameter space
($\Omega_m$, $\Omega_v$, $h$, $\sigma_8$, $\Omega_b$, $w_0$,
$w_a$, $n_s$, $\tau$, $\alpha_n$, $r$). Other parameters for the CMB calculation are
marginalized over.}
 \label{speparams}
\end{figure*}

The effect of the geometric tests constraints within an
11-dimensional parameter space can be seen in Figure
\ref{speparams}. All other parameters are marginalized over. Even
though the geometric test does not place any direct constraint on
the non-geometric parameters, we note that there is improvement in
the normalization of matter perturbations, $\sigma_8$. This arises
because $\sigma_8$, measured from the CMB is dependent on
the parameters. Hu \& Jain (2004) show the dependence of
$\sigma_8$ on other cosmological parameters, and in particular a
constant value of $w$. In calculating the value of $\sigma_8$ using
dark energy dependent growth factors they find that the value of
$\sigma_8$ depends on a combination of dark energy parameters, they
find an analytic expression in the special case of a flat Universe
with constant $w$. These general arguments can be generalized to $w_0$
and $w_a$ using the growth factors given in Linder (Linder, 2003). An
alternative parameter 
would be to use the horizon-scale amplitude of matter perturbations,
$\delta_\xi$, which is an independent parameter. We have chosen to
use $\sigma_8$ to compare with other analysis. The improvement on
CMB parameters are summarized in Table \ref{speimprov}.

\begin{table}
\begin{center}
\begin{tabular}{|l|c|c|}
\hline
Parameter&Planck only&Combined\\
\hline
$\Omega_m$&$0.0058$&$0.0042$\\
\hline
$\Omega_v$&$0.0024$&$0.0020$\\
\hline
h&$0.0088$&$0.0070$\\
\hline
$\sigma_8$&$0.1002$&$0.0383$\\
\hline
$\Omega_b$&$0.0011$&$0.0008$\\
\hline
$w_0$&$0.5015$&$0.0751$\\
\hline
$w_a$&$1.8618$&$0.3256$\\
\hline
$n_s$&$0.0034$&$0.0034$\\
 \hline
$\alpha_n$&$0.0062$&$0.0056$\\
 \hline
$\tau$&$0.0208$&$0.0204$\\
 \hline
$r $&$0.0079$& $0.0077$\\
 \hline
\end{tabular}
\caption{Improvements on CMB Planck one parameter 1-$\sigma$,
  constraints by 
  adding the geometric test from a 10,000 square degree lensing
  survey to a median depth of $z_m=0.7$.}
\label{speimprov}
\end{center}
\end{table}

\subsubsection{Comparing and combining lensing with CMB, BAO and SNIa
experiments}
\label{COMSBS}
\begin{figure}
 \psfig{figure=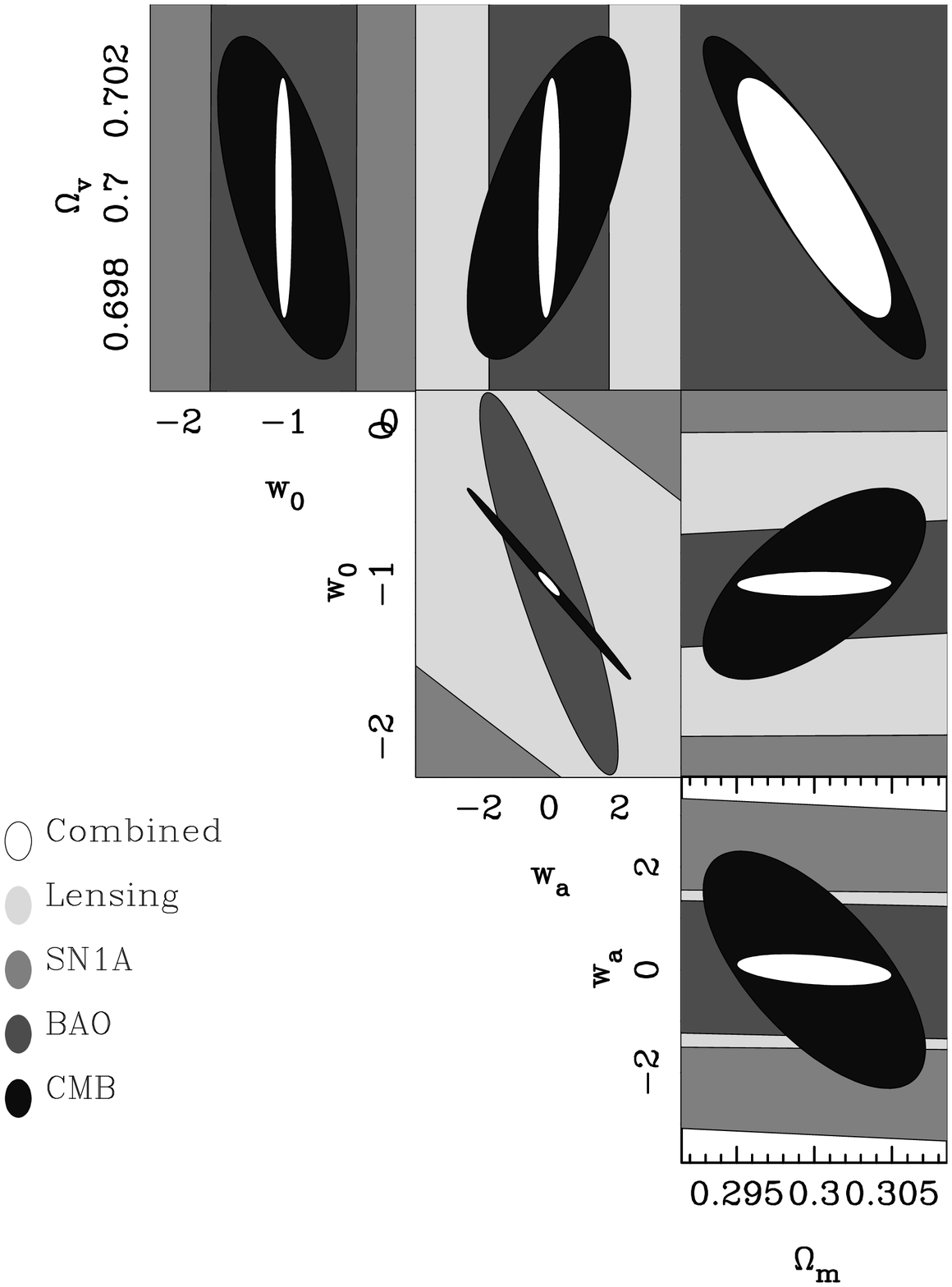,width=\columnwidth,angle=0,clip=}
 \caption{Two-parameter, 1-$\sigma$ (68.3\% confidence) likelihood contours
 for geometric parameters  for a 10,000 square degree
 lensing survey to a median depth of $z_m=0.7$, combined with a 14-month
 Planck experiment, a
 WFMOS BAO experiment and a SNAP SNIa experiment. 1-parameter marginalized
 results are tabulated in Table \ref{fullresults}.}
 \label{ALLLENS}
\end{figure}

Figure \ref{ALLLENS} shows comparisons between the geometric
lensing, CMB, SNIa and BAO experiments for the geometric parameter
set $(\Omega_m,\Omega_v,w_0,w_a)$. The broad, second lightest grey
ellipses are for a SNAP-like SNIa experiment, the closed darker
ellipses are for a WFMOS-like BAO 
experiment, the lightest grey is for the lensing geometric
test, while the darkest ellipse is for a 14-month Planck CMB
experiment. The small white ellipse at the centre is the
combined uncertainty. We have scaled the axes so that the full
parameter degeneracies can be seen. It is clear that allowing for
spatial curvature and evolution of the dark energy opens up large
degeneracies in many of the experiments. Because of the large-data
set and sensitivity of the CMB to parameters, the CMB provides the
strongest constraints alone. In particular we can see very similar
degeneracies between experiments in the $(\Omega_v,w_0)$ plane,
while there is some orthogonality between experiments in the
$(w_0,w_a)$ plane. Combining experiments improves the constraints
on all of the parameters. In particular, allowing for spatial
curvature we find $\Delta w_0=0.043$, and $\Delta w_a=0.108$. We
shall study the combination of experiments in more detail in
Section \ref{Comparing and combining experiments for dark energy parameters}.
\begin{figure}
 \psfig{figure=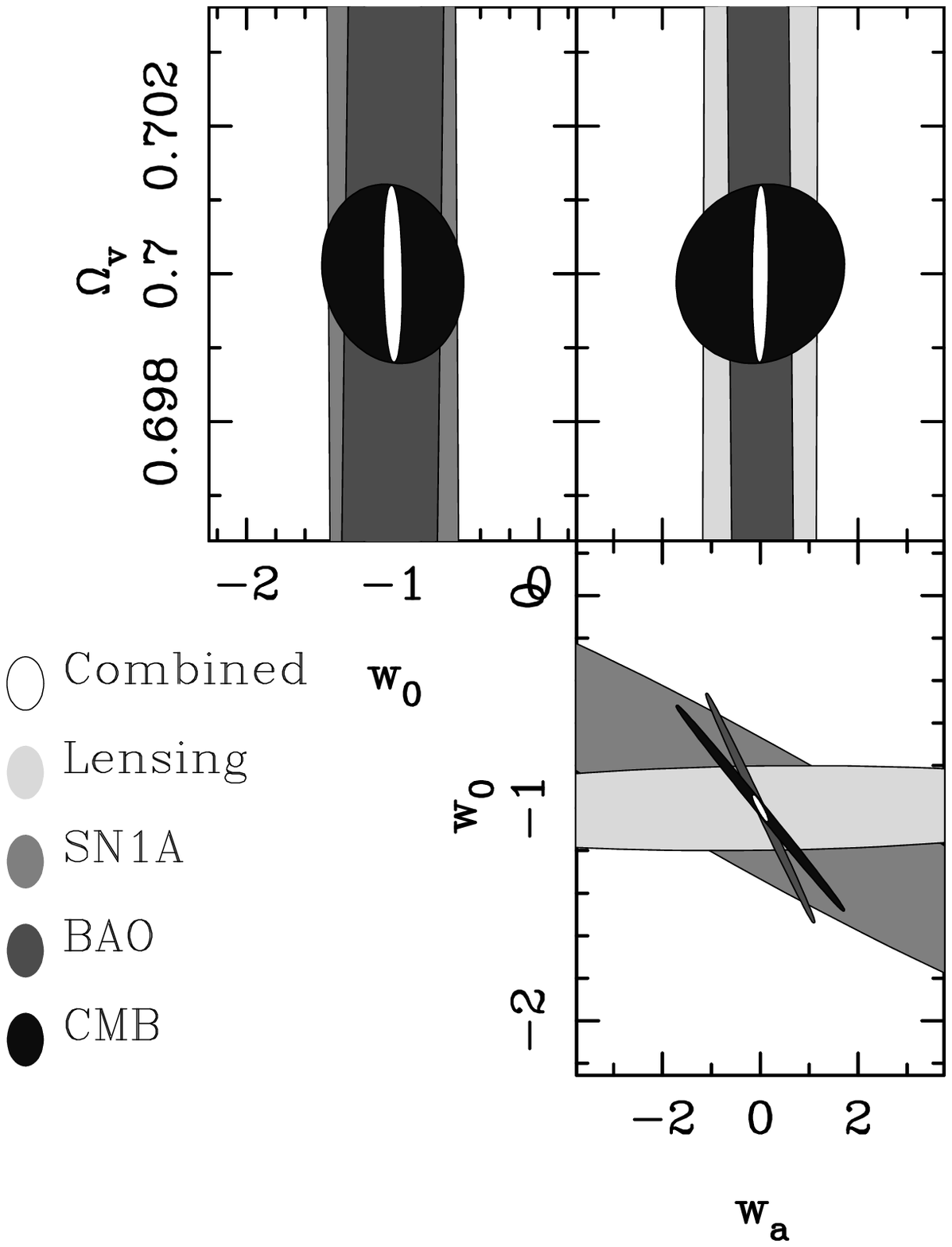,width=\columnwidth,angle=0,clip=}
 \caption{Two-parameter, 1-$\sigma$ (68.3\% confidence) likelihood contours
 for geometric parameters  for a 10,000 square degree
 lensing survey to a median depth of $z_m=0.7$, with a
 14-month Planck experiment, a
 WFMOS BAO experiment and a SNAP SNIa experiment, assuming
 spatial flatness with
 $\Omega_m+\Omega_v=1$.}
 \label{Flat}
\end{figure}

In Figure \ref{Flat} we show the same set of parameters, but this
time assuming spatial flatness. Again many of the largest
degeneracies in each of the experiments remain. We also see
clearly the insensitivity of the geometric lensing to $w_a$,
rendering it nicely orthogonal to the other experiments. Also note
that the two geometric methods considered, the geometric test and the
supernovae test constrain similar regions in the ($w_0$, $w_a$)
plane. Comparing
Figure \ref{ALLLENS} and Figure \ref{Flat} it is clear that the
assumption of flatness improves the marginal errors of the
lensing, BAO and SNIa significantly, however since the CMB
experiment constrain flatness to a high degree the overall
combined constraints are broadly the same. This
highlights the danger of assuming flatness, given that the
marginal errors without a CMB experiment are drastically altered
by this assumption. Given that some dark energy models involve
variations to the Friedmann equation in non-flat geometries it is
prudent to marginalize over spatially curved models.

\begin{figure}
 \psfig{figure=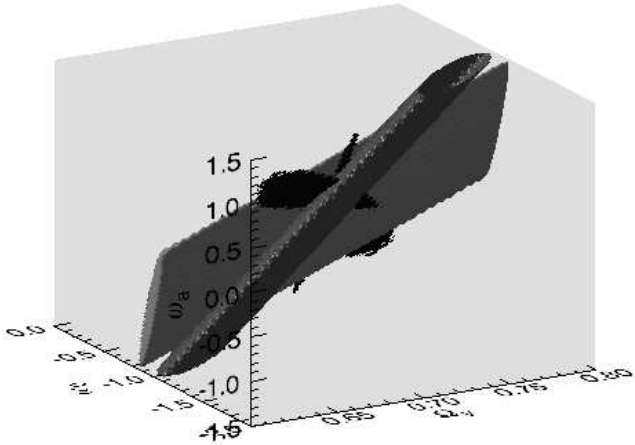,width=\columnwidth,angle=0,clip=}
 \caption{Likelihood contours in the
 3-dimensional $\Omega_v$, $w_0$, $w_a$ parameter space
 for geometric parameters for a 10,000 square degree
 lensing survey to a median depth of $z_m=0.7$, with a 14-month
 Planck experiment, a
 WFMOS BAO experiment and a SNAP SNIa experiment, assuming spatial
  flatness, $\Omega_m+\Omega_v=1$. 1-parameter, 1-sigma contours are
 used for clarity.} 
 \label{Flat3d}
\end{figure}

To illustrate further the orthogonality of the constraints from lensing,
the CMB, BAO and SNIa,  Figure \ref{Flat3d} shows a 3-dimensional
plot of the likelihood contours in the $(\Omega_v,w_0,w_a)$
parameter space, marginalizing over all other parameters. We have
plotted the 1-parameter, 1-sigma contours for clarity.

\subsection{Synergy of dark energy experiments}
\begin{figure}
 \psfig{figure=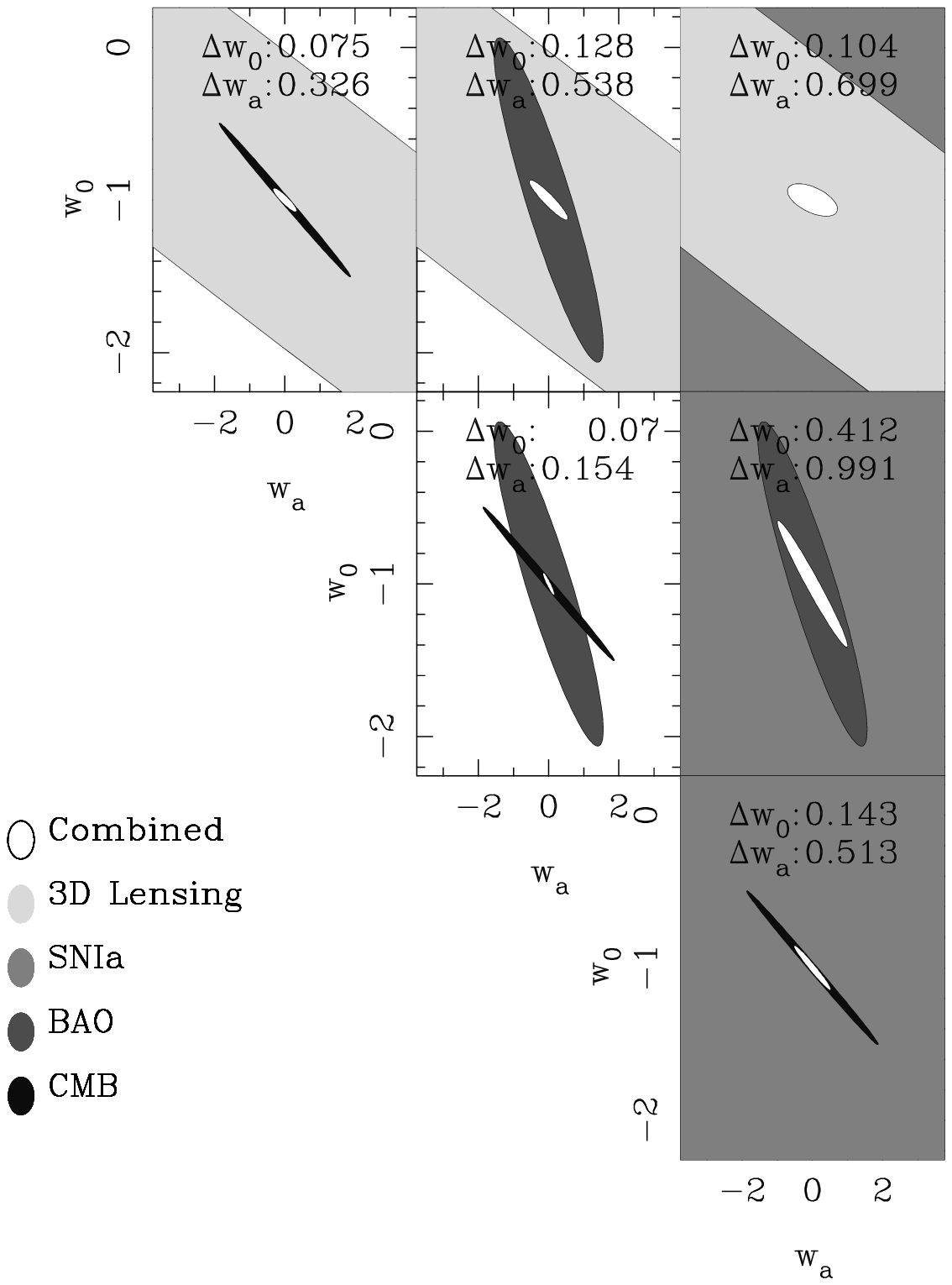,width=\columnwidth,angle=0,clip=}
 \caption{The combined marginal
$w_0$, $w_a$ constraints for two pairs of experiments. The experiments
are a darkCAM lensing experiment and a CMB 14-month Planck
experiment, a BAO WFMOS experiment and a SNIa SNAP experiment.
Note that only the SNIa analysis contains terms for systematic
effects. See Section 4.9 for details.}
 \label{TwoExperi}
\end{figure}

\label{Comparing and combining experiments for dark energy
parameters}
It is interesting to compare the results of each of the dark
energy experiments under the same conditions. In Figure
\ref{TwoExperi} we show the dark energy equation of state
parameters $(w_0,w_a)$, marginalized over all other parameters
including spatial curvature for each experiment in combination.

Of all of the individual experiments considered the Planck CMB
experiment on its own provides the strongest constraint on the
($w_0, w_a$) plane, with the majority of the signal coming from
the low-redshift ISW effect. However the marginalized
uncertainties are still $\Delta w_0=0.502$ and $\Delta w_a=1.86$. A
SNAP-like SNIa experiment on its own provides poor constraints
in the $w_0$, $w_a$ plane, due to the large degeneracy in
$\Omega_m$ and $\Omega_v$ in models allowing curvature. This can
be seen by comparing with Figure \ref{Flat}, but note that of the
experiments considered, the supernova estimates are the only ones
which include terms for extra systematic effects. Removing the extra
systematic terms from the supernova estimates improves the
constraints, when combined with a Planck CMB prior, by a factor of
approximately 
$1.5$ to $\Delta w_0=0.094$ and $\Delta w_a=0.318$. A WFMOS-like BAO 
experiment provides a narrow, but highly degenerate ellipse in the
$(w_0,w_a)$ plane. This is due to the BAO experiment mainly
constraining $w(z)$ at the redshift of the nearest redshift bin
(in this case $z=1.0$). Interestingly the BAO degeneracy is in a
similar direction to the CMB degeneracy, presumably because a
similar geometric effect is being measured. Finally, the geometric
lensing again has a large degeneracy in this plane, but one which
is different from the other experiments.
The combination of pairs of experiments is very interesting. The
combination of the geometric lensing and CMB puts very strong
constraints on the dark energy equation of state and its
evolution, reducing the uncertainty to $\Delta w_0=0.075$ and
$\Delta w_a=0.326$. Geometric lensing and SNIa yields $\Delta
w_0=0.104$ and $\Delta w_a=0.699$ while geometric lensing and BAO
yields $\Delta w_0=0.128$ and $\Delta w_a=0.538$. This provides us
with three cross-checks with similar accuracy. Looking at the
dependency of each method, we see that both the geometric lensing,
BAO and SNIa are all dependent on the geometry of the Universe,
and so should give the same result, assuming that the $w_0, w_a$
parameterization is valid. The CMB combines geometry with
evolution of the potential field, particularly in the ISW effect.

Looking at the other possible combinations without lensing we see
there is a similar sensitivity to dark energy. We have already pointed
out the degeneracy between the BAO and CMB constraints, and so
their combination only marginally improves on the CMB alone. A
similar result is found for combining CMB and SNIa. Finally BAO
and SNIa provides an uncertainly similar to CMB alone.

From this study, we conclude that the best pair combinations come
from combining geometric lensing with any of CMB, BAO or SNIa
experiments, with $\Delta w_0\sim0.10$ and $\Delta w_a\sim0.50$, and
also the BAO and CMB combination. Multiple
combinations will also allow a degree of cross-checking for
consistency. Other combinations are a factor of up to 5 times poorer
due to similar degeneracies between $w_0$ and $w_a$.
\begin{figure}
\psfig{figure=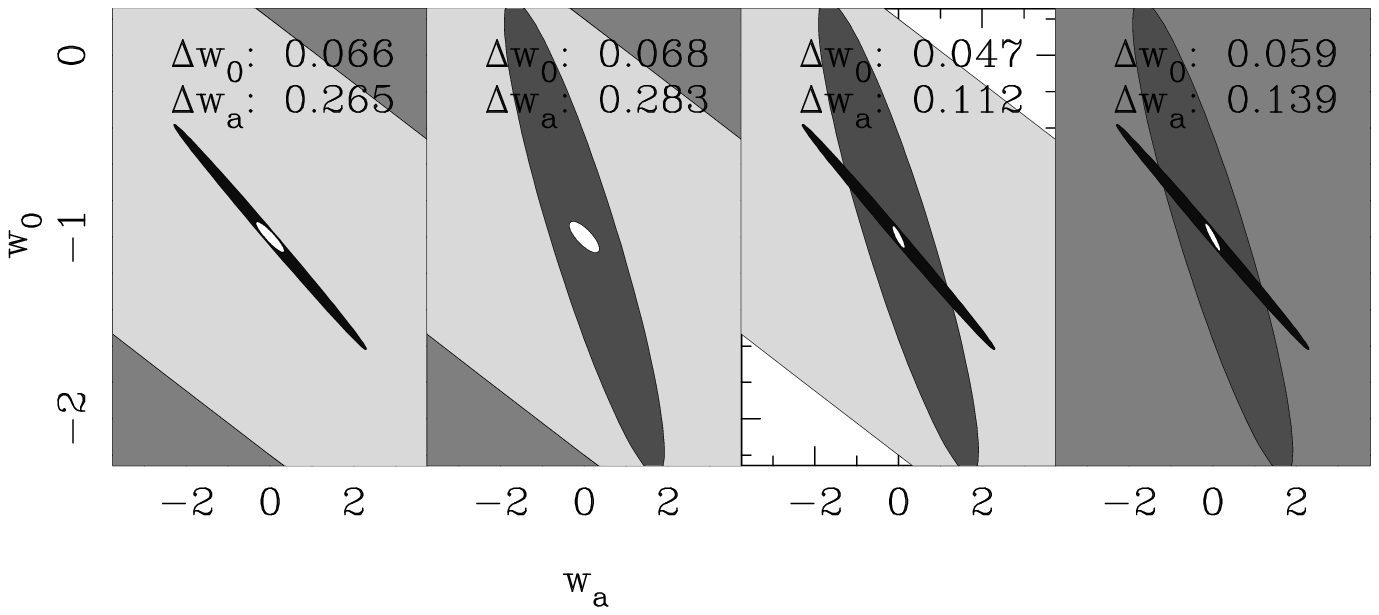,width=\columnwidth,angle=0,clip=}
 \caption{The marginal $w_0$, $w_a$ constraints for a combination of
 any three of the dark energy experiments. A darkCAM lensing
 experiment, CMB 14-month Planck experiment, BAO WFMOS
 experiment and a SNIa SNAP experiment.}
 \label{LCB}
\end{figure}
Combining three experiments, in Figure \ref{LCB} we again see that 
strongest 
measurement of $(w_0,w_a)$ comes from combining the geometric
lensing analysis with the CMB and BAO experiments with the uncertainty
on $w_0$ and $w_a$ pushed down to $\Delta w_0=0.047$ 
and $\Delta w_a = 0.112$. Adding the SNIa results to this makes no a
small difference (see Figure \ref{ALL}, but again recall that
the SNIa is the only estimate to contain systematic effects). Again
the all the three experiment combinations all provide complimentary
constraints in the ($w_0$, $w_a$) plane.
\begin{figure}
  \epsfig{figure=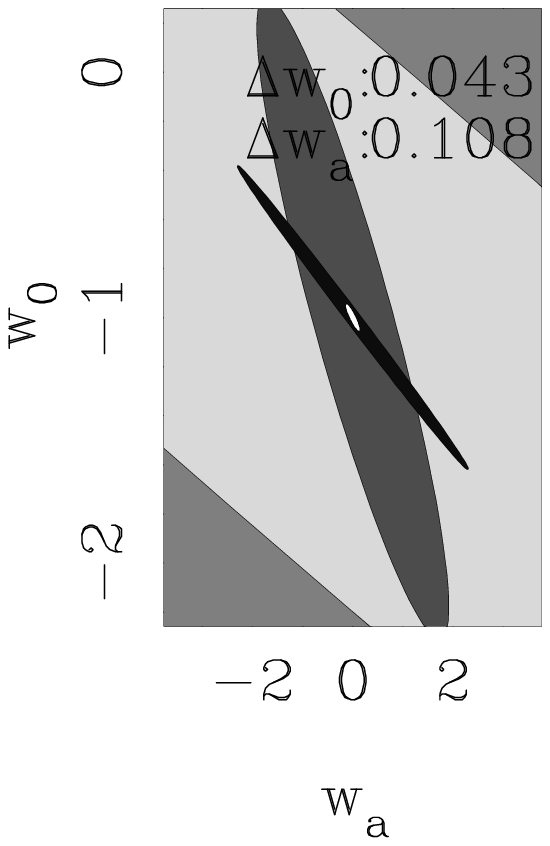,angle=0,width=0.7\columnwidth}
  \caption{The marginal $w_0$, $w_a$ constraints for a darkCAM lensing
    experiment and a CMB 14-month Planck experiment, a BAO WFMOS
    experiment and a 
    SNIa SNAP experiment.}
  \label{ALL}
\end{figure}

\begin{table*}
\begin{center}
 \begin{tabular}{|l|c|c|c|c|c|c|c|c|}
 \hline
 Survey & Area sqdeg &$\rm z_{{\rm median}}$& $N_{\rm Bands}$ &
 $\Delta w_0$&$\Delta w_a$ &
 ${\rm z_{{\rm pivot}}}$&
 $\Delta w({\rm z_{{\rm pivot}}})$&
 $\Delta w({\rm z_{{\rm pivot}}})\Delta w_a$\\
\hline \hline
{\bf Lensing}&$$&$$&$$&$$&$$&$$&\\
\hline
darkCAM &$10000$&$0.7$&$5$&$5.546$&$31.132$&$0.21$&$0.972$&$30.2471$\\
\hline
darkCAM + Planck&$10000$&$0.7$&$5$&$0.075$&$0.326$&$0.27$&$0.030$&$0.0097$\\
\hline
darkCAM + BAO darkCAM&$10000$&$0.7$&$5$&$0.459$&$1.668$&$0.01$&$0.419$&$0.6984$\\
\hline
darkCAM, $9$ bands + Planck&$10000$&$0.7$&$9$&$0.071$&$0.311$&$0.26$&$0.029$&$0.0089$\\
\hline
SNAP Lensing + SNIa + Planck&$1000$&$1.38$&$9$&$0.073$&$0.293$&$0.31$&$0.024$&$0.0071$\\
\hline
All-Sky Space + Planck&$40000$&$1.00$&$9$&$0.023$&$0.146$&$0.16$&$0.012$&$0.0017$\\
\hline
darkCAM+Planck+BAO+SNIa&$10000$&$0.7$&$5$&$0.043$&$0.108$&$0.58$&$0.018$&$0.0019$\\
\hline
VST-KIDS+WMAP4&$1400$&$0.6$&$5$&$0.227$&$0.888$&$0.19$&$0.176$&$0.1562$\\
\hline
CFHTLS(Wide)+WMAP4&$170$&$1.17$&$5$&$0.282$&$1.014$&$0.29$&$0.1663$&$0.1687$\\
\hline \hline
{\bf CMB}&$$&$$&$$&$$&$$&$$&\\
\hline
4-year WMAP&$$&$$&$$&$2.060$&$3.612$&$1.18$&$0.758$&$2.7379$\\
\hline
14-Month Planck&$$&$$&$$&$0.501$&$1.873$&$0.367$&$0.035$&$0.0655$\\
\hline \hline
{\bf BAO}&$$&$$&$$&$$&$$&\\
\hline
BAO WFMOS+Planck&$2000$&$1.0$&$$&$0.070$&$0.154$&$0.78$&$0.019$&$0.0029$\\
\hline
\hline
{\bf SNIa}&$$&$$&$$&$$&$$&$$&$$\\
\hline
  SNIa SNAP+Planck&$$&$$&$$&$0.142$&$0.513$&$0.37$&$0.028$&$0.0144$\\
\hline
 \end{tabular}
 \end{center}
 \caption{The table gives experimental parameters and marginalized
 cosmological parameter error forecasts for various surveys.}
 \label{fullresults}
\end{table*}

\subsection{Complementary Figures of Merit and Pivot Redshifts}
\begin{figure}
\psfig{figure=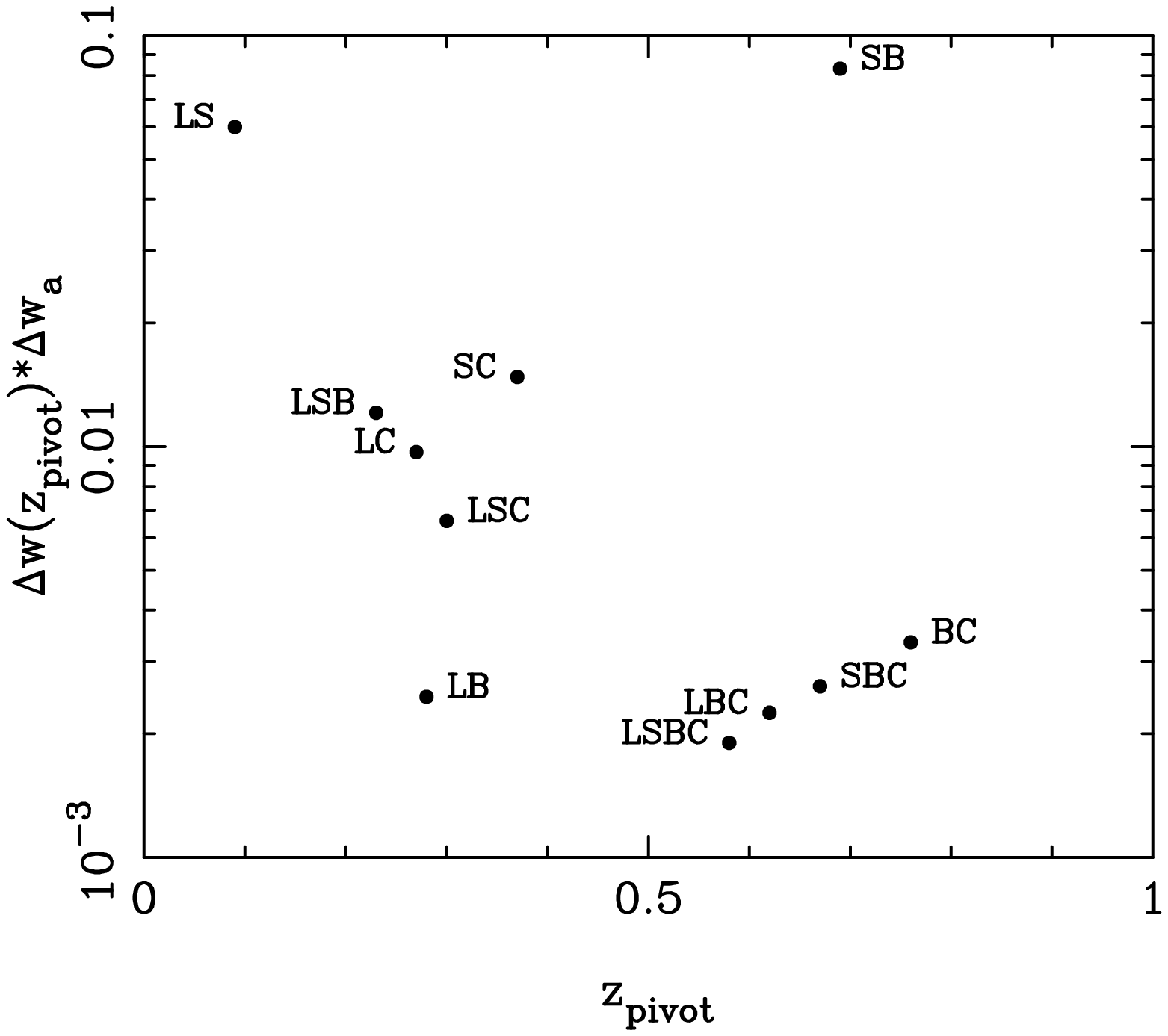,width=\columnwidth,angle=0}
\caption{The figure of merit and pivot redshift for various
  experimental combinations. The combinations are labeled as
  L=Lensing, B=BAO, S=SNIa, C=CMB.
  Combinations of letters represent combinations of experiments.}
\label{fompiv}
\end{figure}

The figure of merit and pivot redshift information can be
represented as in Figure \ref{fompiv} so that both values can be
seen simultaneously. The Figure shows a number of broad
characteristics. As more experiments are added in combination both
the pivot redshift converges to one mean value and the figure of merit
decreases. The geometric test constraint forces the pivot redshift
to lower values due to its unique degeneracy whilst the BAO
constraint forces the pivot redshift to higher values. It is also
evident that the CMB constraint is not necessary for a low figure of
merit (for example LBS). This Figure also shows how different
combinations of surveys
can probe dark energy at significantly different redshifts. For
example the BC and LSC combinations both have a similar figure of
merit with the BC combination $z_{\rm pivot}=0.76$ and the LB
combination $z_{\rm pivot}=0.28$.

\subsection{The effect of changing the fiducial dark energy model}
\label{The effect of changing the fiducial dark energy model}
The assumed fiducial cosmology has so far been a $\Lambda$CDM
cosmology in which any derivatives in the Fisher matrix
calculations have been about $w_0=-1.0$ and $w_a=0.0$ for the
equation of state parameters. The effect of altering this
assumption is investigated here. We consider two alternative
extremes which are just allowable by present constraints, dark
energy models: a SUGRA (Super Gravity) model proposed by Weller \&
Albrecht (2002) represented by $w_0=-0.8$
and $w_a=+0.3$; and a phantom model proposed by Caldwell et al.
(2003) with $w_0=-1.2$ and $w_a=-0.3$. To test the effect of
changing our default dark energy model we re-run our Fisher
analysis. 

Aswell as changing the point in parameter space about which
the signal ratio is expanded in the Fisher matrix calculations the
assumed fiducial dark energy model also affects the SIS to 
NFW scaling as a function of redshift and mass, as shown in Figure
\ref{testNFW}. It also affects the number density distribution of haloes as a
function of redshift and mass given by equation (\ref{NMZ}), when extending
to arbitrary dark energy models we exchange the growth factor in
equation (\ref{GROWTH}) to the one given in Linder \& Jenkins (2003).  
\begin{figure}
\psfig{figure=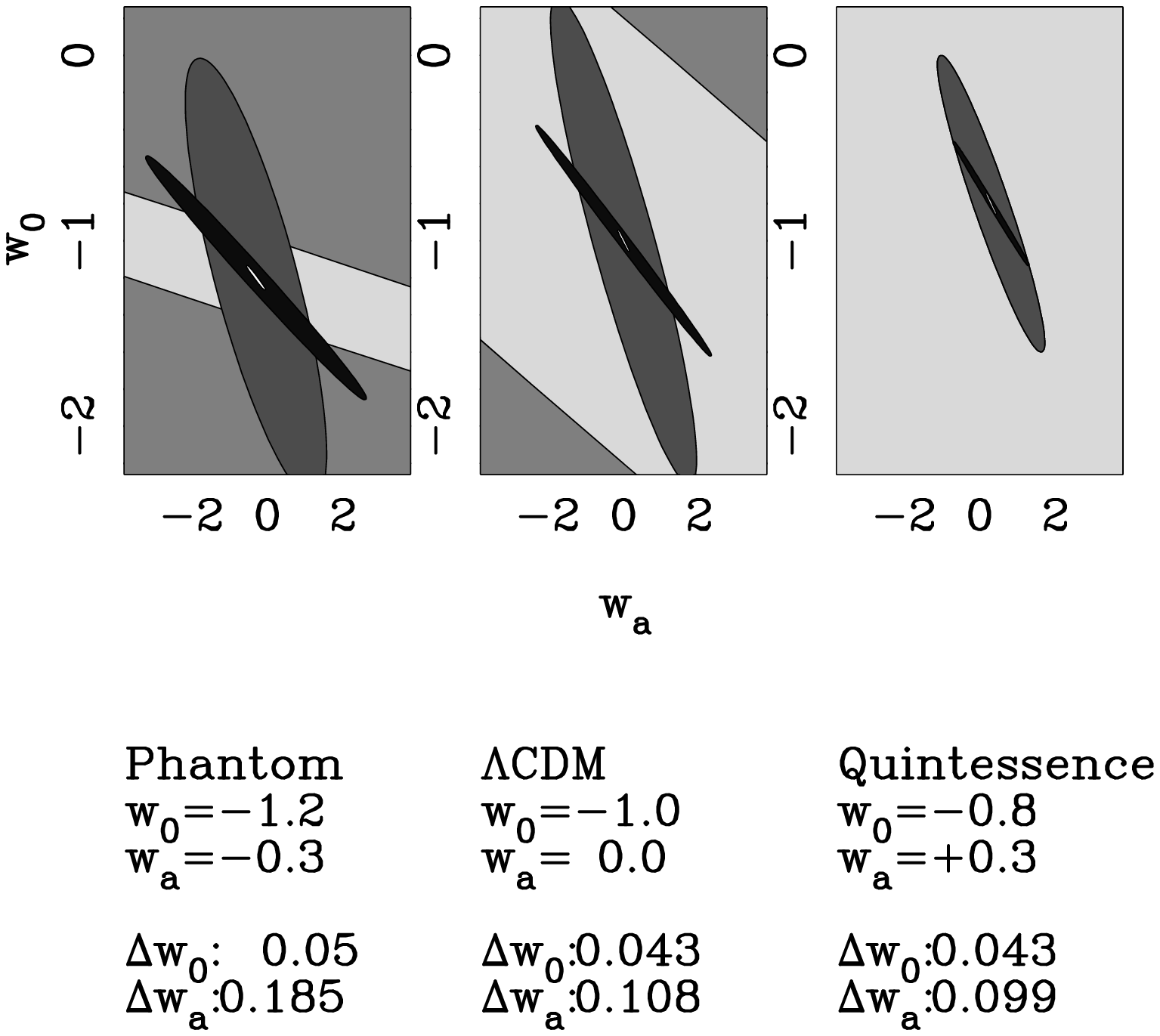,width=\columnwidth,angle=0,clip=}
\caption{The dependence of the marginal error on $w_0$ the assumed
  dark energy model, for a  10,000
  square degree
  survey to a median depth $z=0.7$, with a 14-year Planck prior, a BAO
  WFMOS prior and a SNIa SNAP prior. The errors quoted are marginal
  over all parameters.}\label{DEScenarios}
\end{figure}

The effects of changing the default dark energy model are shown in
Figure \ref{DEScenarios}, where we plot the $(w_0,w_a)$
plane, fully marginalizing over all other parameters. The marginal
errors for each experiment, and its degeneracy in parameter space
does indeed depend on the dark energy model. But the combined
marginal errors do not significantly change. The main difference is
manifest in the error on $w_a$ which increases for all methods as its
value becomes more negative. This is simply due to the fact that a
negative $w_a$ represents a dark energy scenario in which the dark
energy density was less in the past (increasing in the future); so
that the effect of dark energy on the expansion rate on observed
galaxies (in the past) is less in these scenarios (and similarly the
opposite effect for a positive $w_a$).

\section{Additional Systematic Effects for Lensing}
In this Section we consider come potential systematic effects for weak
lensing. To realize a $\sim1\%$ measurement of $w_0$ and $w_a$ from
shear ratios not only do
we need a large enough survey, but we must also be able to control
systematics in a weak lensing survey to a high level. This will
require controlling the 
systematics in the measurement of lens shear to $\Delta \gamma
\sim10^{-5}$.

\subsection{Image Shear Analysis}
The current generation of lensing surveys, with
telescopes not specifically designed for lensing, induce $10\%$
distortions, which can be corrected down to a net systematic of
$0.01\%$ ($\Delta \gamma \sim 10^{-4}$; Heymans et al., 2005). There
has been extensive 
work into methods that can both diagnose and remove systematic errors
from both intrinsic galaxy alignments (Hirata et al., 2004) and shear
calibration errors. Mandelbaum et al. (2005) use a geometric test to
diagnose systematic errors in the Sloan Digital Sky Survey
(SDSS).

\subsection{Strong Lensing Effects}
There is a systematic effect in the strong lensing r\'egime, where
the reduced shear $e_i$, defined as
 \be
    e_i=\frac{\gamma_i}{1-\kappa_i}
 \ee
is measured from galaxy ellipticities. The ratio $R_{ij}$ now
becomes
 \be
    R_{ij}=\frac{e_i}{e_j}=\frac{\gamma_i(1-\kappa_j)}{
            \gamma_j(1-\kappa_i)}
 \ee
which, for the mildly non-linear lensing r\'egime can be
approximated as
 \be
        R_{ij}\approx\frac{\gamma_i}{\gamma_j}(1+\kappa_i-\kappa_j)
 \ee
Furthermore for a SIS $\kappa \approx \gamma$. This was
numerically tested, using observable clusters, and the amplitude
of the correction was found to be
$\rm{max}(\kappa_i-\kappa_j)\sim0.15$ and
$\rm{mean}(\kappa_i-\kappa_j)\sim1.5\times10^{-3}$. This numerical
analysis implicitly assumes a radius of $1$ arcmin from equation
\ref{gammatheta}, which is relatively narrow: tangential shear can
be measured out to radii of at least $200$ arcseconds see Gray et
al. 2004. The largest source of this systematic error will be from
the largest clusters, those producing the largest convergence, and
as shown in Section \ref{ClusterDep} the majority of the $w_0$
signal comes from clusters of intermediate mass for which we would
expect this systematic to be smaller.

Alternatively, one can construct a statistic which eliminates the
mass-dependence of $\gamma/(1-\kappa)$, such as the three-point
statistic suggested by Gautret, Fort \& Mellier (2000). This could
be applied in the strong-lensing r\'egime, again independent of the
lens strength, and combined with the two-point geometric ratio
test in the weaker lensing r\'egime. We shall investigate this
elsewhere.

\subsection{Cluster Substructure}
A further expected systematic is that arising from cluster
sub-structure, which we assume is averaged over. The effect of
including sub-structure can only increase the signal; as long as the
mass map of a cluster can be accurately measured the expected
tangential shear signal can be modelled. One promising avenue which
may yield information on sub-structure is flexion (see Bacon et al.,
2005). In the low signal-to-noise r\'egime (low galaxy number
counts) in which a mass model may have to be assumed for a
cosmological signal to be extracted then this systematic source of
error will become important and the mass model will need to be
accurately reproduced. However, in the high signal-to-noise r\'egime
where the number of available galaxies is such that ratio of the
shears from the data can simply be taken this systematic source of
error will not affect the analysis.

\subsection{The effect of photometric redshift outliers}
In any weak lensing photometric redshift survey there will be a
sample of imaged galaxies that will not have photometric redshifts
assigned. There will be several classes of objects, for some of
which determining a photometric redshift will be difficult. We
test the effect of such `outliers' here by assuming a population
within the survey $p2$ that have photometric redshifts
$\sigma^{p2}_z(z)=0.5$, that is they have practically no redshift
information.

There are two ways in which such a population can be used, either
they are included in the sample of galaxies used somehow, or they are
discarded. If 
they are used then the galaxies can be treated as a seperate
population, with $\sigma^{p2}_z(z)=0.5$ and $n^{p2}=(1-A^{p2})n_0$
where $A^{p2}$ is the fraction of outliers in the total population,
analysed independently and the 
constraints from the 
outliers added to the constraints from galaxies with good
redshifts. Or, the effective redshift error 
distribution at a 
particular redshift $z$ can be modelled by the sum of two Gaussian
distributions, with errors $\sigma^{p1}_z(z)$ and
$\sigma^{p2}_z(z)$, the relative amplitudes of the Gaussians
constrained so that $A^{p1}+A^{p2}=1$. Such a sum of Gaussians can
be accurately modelled as an effective Gaussian, see Blake \& Bridle
(2005), with an effective width
$\sigma_{eff}=\sqrt{A^{p1}[\sigma^{p1}_z(z)]^2+A^{p2}[\sigma^{p2}_z(z)]^2}$.
We investigated varying the relative amplitudes of two Gaussians
with $\sigma^{p1}_z(z)=\sigma_z(z)$, the original photometric
redshift error from equation (\ref{intdz}) and a second with
$\sigma^{p2}_z(z)=0.5$. 

\begin{figure}
\psfig{figure=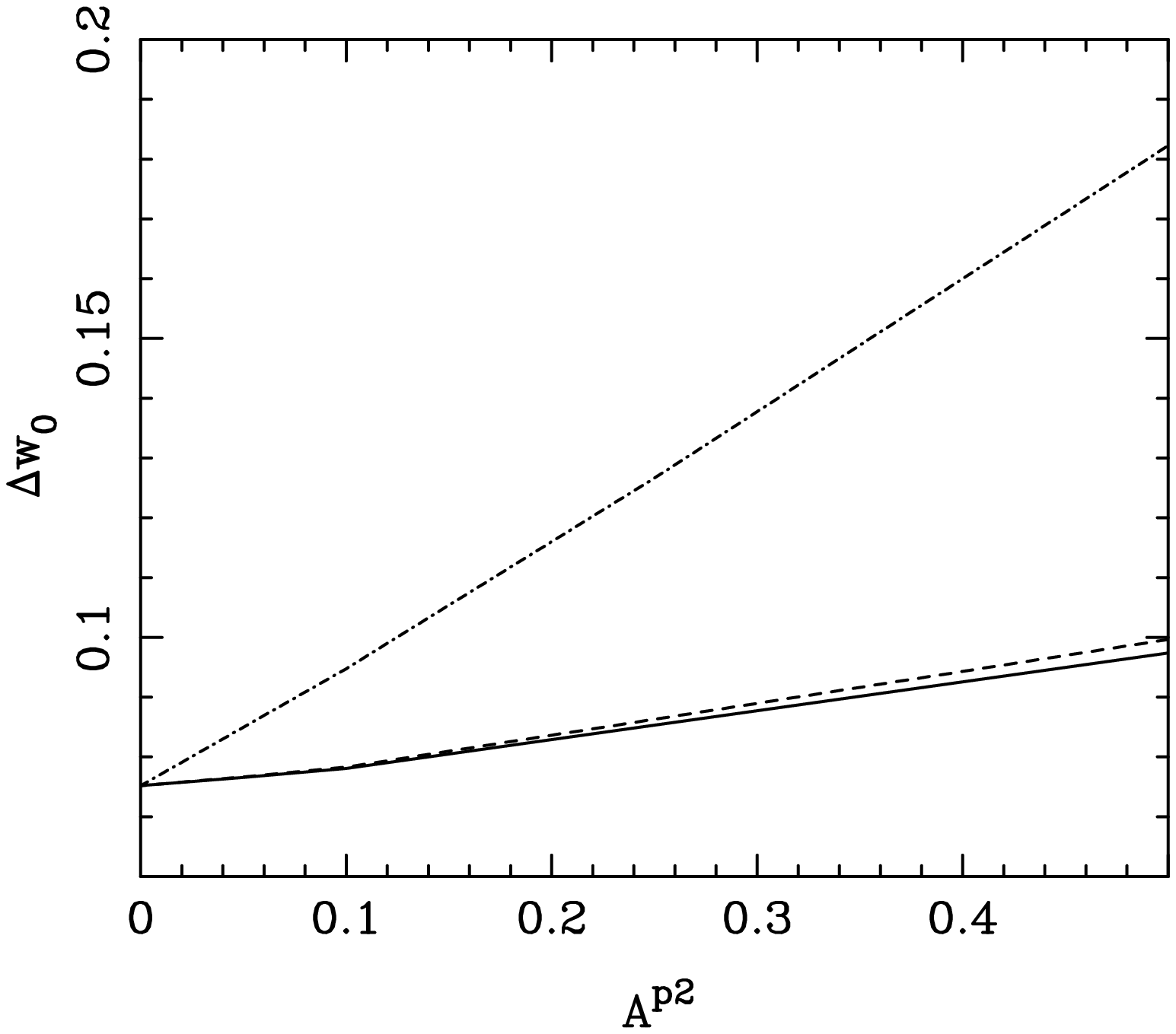,width=\columnwidth,angle=0} \caption{The
dependence of the marginal error on $w_0$ on the
  amplitude of outliers with a $\sigma^{p2}_z(z)=0.5$ for a  10,000
  square degree
  survey to a median depth $z=0.7$, with a 14-year Planck prior. The
  solid line
  combines the outliers constraint with the galaxies with good
  photometric redshift errors by adding the constraints. The
  dot-dashed line shows the effect of combining the outliers with the
  rest of the population using an effective Gaussian. The dashed line
  discards the 
  outlying sample of galaxies.}\label{twog}
\end{figure}

Figure \ref{twog} shows the effect of
varying the fraction of total population in oultiers, combined with a
14-month Planck prior, the solid line shows the constraints from
treating the outliers as a seperate population analysed seperately,
the dot-dashed line 
includes the outliers into a degraded population using an effective
Gaussian. The second 
possibility, that of discarding 
the outlying sample, is investigated by simply reducing the surface
number density by $n_0\rightarrow (1-A^{p2})n_0$, this is also shown
in Figure \ref{twog} as the dashed line.

All the methods for dealing with the outliers result in an increase in
the marginal error on $w_0$. 
As expected
using all the galaxies and treating the outliers as a seperate
population has the smallest effect on the marginal error, 
by using the outliers the marginal error is less than when they
discarded.
The effective Gaussian method has the effect of decreasing the
number of redshift bins in the survey that can be used while
retaining the surface number density, thus decreasing the signal.
By discarding the outliers the number of redshift bins is retained
while the surface number density is uniformly degraded. This shows
that the signal is more dependent on the number of redshift bins,
than the number density and that the strategy for dealing with
outliers will be an important issue in future surveys. 

\subsection{CMB Lensing}
When combining the shear ratio analysis with CMB measurements we
have assumed that the weak lensing of the CMB by large scale
structure and galaxy clusters can effectively be ignored. Since
the shear ratios do not contain any information about structure,
there can be no correlation due to this. There may, however, be
some correlation between lensing of the CMB and the noise term in
the shear ratio method. We shall explore this elsewhere.

\section{Conclusions}
In this paper we have set out a new method for the analysis of the
geometric shear ratio test for measuring the dark energy equation
of state, based on the measurement of shear ratios around
individual galaxy groups and clusters. The shear ratio test is
insensitive to the growth of structure, but sensitive to the
geometry of the Universe, via the matter and dark energy density
and the dark energy equation of state. This approach allows
one to apply the method to individual objects,
rather than requiring the measurement of some other statistic
such as the galaxy-shear cross-correlation function which may be
noisy for small data-sets. The down-side is that the method is now
contaminated by structure along the line of sight, which can be
overcome by using many independent lines of sight.

Of the parameters which govern the geometry of the Universe, or more
properly the photon distance-redshift relation, the shear ratio is
most sensitive to a constant dark energy equation of state, $w_0$,
and very insensitive to evolution, parameterized here by $w_a$. This
can be understood as due to the shear ratios being sensitive only to
the change in shape of the shear signal as a function of redshift.
As $w_a$ parameterizes the high-redshift effect of the dark energy
equation of state, its effects are ``renormalized'' away. This
behavior is very different to other probes of dark energy, and so
helps to break parameter degeneracies when combined with other
probes.

It must be emphasised that the Fisher matrix framework used in this
paper may result in overly optimistic constraints. Since the errors
are calculated by expanding about a fiducial point in parameter space
any higher order effects that may change the shape of the likelihood
surface cannot be taken into account. The effect of varying the
fiducial dark energy model, in Section \ref{The effect of changing the
  fiducial dark energy model}, demonstrates that the 
errors are sensitive to the choice of the fiducial model. A concrete
example of higher order likelihood effects can be seen in a 3D cosmic
shear analysis by comparing
Fisher matrix calculations of the ($\sigma_8$, $\Omega_m$) plane (for
example in Heavens et al., 2006) (predicting an ellipse) with the
measured constraints from data (for example Kitching et al., 2006)
which measure an extended curved constraint. These effects can be
investigated by large simulations or by a Monte-Carlo type exploration
of the likelihood surface, we leave such investigations for future
work. 

To account for many of the sources of uncertainty in the method,
we have developed a halo decomposition analysis of the lensing
dark matter distribution to model the signal from dark matter
haloes over a range of mass scales and redshifts. We have also
included the effects of shot-noise due to the random intrinsic
orientation of each galaxy, photometric redshift errors and the
contribution of large-scale structure lensing to the error budget.
We have also investigated in detail a model for the photometric
redshift error, based on studies of the COMBO-17 data-set, as a
function of redshift, number of imaging bands and limiting
magnitude. The effect of a bias in the calibration and
distribution of photometric redshifts with spectroscopic redshifts
is also studied, and we find that we require some $10^4$ galaxies
with spectroscopic redshifts to control calibration issues. The
limitations of observing the shear signal from the ground and
space are also discussed, and we argue that without adaptive
optics ground-based lensing studies are seeing limited, suggesting
that it will be difficult to use galaxies beyond $z=1.5$.

The halo decomposition analysis of the dark matter lenses has
allowed us to probe the origin of the shear signal in different
types of survey, taking a 4-metre telescope with a 2 square degree
field of view as our default survey. These results can be scaled
to any other telescope parameters.

For targeted observations, where the time-limitation translates
into the number of clusters and groups one can observe to a given
depth, we have shown that we only require around 60 of the largest
clusters in a celestial hemisphere to constrain $w_0$ to around
$\Delta w_0\sim 0.50$, marginalizing over all other parameters,
including $w_a$, a factor of $3$ improvement on 4-year WMAP
given a marginalization over $w_a$.
To achieve a higher accuracy requires the imaging of an unfeasible
number of haloes, and instead one should turn to a wide-field
imaging and photometric redshift survey. We find for a 4-meter
class telescope with a 2 degree field of view that with a 10,000
square degree, 5-band photometric redshift survey with median
redshift $z_m =0.7$ ($r=23.8$), we can expect to reach an accuracy
of $\Delta w_0\sim 0.07$, again marginalizing over all other
parameters including $w_a$. Our results can be easily rescaled to
other telescope types, and survey strategies.

The halo decomposition allows us to deduce where the main signal
comes from in both the targeted and surveying modes. In both cases
a significant fraction of the signal comes from the largest hundred
clusters in each survey, reaching a sensitivity of $\Delta w_0
\sim 0.5$, however the majority of the signal comes from the numerous
($\sim 10^{5-6}$) $M>10^{14}M_\odot$ haloes which can push the
accuracy up to $\Delta w_0\sim 0.07$.

Having determined where the majority of the dark energy signal
will come from in a geometric shear ratio test, we then
investigate the optimization of such a survey, when combined with
the expected results from the Planck Surveyor experiment. We find
that for our fiducial telescope for a fixed-time survey, going
shallower ($z_m<0.7$) over a wider area decreases the accuracy due
to the drop in the number of available background sources and
corresponding increase in shot-noise. Going deeper ($z_m>0.7$)
over a smaller area increases the clustering noise, since we now
have fewer clusters to average over.

We have also studied the effect of varying the number of imaging
bands to increase or decrease the photometric accuracy. We find
that when combined with Planck an increase from 5, 9 or 17 optical
bands 
makes little difference to the optimal survey. The reason for this
insensitivity to higher accuracy photometric redshifts is due to
the integral nature of the lensing effect, and the weak effect
when combined with another data-set. However decreasing the number
of bands is expected to have a strong effect on the accuracy of
the lensing survey as redshift information is lost. We discuss how
our results can be scaled to other telescope classes and survey
parameters.

The dark energy parameters $w_0$ and $w_a$ can be combined to give
an uncertainty on $w(z)=w_0+w_a z/(1+z)$, at some optimal
redshift. This combination helps distinguish where the survey is
most sensitive to the dark energy equation of state. In the case
of our optimal lensing survey this is at $z=0.27$ with $\Delta
w(z=0.27)=0.0298$. Again, the reason for the low-redshift
sensitivity to $w(z)$ is due to the insensitivity of the shear
ratio test to $w_a$.

Having optimized the lensing survey for the geometric test in
combination with the expected results from the CMB, we have
investigated the effect on the full set of cosmological parameters
for the CMB and lensing. The geometric test constrains a narrow
sheet in the $(\Omega_m,\Omega_v,w_0,w_a)$ parameter-space, which
is nicely orthogonal to the CMB parameter constraints. Here we
show that the CMB mainly constrains the curvature of the model,
while the geometric test constrains $w_0$, and the combination
constrain $w_a$.

We have also compared and combined the geometric shear ratio test
with the expected results from an Baryon Acoustic Oscillation
(BAO) experiment, such as proposed for WFMOS, and a supernova Type
Ia survey, such as that proposed for SNAP. Here we have put all of
the surveys (lensing, CMB, BAO and SNIa) on an equal footing, using
the same curved background cosmology and the same dark energy
model parameterization. We find that the degeneracies in the
geometric test, in particular the insensitivity to $w_a$, are
nicely orthogonal to all these other probes. Combining the
geometric test with the CMB, BAO or SNIa will yield accuracies of
a $\Delta w_0\approx 0.10$ and $\Delta w_a\approx 0.5$, and can be
compared for systematics. An optimal combination is a geometric
lensing test, with the Planck CMB and WFMOS BAO experiment,
yielding an expected accuracy of $\Delta w_0=0.047$ and $\Delta
w_a=0.11$.

Finally we discuss some of the potential systematic effects which
could affect the predicted accuracy of lensing.

In summary, the prospects of accurately measuring the dark energy
equation of state and its evolution to high accuracy over the next
decade are very good. The key to this is the gravitational lensing
geometric shear ratio test, which, due to its orthogonal
degeneracies, can be optimally combined with a large range of
other dark energy probes, such as the CMB, BAO or SNIa. In
addition, gravitational lensing can also be a probe via two-point
analysis, either from correlation functions or power spectra in
redshift-space (Heavens et al. 2006 and Castro et
al. 2005). Just as 
with lensing of the CMB, since the shear ratio analysis does not
contain any information on structure, we can expect there to be
little correlation between the two methods, even for the same
survey. However, the shear ratio covariance may be correlated with
the shear power. We shall explore combining these methods elsewhere.

\section*{Acknowledgements}
ANT was supported by a PPARC Advanced Fellowships for part of this
work, while DJB is a PPARC Advanced Fellow. TDK acknowledges a
PPARC studentship. We thank Lance Miller, David Goldberg, Bhuvnesh
Jain, John Peacock, Chris Wolf, Klaus Meisenheimer, Meghan Gray,
Konrad Kuijken and Eric Bell for useful discussion, and Masahiro
Takada for discussions concerning the CMB and BAO predictions. ANT thanks the
Max-Planck Institut fur Astronomie in Heidelberg for its
hospitality where parts of this paper was first discussed.

\onecolumn

\section*{Appendix A: Covariance of tangential cosmic shear}

The cosmic tangential shear covariance averaged over a circular
aperture is given by
\ba
C^{\gamma_t \gamma_t} (\theta) &=& \frac{1}{A^2} \int d^2\!\theta
\int\! d^2 \theta' \,\lgl \gamma_t(\thetab) \gamma_t(\thetab') \rgl
\nn
&=&\frac{1}{A^2} \int \!d^2\theta
\int \!d^2 \theta'\,
\left( \lgl \gamma_1(\thetab) \gamma_1(\thetab') \rgl
\cos 2 \varphi \cos 2 \varphi' +
\lgl \gamma_2(\thetab) \gamma_2(\thetab') \rgl
\sin 2 \varphi \sin 2 \varphi' \right)\nn
&=& \frac{1}{A^2} \int \!d^2\theta
\int \!d^2 \theta'\, \left[ \int \!\frac{d^2 \ell}{(2 \pi)^2}\,
  C^{\kappa \kappa}_\ell
  (\cos^2 2 \varphi_\ell \cos 2 \varphi \cos 2\varphi'+
  \sin^2 2 \varphi_\ell \sin 2 \varphi \sin 2 \varphi' )
  e^{i \lb . (\thetab-\thetab')}\right] \nn
&=&\int \!\frac{d^2 \ell}{(2 \pi)^2}\,
C^{\kappa \kappa}_\ell \cos^2 2 \varphi_\ell
\left( \frac{1}{A} \int \!d^2\theta
  e^{i \lb . \thetab} \cos 2 \varphi \right)^2 \nn
&=& \int_0^{\infty} \!
\frac{\ell d \ell}{ \pi} \, C^{\kappa \kappa}_{\ell}
\left\{ \frac{2[1-J_0(\ell \theta)]}{\ell^2 \theta^2}-
  \frac{J_1(\ell\theta)}{\ell \theta}\right\}^2
\ea
where $\cos \varphi_\ell = \lbh . \thetabhat_1$ and
$\thetabhat_1$ is the unit vector along one axis.

\section*{Appendix B: Bias in assumed parameters}
In this Appendix we show that for a Gaussian distributed
likelihood function, the linear bias in a parameter, which we
shall call $\delta \theta_i$, due to a bias in a fixed model
parameter (i.e., one whose value we have assumed and is not being
measured), which we shall call $\delta \psi_j$, is given by (e.g.,
Knox, Scoccimarro and Dodelson, 1998; Kim et al., 2004)
 \be
    \delta \theta_i = - [F^{\theta\theta}]^{-1}_{ik} F^{\theta
  \psi}_{kj} \delta \psi_j,
   \label{bias_fish}
 \ee
where $\F^{\theta \theta}$ is the parameter Fisher matrix defined as
\be 
F^{\theta\theta}_{ij}=\frac{1}{2}{\rm
  Tr}(C^{-1}\partial_i^{\theta}CC^{-1}\partial_j^{\theta}C
  +2\partial_{(i}^{\theta}\mu C^{-1}\partial_{j)}^{\theta}\mu^T)
\ee
and
$\F^{\theta \psi}$ is a column matrix (for one bias parameter) defined
as
\be
F^{\theta\psi}_{ij}=\frac{1}{2}{\rm
  Tr}(C^{-1}\partial_i^{\theta}CC^{-1}\partial_j^{\mu}C
  +\partial_{i}^{\theta}\mu C^{-1}\partial_{j}^{\psi}\mu^T
  +\partial_{i}^{\psi}\mu C^{-1}\partial_{j}^{\theta}\mu^T)
\ee
which we
will refer to as a pseudo-Fisher matrix between measured and assumed
parameters. 

We begin with a Likelihood function,
$       \ln L (\thetab|\psib),
$
which depends on a set of free parameters to be determined by the
data, $\thetab$, and a set of fixed parameters which we assume are
known, $\psib$. If the $\thetab$ are at their maximum likelihood
values, $\thetab_0$, then
\be
\lgl \de_i \ln L (\thetab_0|\psib) \rgl = 0
\ee
where the derivative is in parameter space, and we have ensemble
averaged over all possible data.

We now ask what is the effect of displacing the fixed parameters.
Expanding both $\psib$ and $\thetab$ to first-order we find
\be
\ln L (\thetab|\psib) = \ln L (\thetab_0|\psib_0) +
\delta \theta_i \de_i  \ln L (\thetab_0|\psib_0)
+ \delta \psi_j \de_{\psi,j} \ln L (\thetab_0|\psib_0),
\ee
were $\de_{\psi,i}$ is a derivative in the $\psib$-parameter
space. This displaced likelihood now maximizes when
\be
\lgl \de_i \ln L (\thetab|\psib) \rgl=
\lgl \de_i \ln L (\thetab_0|\psib_0) \rgl
+
\delta \theta_j \lgl\de_i \de_j  \ln L (\thetab_0|\psib_0)\rgl
+  \delta \psi_j \lgl\de_i \de_{\psi,j} \ln L (\thetab_0|\psib_0)\rgl
= 0.
\ee
We know that the unperturbed likelihood peaks at the maximum
likelihood values, and by inspection we can see that the averaged
second derivatives of the likelihood are the Fisher matrices.
Hence we see
\be
\delta \theta_j \lgl\de_i \de_j  \ln L (\thetab_0|\psib_0)\rgl
= -  \delta \psi_j \lgl\de_i \de_{\psi,j} \ln L (\thetab_0|\psib_0)\rgl
\ee
which with the definition of the Fisher matrices yields equation
(\ref{bias_fish}).

 \label{lastpage}
\end{document}